\pdfoutput=1

\documentclass{ucetd}

\usepackage{times}
\usepackage{latexsym}
\usepackage[T1]{fontenc}
\usepackage[utf8]{inputenc}
\usepackage{microtype}
\usepackage{inconsolata}

\usepackage{amsmath,amsthm,amssymb,amsfonts}
\usepackage{multirow, hhline, array, float, tabularx, enumitem}
\usepackage[square,sort,comma,numbers]{natbib}
\usepackage{url, hyperref}
\hypersetup{
  bookmarksnumbered,
  unicode,
  linktoc=all,
  pdftitle={Computational Modeling of Antibody-Antigen Complexes: PLM-Based and MSA-Based Approaches},
  pdfauthor={Xiao Luo},
  pdfborder={0 0 0}
}

\usepackage{bbm}
\usepackage{threeparttable}
\usepackage{booktabs, multirow, xspace}

\usepackage{graphicx}
\usepackage{palatino, booktabs}
\usepackage{tablefootnote}

\newcolumntype{P}[1]{>{\centering\arraybackslash}p{#1}}

\usepackage{algorithm}
\usepackage{algpseudocode}
\usepackage{adjustbox, subcaption}
\usepackage{xcolor}

\title{Computational Modeling of Antibody-Antigen Complexes:\\
PLM-Based and MSA-Based Approaches}
\author{Xiao Luo}
\department{Toyota Technological Institute at Chicago}
\degree{Doctor of Philosophy in Computer Science}
\date{June, 2026}

\committee{
Professor Jinbo Xu (Thesis Advisor) \\
Professor Avrim Blum \\
Professor Aly Azeem Khan
}

\begin{document}

\maketitle
\makecopyright

\tableofcontents
\listoffigures
\listoftables

\acknowledgments
I am profoundly grateful to my advisor, Professor Jinbo Xu, for giving me the intellectual freedom to pursue questions I found compelling, and for his patience and encouragement throughout these years. I have benefited enormously from his deep insight into computational biology and his instinct for what matters in a problem.

To my committee members, Professor Avrim Blum and Professor Aly Khan: thank you for your thoughtful questions and constructive feedback on this work.

I thank Fandi Wu and Xiaoyang Jin for our collaboration and their help over the years.

To my fellow students and friends at TTIC and the University of Chicago---Ben Lai, Jiading Fang, Lifu Tu, Lingyu Gao, Matt McPartlon, Mingda Chen, Qingming Tang, and Ziwei Xie---thank you for the many helpful discussions, advice, and friendship that made graduate school a community rather than a solitary pursuit.

Finally, my deepest thanks go to my parents, Yongjun Luo and Yu Gao, for their unconditional love and unwavering support; everything I have achieved rests on the foundation they built.

\abstract
Antibodies are a specialized class of proteins that play a central role in the immune response by specifically recognizing and neutralizing antigens. Therapeutic antibodies have become major drugs for cancer and autoimmune diseases, yet their discovery still depends on extensive in vitro screening. Accurate computational modeling of antibody structures and their interactions with antigens can prioritize candidates, reduce experimental burden, and accelerate the rational design of therapeutic antibodies.

Recent advances in high-accuracy protein and complex prediction have transformed computational protein design. However, a persistent performance gap remains for antibodies and antibody-antigen complexes compared with general protein-protein interactions, limiting downstream design. This thesis investigates why antibody-related tasks are harder and proposes improvements to general-purpose algorithms that better capture antibody-specific constraints and interaction patterns. We further analyze model behavior to provide insights into how these algorithms work and to suggest future research directions.

We first investigate protein language model (PLM)-based methods for antibody and antibody-antigen structure prediction. Using embeddings from multiple PLMs to enrich structural patterns, our PLM-based approach achieves the best CDR-H3 accuracy among the compared PLM-based methods on antibody monomer prediction. Extending the same approach to antibody-antigen complex prediction, however, does not generalize: without co-evolutionary signals between antibody and antigen, single-sequence PLM representations do not reliably identify binding interfaces, motivating a return to MSA-based approaches.

We then develop two MSA-based interventions for antibody-antigen complex prediction: MSA refinement, which combines CDR-focused filtering with depth recovery from a larger sequence database, and convergence-aware recycling, which selects a stable intermediate recycle state for final diffusion sampling. Together, these interventions provide consistent gains over the AlphaFold3 baseline on a held-out antibody-antigen test set, at both acceptable- and medium-quality DockQ thresholds. Because the methods modify MSA construction and recycling behavior rather than model parameters, they can be applied without retraining or weight access whenever these stages are controllable.

Together, this study clarifies the capabilities and limitations of current methods for antibody-antigen structure prediction and offers practical guidance for future model development. The proposed methods improve accuracy and reliability, integrate easily into existing modeling pipelines, and help reduce experimental screening to accelerate therapeutic antibody design and engineering.

\mainmatter

\chapter{Introduction}
\label{ch:introduction}

Antibodies, or immunoglobulins, play a vital role in the immune response by specifically recognizing and neutralizing antigens~\cite{schroeder2010structure}. They have a Y-shaped structure with two heavy chains and two light chains. The specific ability of antibodies to target distinct pathogens comes from six highly variable loop regions, called complementarity-determining regions (CDRs). Their ability to target specific pathogens makes antibodies essential for diagnostic and therapeutic applications. Accurate structures of antibodies and antibody-antigen complexes offer helpful information about immune recognition and aid in the design of vaccines and therapies; they also inform assessments of developability and engineering decisions~\cite{raybould2019five} and guide analyses of epitope-paratope relationships relevant to docking~\cite{krawczyk2014improving}.

Computational antibody design follows a pipeline of sequence generation, structure prediction, binding assessment, and experimental validation~\citep{norman2020computational,kim2023computational}. Within this workflow, accurate structure prediction serves as a critical filtering step: it provides geometric assessment of antibody-antigen complementarity to identify promising candidates, localizes interface residues to guide targeted optimization, and enables scalable in silico screening before costly experimental validation~\citep{hummer2022advances,mason2021optimization}. The methods developed in this thesis advance the structure prediction capabilities that underlie these design applications.

Recent deep learning methods have boosted the performance of general protein structure prediction~\cite{jumper2021highly,lin2023evolutionary,baek2021accurate}; however, antibody modeling remains underdeveloped, providing room for methodological enhancements~\cite{yin2024evaluation,yin2022benchmarking}.

\section{Current Approaches in Structure Prediction}

There are two main strategies used in protein structure prediction. The MSA-based method tries to extract co-evolutionary signals from multiple sequence alignments to learn the potential spatial contact of residue pairs. Early advances, such as RaptorX~\cite{wang2017accurate} and trRosetta~\cite{yang2020trrosetta}, show that deep learning methods are capable of efficient learning from the MSA to enhance protein folding predictions. Further developments in algorithms, like AlphaFold2~\cite{jumper2021highly} (building on AlphaFold1~\cite{senior2020improved}) and RoseTTAFold~\cite{baek2021accurate}, utilized deeper transformer-based neural networks and increased data to further improve the accuracy. These techniques are now widely used in structural biology~\cite{akdel2022structural}, with ColabFold making AlphaFold-style pipelines broadly accessible~\cite{mirdita2022colabfold}, though caution is required for predicting variant effects~\cite{buel2022can}. Recently, diffusion-based methods like AlphaFold3 have generalized the structure prediction ability to other types of biomolecules, including DNA and RNA~\cite{abramson2024accurate}. These approaches have reached nearly experimental accuracy for many proteins, as validated by CASP assessments~\cite{kryshtafovych2021critical,moult1995large}, but require lots of computational resources for MSA creation and processing.

The success of the MSA-based method provides a large amount of high-quality predicted structures for the known sequences that lack a resolved structure. This makes the learning of sequence-structure relations without MSAs possible. Consequently, PLM-based methods predict protein structure directly from the primary sequence using large-scale protein language models (PLMs)~\cite{elnaggar2021prottrans,lin2023evolutionary,jing2024single}. Benchmarks such as TAPE~\cite{rao2019evaluating} and analyses of Transformer attention maps~\cite{rao2020transformer} demonstrated that PLM representations encode structural constraints even without MSAs. In addition to predicting structure, generative PLMs have also shown the capability of sequence design~\cite{madani2023large,ferruz2022protgpt2}. For this type of method, the evolutionary information is embedded in model parameters during training, removing the need for explicit MSA input while maintaining competitive accuracy with faster inference and lower computational cost.

\section{Challenges in Antibody-Antigen Modeling}

Despite the overall success in protein structure prediction, a notable performance gap exists between general proteins and antibody-related tasks. In our study, we focus on tackling specific challenges in the prediction of antibody-antigen complexes:

\begin{enumerate}

  \item \textbf{Limited evolutionary information}: General proteins that evolve through gradual mutation provide rich evolutionary information in the sequence database. Antibodies are created through V(D)J recombination and somatic hypermutation~\cite{chi2020v,litman1993,muramatsu2000,victora2012germinal,roth2015v}. This process makes the MSA of the antibody less informative, with different co-evolutionary patterns that current methods struggle to use efficiently, especially given the immense repertoire diversity observed across individuals~\cite{miho2018computational}.

  \item \textbf{Underrepresentation in training data}: Among proteins with resolved structures recorded in PDB (Protein Data Bank), fewer than $5\%$ of structures are antibody-related. This results in little exposure of the model to immune recognition patterns during training~\cite{wwpdb2019protein,dunbar2014sabdab}.

\item \textbf{Unique structural characteristics}: The hypervariable CDR regions, especially CDR H3 for the heavy chain, exhibit extreme diversity in sequence and structure, which challenges the model's ability to learn the precise sequence-structure relation~\cite{chen2023h3,ruffolo2023fast,jing2024accurate,regep2017h3,north2011new}.

\end{enumerate}

These challenges emphasize that antibody-antigen modeling remains a field for further improvement.

\section{Our Approach}

We organize our approach into two complementary directions. First, we explore PLM-based methods because they require less computation and allow rapid assessment of potential and limitations for antibody modeling. Second, we investigate MSA-based methods, with a focus on training-free refinements (e.g., CDR-focused MSA filtering, convergence-aware recycling) that improve prediction quality without retraining or direct modification of model weights.

We explore the following areas:

\begin{enumerate}

    \item \textbf{PLM-based modeling (monomer and complex)}: We integrate PLM models to enhance sequence-based predictors without MSAs. For monomers, we assess antibody-specific adaptation (e.g., improved CDR H3). For complexes, we extend PLM-based prediction to antibody-antigen docking, analyzing Blind (no epitope prior) and Epitope-Guided (binding-site prior) settings.

    \item \textbf{MSA-based methods for antibody-antigen complexes}: To address the performance gap with general protein complexes, we propose training-free refinements (CDR-focused MSA filtering, MSA depth recovery, and convergence-aware recycling) for modern diffusion-style architectures, providing improvements when input generation and inference settings can be controlled.

\end{enumerate}

We aim to improve computational tools for modeling antibody-antigen interactions, which may aid therapeutic development and understanding of diseases.

Through these explorations, we find that PLM-based models can predict antibody structures fairly well but have difficulties with antigen interactions~\cite{yin2024evaluation,hitawala2025does,gaudreault2023enhanced,harmalkar2025reliable}, while the integration of binding-site knowledge boosts performance, suggesting an appealing direction to integrate experimental information, such as cross-linking or alanine scanning, to help correct complex modeling. For MSA-based approaches, our research shows that the differences between antibody-antigen and general protein-protein complex predictions—such as the unique features of hypervariable CDR regions and the absence of co-evolutionary signals—may guide targeted improvements. 

\section{Thesis Organization}

The chapters and appendices are organized as follows:

\begin{itemize}

    \item Chapter~\ref{ch:background} covers the biological and computational background, including antibody structure and function, protein structure prediction methods, deep learning approaches, datasets, and evaluation metrics.

    \item Chapter~\ref{ch:plm_modeling} investigates protein language model-based approaches to antibody modeling at two levels: monomer prediction (Section~\ref{sec:monomer_prediction}), where PLM representations approach MSA-based accuracy, and antibody-antigen complex prediction (Section~\ref{sec:antibody_antigen_complex}), where PLMs are limited by the absence of co-evolutionary signals between antibody and antigen.

    \item Chapter~\ref{ch:pipeline_adaptation} develops two MSA-based interventions for antibody-antigen complex prediction: MSA refinement (CDR-focused filtering combined with depth recovery via the BFD database) and convergence-aware recycling, implemented and evaluated with AlphaFold3. These methods provide improvements on antibody-antigen complexes without retraining or direct modification of model weights.

    \item Chapter~\ref{ch:conclusion} summarizes our contributions and discusses future directions.

    \item Appendix~\ref{app:plm_details} provides implementation details for the PLM-based monomer and complex predictors in Chapter~\ref{ch:plm_modeling}.

    \item Appendix~\ref{app:lora} documents LoRA fine-tuning of AlphaFold2-Multimer as a model-level adaptation baseline for comparison with Chapter~\ref{ch:pipeline_adaptation}.

    \item Appendix~\ref{app:pipeline_details} provides evaluation protocols, validation ablations, algorithm pseudocode, and evaluation-convention notes for Chapter~\ref{ch:pipeline_adaptation}.

\end{itemize}

\chapter{Background}
\label{ch:background}

This chapter covers the background for computational antibody-antigen structure prediction. We introduce antibody biology and why antibodies pose unique challenges for structure prediction, machine learning methods for protein modeling, and evaluation metrics for assessing prediction quality.

Section~\ref{sec:bg_biology} covers antibody structure, CDRs, numbering schemes, and the somatic mechanisms that generate antibody diversity. Section~\ref{sec:bg_ml} gives an overview of MSA-based and PLM-based methods for structure prediction. Section~\ref{sec:bg_evaluation} defines the datasets and metrics used throughout this thesis.

\section{Antibody Biology}
\label{sec:bg_biology}

\subsection{Antibody Structure}

\begin{figure}[htbp]
\centering
\includegraphics[width=0.85\textwidth]{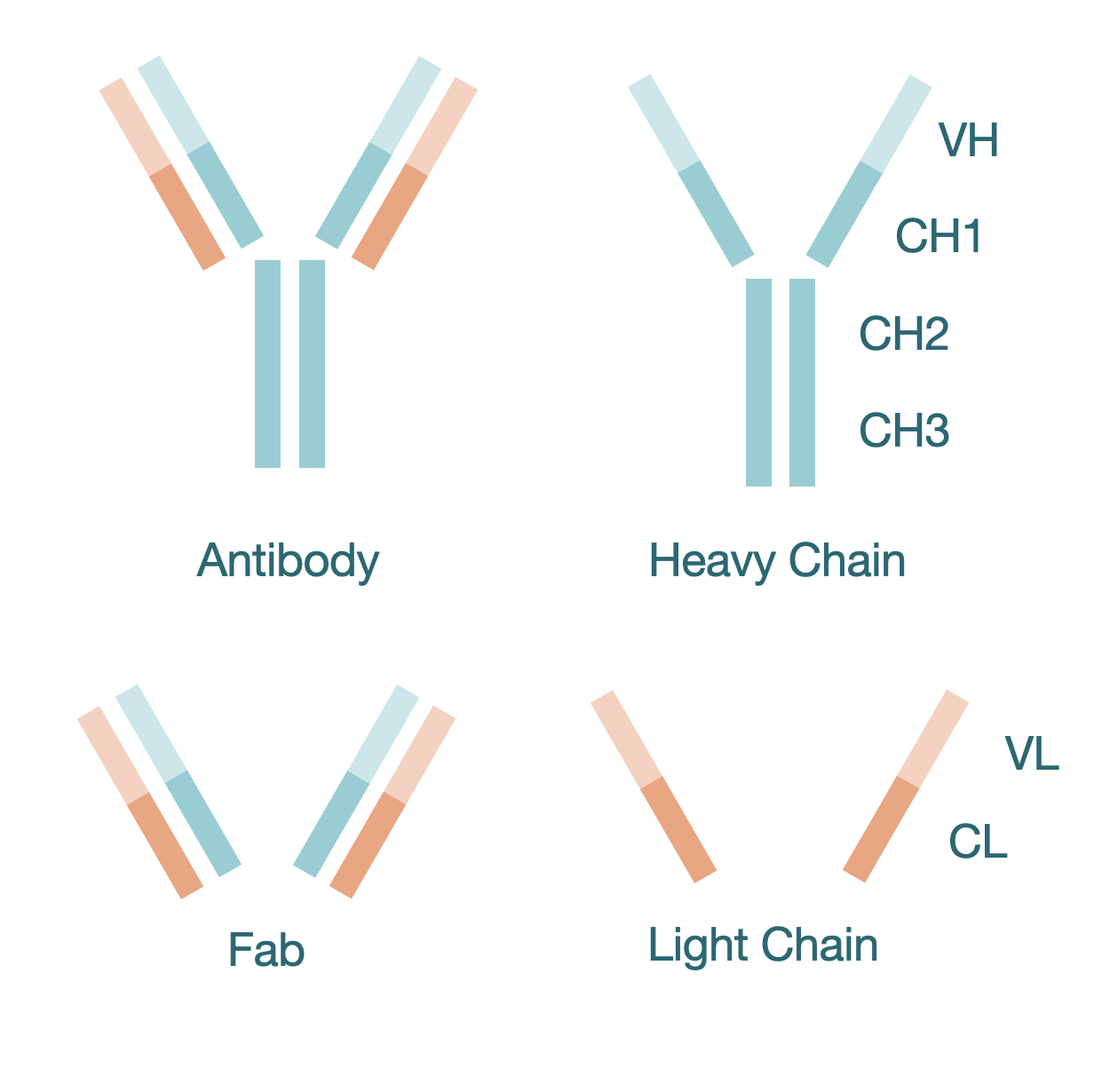}
\caption{Schematic representation of antibody domain organization. (Top left) The complete antibody consists of two heavy chains and two light chains arranged in a Y-shaped configuration. (Top right) The heavy chain contains one variable domain (VH) followed by three constant domains (CH1, CH2, and CH3). (Bottom left) The Fab fragment comprises the VH-CH1 portion of the heavy chain paired with the complete light chain. (Bottom right) The light chain consists of one variable domain (VL) and one constant domain (CL). The variable domains (VH and VL) together form the antigen-binding site.}
\label{fig:antibody_schematic}
\end{figure}

\begin{figure}[htbp]
\centering
\includegraphics[width=0.7\textwidth]{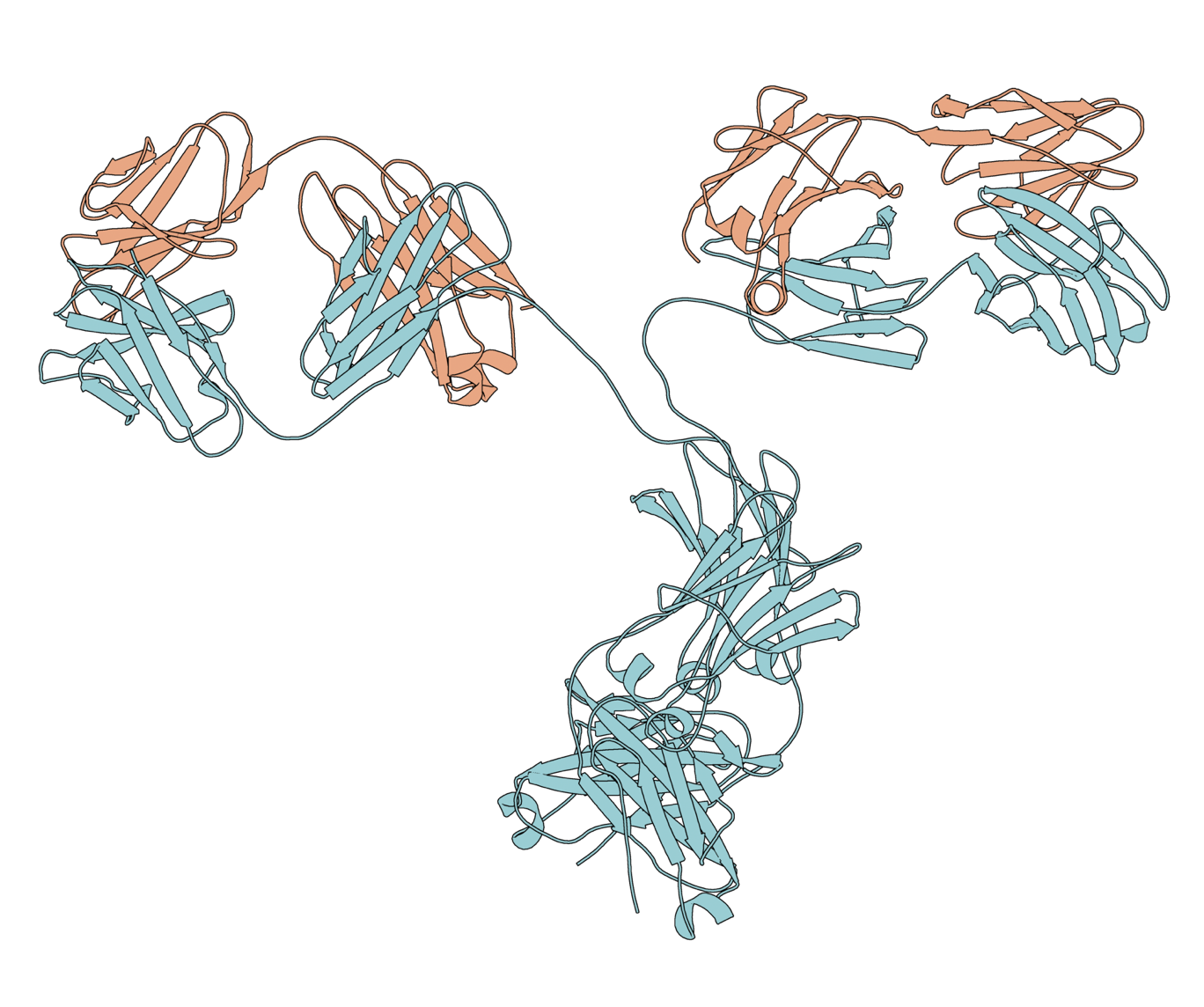}
\caption{Three-dimensional ribbon structure of an IgG antibody showing the characteristic Y-shaped architecture. The antibody consists of two identical heavy chains (cyan) and two identical light chains (orange). The two Fab (fragment antigen-binding) regions in the upper arms contain the variable domains that form antigen-binding sites, while the Fc (fragment crystallizable) region in the stem mediates effector functions. The flexible hinge region connects the Fab arms to the Fc stem.}
\label{fig:antibody_structure}
\end{figure}

Antibodies are Y-shaped glycoproteins produced by B cells that recognize and neutralize foreign molecules (antigens)~\citep{schroeder2010structure}. The canonical IgG antibody ($\sim$150~kDa) consists of four polypeptide chains: two identical heavy chains ($\sim$50~kDa each) and two identical light chains ($\sim$25~kDa each), linked by disulfide bonds (Figures~\ref{fig:antibody_structure} and~\ref{fig:antibody_schematic})~\citep{alzari1988three}.

Each chain contains immunoglobulin domains ($\sim$110 residues each). Heavy chains have one variable domain (VH) and three constant domains (CH1, CH2, and CH3); light chains have one variable domain (VL) and one constant domain (CL). The variable domains form the antigen-binding site, while constant domains mediate effector functions.

\subsection{Complementarity-Determining Regions}

\begin{figure}[htbp]
\centering
\begin{subfigure}[b]{0.48\textwidth}
    \centering
    \includegraphics[width=\textwidth]{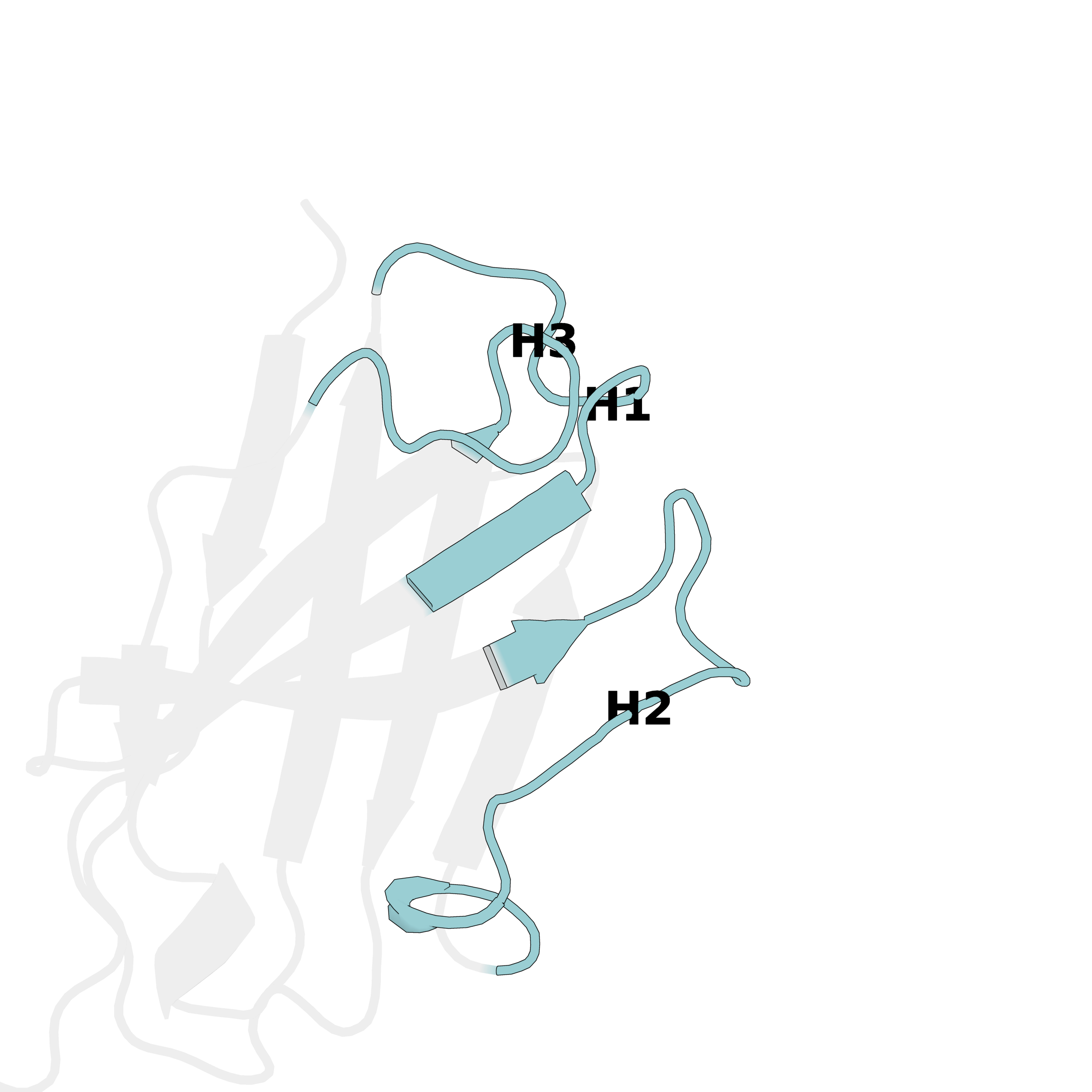}
    \caption{Heavy chain variable domain (VH)}
    \label{fig:heavy_chain_cdr}
\end{subfigure}
\hfill
\begin{subfigure}[b]{0.48\textwidth}
    \centering
    \includegraphics[width=\textwidth]{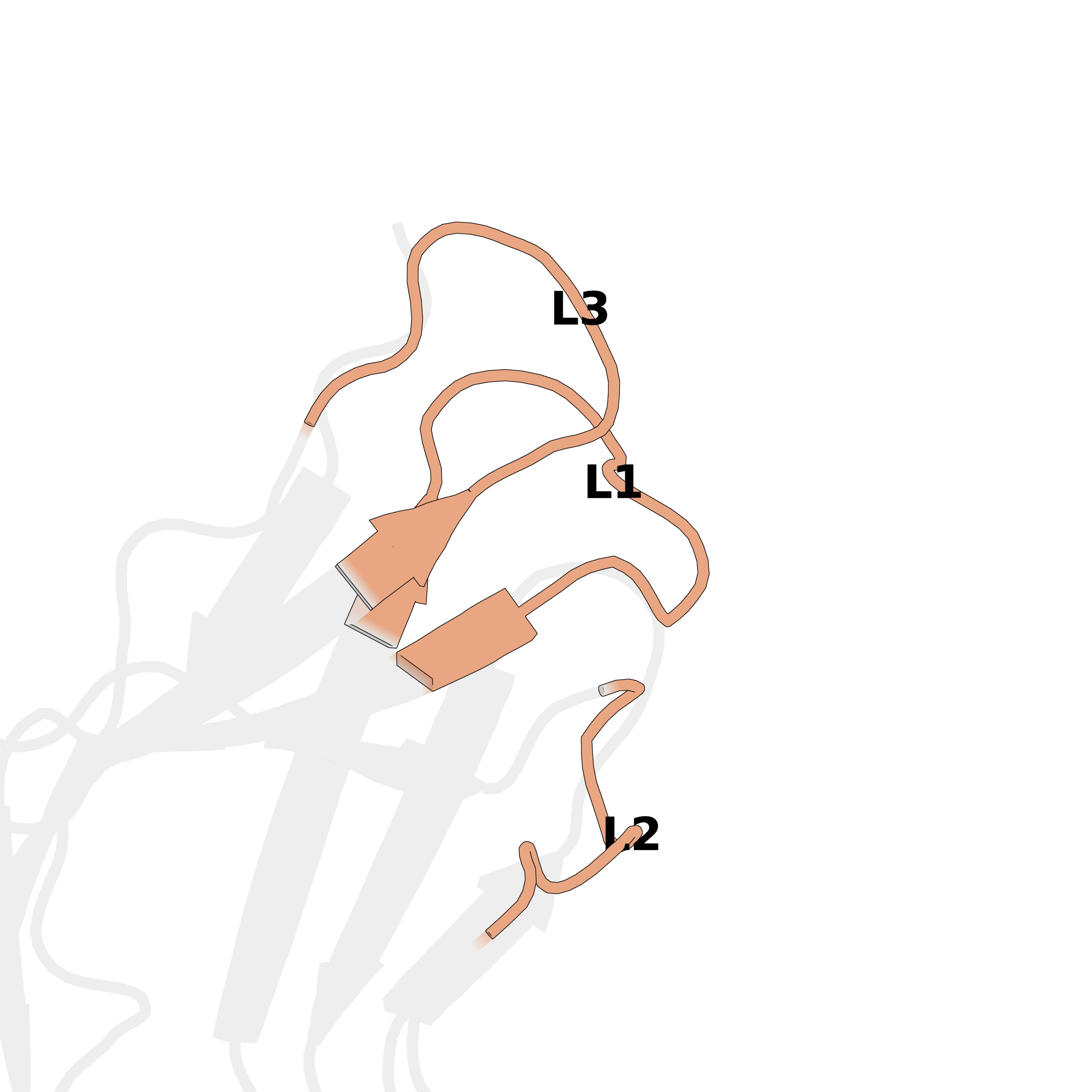}
    \caption{Light chain variable domain (VL)}
    \label{fig:light_chain_cdr}
\end{subfigure}
\caption{Variable domain structures showing CDR loops (colored) extending from the conserved framework scaffold. (a) Heavy chain with CDR-H1, CDR-H2, and CDR-H3. CDR-H3, located at the binding site center, shows the highest diversity (typically 7--26 residues). (b) Light chain with CDR-L1, CDR-L2, and CDR-L3. Light chain CDRs show lower diversity and more frequently adopt canonical conformations~\citep{chothia1987canonical,north2011new}.}
\label{fig:cdr_structures}
\end{figure}

\begin{figure}[htbp]
\centering
\includegraphics[width=0.55\textwidth]{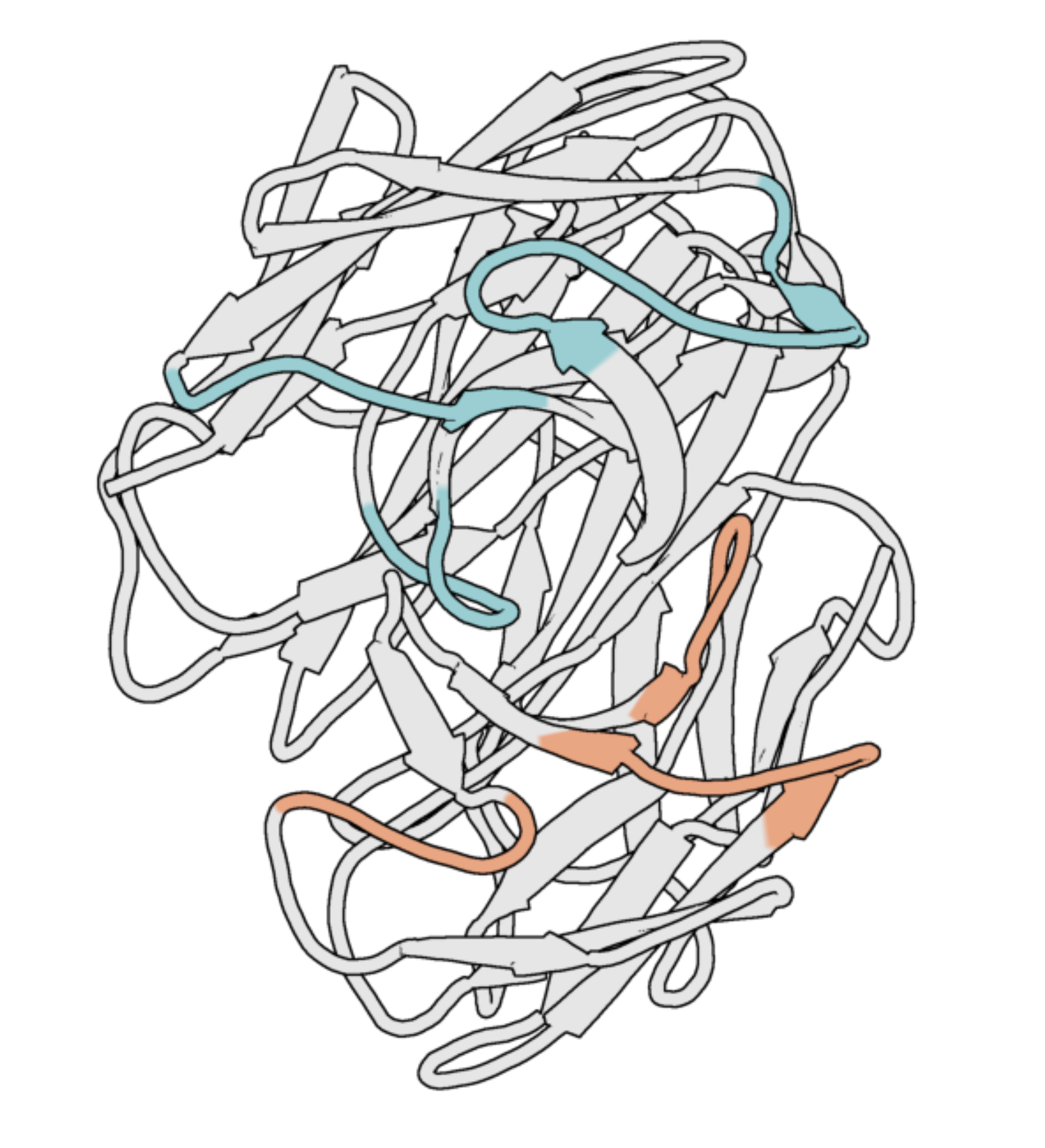}
\caption{Side view of the Fv region showing the spatial clustering of CDR loops. Heavy chain CDRs (cyan) and light chain CDRs (orange) extend from the conserved framework (gray) and converge to form the antigen-binding surface (paratope).}
\label{fig:cdr_sideview}
\end{figure}

Within each variable domain, there are six hypervariable loops called \emph{Complementarity-Determining Regions} (CDRs): three per chain (H1, H2, H3 for heavy; L1, L2, L3 for light). These loops are supported by conserved $\beta$-strand scaffolds called \emph{framework regions} (Figures~\ref{fig:heavy_chain_cdr} and~\ref{fig:light_chain_cdr})~\citep{chothia1989conformations}. The six CDRs cluster spatially to form the \emph{paratope}---the surface that binds the antigen's \emph{epitope} (Figure~\ref{fig:cdr_sideview}).

\textbf{CDR-H3} is the most variable and challenging region to model in antibodies.
It shows diversity in both length and sequence and often plays a dominant role in binding specificity.
This exceptional variability largely arises from its location at the V(D)J junction, making accurate CDR-H3 conformation prediction a central challenge for computational methods~\citep{weitzner2015origin,regep2017h3}.
Across antibody-antigen complexes, CDR-H3 frequently makes a large contribution to the paratope buried surface area~\citep{ramaraj2012antigen}.

\subsection{Antibody Numbering Schemes}

Accurate annotation of antibody sequences requires standardized numbering schemes that align residue positions across diverse antibodies despite insertions and deletions, particularly in CDR regions. Three major schemes---Kabat~\citep{kabat1992sequences}, Chothia~\citep{chothia1987canonical,al1997standard}, and IMGT~\citep{lefranc2003imgt}---have been developed, each with distinct design principles and CDR boundary definitions.

\textbf{Kabat numbering}~\citep{kabat1992sequences} was the first widely adopted scheme, developed from sequence analysis of antibody variable regions. Kabat numbering assigns position numbers based on sequence variability: positions with high variability across known antibodies define CDR regions. Insertions are handled by appending letters (e.g., 100A, 100B). While historically important, Kabat's sequence-based definitions can misalign structurally equivalent positions when loop lengths vary widely.

\textbf{Chothia numbering}~\citep{chothia1987canonical,al1997standard} refined the Kabat scheme by incorporating structural information. Chothia and Lesk analyzed antibody crystal structures and defined CDR boundaries based on loop conformations rather than sequence variability alone. This structure-based approach better captures the residues that form the antigen-binding loops. Chothia numbering uses the same position numbers as Kabat for most residues but differs in CDR boundary definitions, particularly for CDR-L1 and CDR-H1.

\textbf{IMGT numbering}~\citep{lefranc2003imgt} provides a standardized scheme designed for cross-species comparison of immunoglobulin and T cell receptor sequences. IMGT assigns unique position numbers from 1 to 128 for variable domains, with gaps inserted at fixed positions to maintain structural alignment. CDRs are defined at positions 27--38 (CDR1), 56--65 (CDR2), and 105--117 (CDR3) for both heavy and light chains. The consistent gap placement enables unambiguous alignment across highly divergent sequences, making IMGT particularly suitable for computational analyses.

Table~\ref{tab:numbering_schemes} summarizes the CDR boundary definitions for each scheme. Throughout this thesis, we use IMGT numbering for its consistent alignment properties and widespread adoption in computational antibody analysis~\citep{dunbar2016anarci}.

\begin{table}[htbp]
\centering
\caption{CDR boundary definitions under different numbering schemes~\citep{zhu202450}. Position ranges indicate the residue numbers that define each CDR loop.}
\label{tab:numbering_schemes}
\begin{tabular}{lccc}
\toprule
Region & Kabat & Chothia & IMGT \\
\midrule
CDR-H1 & 31--35 & 26--32 & 27--38 \\
CDR-H2 & 50--65 & 52--56 & 56--65 \\
CDR-H3 & 95--102 & 95--102 & 105--117 \\
CDR-L1 & 24--34 & 24--34 & 27--38 \\
CDR-L2 & 50--56 & 50--56 & 56--65 \\
CDR-L3 & 89--97 & 89--97 & 105--117 \\
\bottomrule
\end{tabular}
\end{table}

\subsection{Generation of Antibody Diversity}

Antibody diversity arises through somatic mechanisms rather than germline inheritance:

\textbf{V(D)J Recombination.} During B cell development, gene segments (Variable, Diversity, Joining) combine through random selection and imprecise joining~\cite{hozumi1976evidence}. CDR-H3 spans the V-D-J junction, explaining its extreme hypervariability.

\textbf{Somatic Hypermutation.} After antigen encounter, point mutations accumulate at high rates (~$10^{-3}$ per base per division) in immunoglobulin variable regions via an AID-initiated mutagenic process~\citep{muramatsu2000,odegard2006targeting}, with beneficial mutations selected through affinity maturation~\citep{victora2012germinal}. Large-scale repertoire sequencing resources such as OAS catalog this diversity across individuals and species~\citep{kovaltsuk2018observed}, and curated compendia like CoV-AbDab summarize coronavirus-targeting antibodies and nanobodies~\citep{raybould2021}.

\subsection{Implications for Computational Modeling}
\label{sec:bg_implications}

The somatic origin of antibodies has direct consequences for sequence-based structure prediction. Because each antibody arises through V(D)J recombination and somatic hypermutation rather than gradual evolution, standard homology searches recover abundant framework matches but few CDR homologs, leaving CDR-H3 regions with sparse co-evolutionary signal even when the framework MSA is deep. A second, complex-specific consequence is that antibodies and antigens evolve in competition rather than cooperatively, so the cooperative cross-chain signals exploited by general protein-protein structure predictors are not reliable for antibody-antigen interfaces. These two effects---sparse intra-chain signal at CDRs and absent cooperative cross-chain signal---frame the technical problems addressed in Chapters~\ref{ch:plm_modeling} and~\ref{ch:pipeline_adaptation}.

\section{Machine Learning for Structure Prediction}
\label{sec:bg_ml}

This section provides a high-level overview of machine learning approaches for protein structure prediction. We first introduce foundational deep learning concepts, then describe the two main approaches for structure prediction, and discuss their shared architectural components. Detailed implementations are presented in subsequent chapters where they are applied.

\subsection{Deep Learning Foundations}

Modern structure prediction methods build on several key deep learning concepts.

\textbf{Transformer architecture}~\citep{vaswani2017attention} has become the foundation of modern deep learning for sequential data. The key innovation is the self-attention mechanism, which enables models to relate all positions in a sequence in parallel, capturing long-range dependencies without the sequential constraints of recurrent networks~\citep{hochreiter1997long}. Given input representations, attention computes query, key, and value projections, then uses scaled dot-product attention to weight contributions from all positions. Modern Transformer blocks typically combine multi-head attention with residual connections~\citep{he2016deep}, Layer Normalization~\citep{ba2016layer}, and position-wise feed-forward networks for stable optimization.

\textbf{Self-supervised learning} enables models to learn meaningful representations from unlabeled data by predicting parts of the input from other parts~\citep{liu2021self}. Masked language modeling, introduced by BERT~\citep{devlin2019bert}, randomly masks tokens and trains the model to predict them from context. This approach has been successfully adapted for proteins, where amino acids serve as tokens and models learn to predict masked residues from their sequence context~\citep{rives2021biological,elnaggar2021prottrans}. Through this pretraining, models capture evolutionary and structural patterns without requiring explicit structure labels.

\textbf{Representation learning} aims to encode inputs into embeddings that capture relevant features for downstream tasks~\citep{bengio2013representation}. For proteins, learned representations can encode evolutionary conservation, structural constraints, and functional properties. The quality of these representations---how well they capture the underlying biology---determines the effectiveness of downstream structure prediction modules.

\subsection{Two Approaches: MSA-Based vs. PLM-Based}

\begin{figure}[htbp]
\centering
\includegraphics[width=0.95\textwidth]{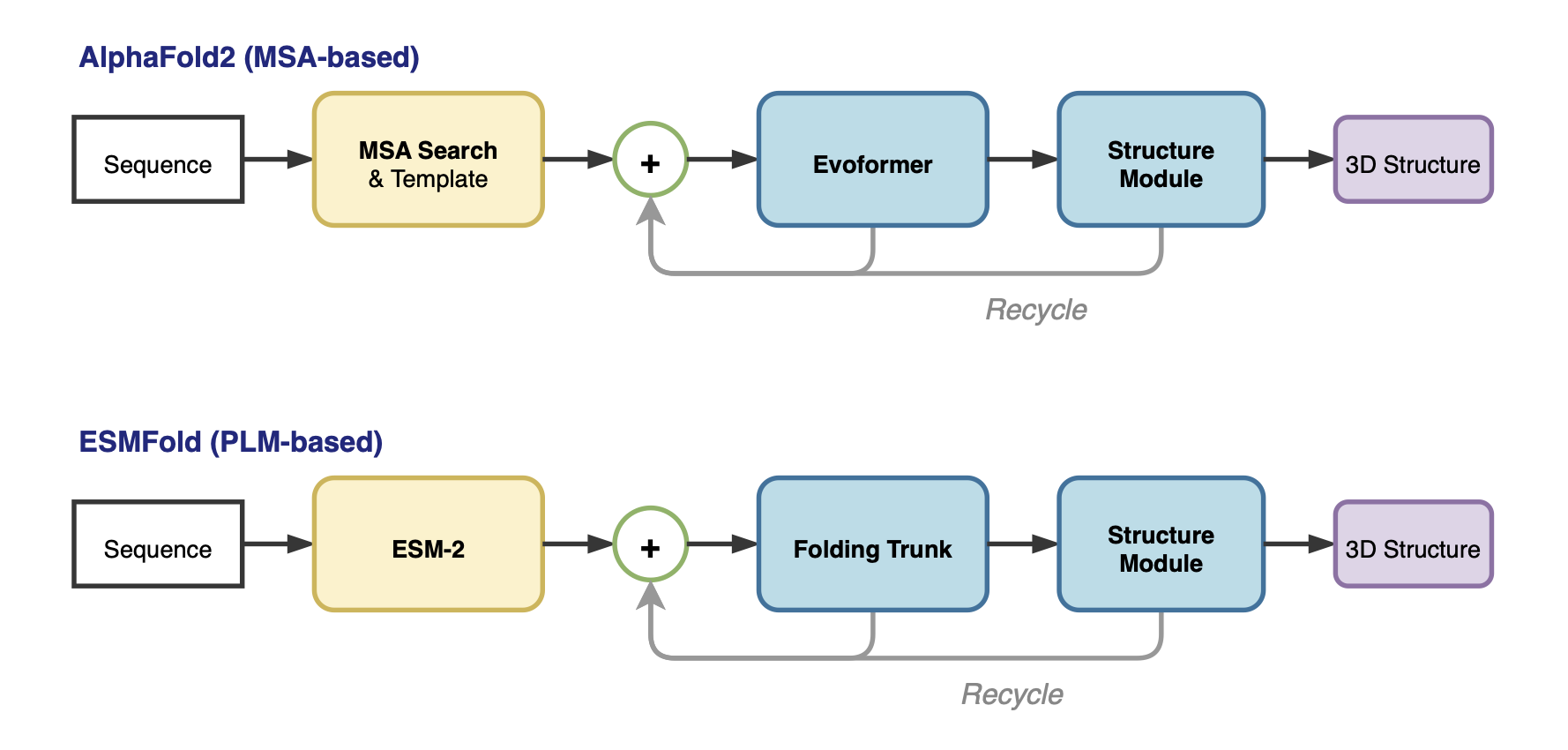}
\caption{Comparison of MSA-based (AlphaFold2) and PLM-based (ESMFold) structure prediction pipelines. \textbf{Top:} AlphaFold2 searches sequence databases to construct MSAs and structural templates, then processes these through the Evoformer to extract co-evolutionary patterns before the Structure Module generates 3D coordinates. \textbf{Bottom:} ESMFold uses the pretrained ESM-2 language model to encode single sequences directly, then processes representations through the Folding Trunk before the Structure Module. Both approaches share key components: the Structure Module for coordinate generation and iterative recycling for refinement. The main difference is how evolutionary information is obtained---explicit database search versus implicit encoding in pretrained model parameters.}
\label{fig:paradigm_comparison}
\end{figure}

There are two main approaches to structure prediction, based on how they use evolutionary information (Figure~\ref{fig:paradigm_comparison}):

\textbf{MSA-based methods} construct multiple sequence alignments (MSAs) by searching sequence databases for homologous proteins. Co-evolutionary patterns in MSAs---where spatially proximate residues mutate together to maintain structural contacts---provide powerful constraints for structure prediction~\citep{morcos2011direct,marks2011protein}. AlphaFold2~\citep{jumper2021highly} exemplifies this approach, achieving near-experimental accuracy on proteins with deep MSAs. However, MSA construction is computationally expensive (requiring minutes to hours of database search) and the approach struggles when homologs are sparse or uninformative.

\textbf{PLM-based methods} predict structure directly from single sequences using protein language models (PLMs) pretrained on large sequence databases~\citep{rives2021biological,elnaggar2021prottrans}. By encoding evolutionary patterns in model parameters through self-supervised learning, these methods eliminate MSA construction while maintaining competitive accuracy~\citep{lin2023evolutionary,wu2022high}. ESMFold~\citep{lin2023evolutionary} and OmegaFold~\citep{wu2022high} demonstrate that PLM embeddings can replace explicit MSAs for many proteins, enabling inference in seconds rather than minutes. Section~\ref{sec:monomer_prediction} provides detailed analysis of PLM architectures and their application to antibody structure prediction.

\subsection{Shared Architectural Components}

As illustrated in Figure~\ref{fig:paradigm_comparison}, despite using different approaches to get evolutionary information, both types share key architectural components for converting sequence representations to 3D structures.

\textbf{Representation processing modules.} Both approaches use specialized modules to process sequence information into pairwise residue representations. In AlphaFold2, the Evoformer module~\citep{jumper2021highly} jointly processes MSA and pair representations through 48 stacked blocks, each containing an MSA track (with row-wise and column-wise attention) and a pair track (with triangle multiplicative updates and triangle attention). In ESMFold, the Folding Trunk serves a similar role, processing ESM-2 embeddings into pair representations through a simpler architecture that operates on single sequences.

\textbf{Structure Module.} Both approaches convert pairwise representations to 3D coordinates using Structure Modules with invariant point attention (IPA)~\citep{jumper2021highly}. IPA respects SE(3) equivariance---predictions remain consistent under rotations and translations of the input frame. The module operates on local frames attached to each residue and updates these frames iteratively, allowing the network to reason about geometry in a coordinate-independent manner. This design draws on principles from equivariant neural networks~\citep{fuchs2020se,satorras2021n,jing2021equivariant}.

\textbf{Iterative recycling.} As shown in Figure~\ref{fig:paradigm_comparison}, both approaches apply recycling, where intermediate structure predictions are fed back into the network for refinement. This iterative process allows information from earlier predictions to guide subsequent updates, improving both local geometry and global arrangement. For antibody-antigen complexes, weaker inter-chain constraints can make recycling less stable, motivating convergence-aware stopping and intermediate-state selection. Chapter~\ref{ch:pipeline_adaptation} develops a convergence-aware recycling strategy for this purpose.

\subsection{Diffusion-Based Structure Prediction}

Recent advances have introduced diffusion models as an alternative approach for generating 3D structures. AlphaFold3~\citep{abramson2024accurate} replaces AlphaFold2's deterministic structure module with a diffusion-based approach: starting from noised coordinates, the model iteratively denoises them through a learned reverse-diffusion process to produce valid atomic conformations.

Diffusion models have several advantages. First, they naturally handle uncertainty by generating multiple diverse samples rather than a single prediction. Second, they can handle diverse biomolecular complexes including proteins, nucleic acids, small molecules, and ions within a unified architecture. Third, the generative formulation may better capture conformational flexibility and induced-fit binding.

For antibody-antigen complexes, diffusion models offer potential advantages in sampling diverse binding modes and exploring conformational changes upon binding. Chapter~\ref{ch:pipeline_adaptation} develops refinements to MSA construction and recycling behavior in modern MSA-based predictors, implemented here in an AlphaFold3-style setting.

\subsection{Relevance to Antibodies}

The MSA-based and PLM-based paradigms trade off differently for antibody modeling. PLM-based methods avoid the per-target MSA construction cost, which matters when antibody engineering workflows evaluate thousands of candidates. MSA-based methods retain access to evolutionary alignments that single-sequence embeddings cannot recover, but their iterative recycling can be less stable when MSAs are sparse, as is typical for antibody-antigen targets. The chapters that follow exploit both observations: Chapter~\ref{ch:plm_modeling} develops PLM-based approaches for antibody monomer and complex prediction, while Chapter~\ref{ch:pipeline_adaptation} develops training-free improvements to MSA construction and recycling behavior in modern MSA-based predictors.

\section{Datasets and Evaluation}
\label{sec:bg_evaluation}

\subsection{Structural Databases}

\textbf{Protein Data Bank (PDB)} is the primary repository for experimentally determined protein structures. As of 2024, the PDB contains over 200,000 structures, predominantly from X-ray crystallography (approximately 85\%), cryo-electron microscopy (approximately 10\%), and NMR spectroscopy (approximately 5\%). Each entry includes atomic coordinates, experimental metadata (resolution, R-factors), and chain information. We also reference the AlphaFold Protein Structure Database as a complementary resource that provides predicted structures covering large swaths of sequence space~\citep{varadi2022alphafold}.

\textbf{SAbDab (Structural Antibody Database)}~\citep{dunbar2014sabdab} curates antibody structures from the PDB. It annotates chains as heavy, light, or antigen; assigns CDR boundaries using standard numbering schemes (Kabat, Chothia, IMGT)~\citep{kabat1992sequences,chothia1989conformations}; and provides germline gene assignments. SAbDab covers several thousand antibody structures and a large subset of antibody-antigen complexes, with the active counts continuing to grow as PDB releases accumulate.

Throughout this thesis, we use \textbf{IMGT numbering}~\citep{lefranc2003imgt} to define CDR boundaries, as it provides consistent alignment of structurally conserved positions across diverse antibodies.

\subsection{Evaluation Metrics for Monomers}

We use three metrics for monomer structure assessment.

\begin{figure}[htbp]
\centering
\includegraphics[width=0.9\textwidth]{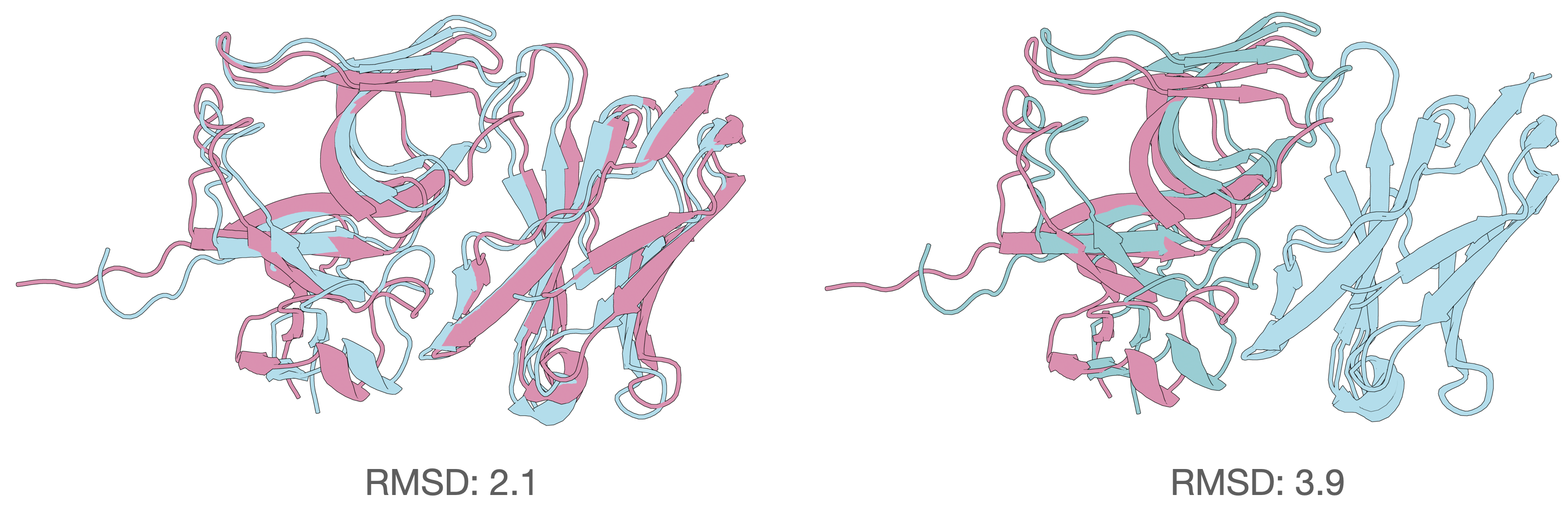}
\caption{RMSD illustration comparing predicted (cyan) and reference (pink) structures after optimal superposition. Left: RMSD 2.1~\AA\ shows good overall alignment with minor deviations. Right: RMSD 3.9~\AA\ shows larger deviations, particularly in peripheral regions where the predicted structure diverges from the reference.}
\label{fig:metric_rmsd}
\end{figure}

\textbf{RMSD} (Root Mean Square Deviation) measures the average distance between predicted and reference atom positions after optimal superposition:
\begin{equation}
\text{RMSD} = \sqrt{\frac{1}{N} \sum_{i=1}^N \| \mathbf{r}_i^{\text{pred}} - \mathbf{r}_i^{\text{ref}} \|^2}
\end{equation}
where $N$ is the number of atoms (typically C$\alpha$), reported in \AA ngstr\"oms. RMSD is non-negative with lower values indicating better agreement between predicted and reference structures; interpretation depends on the size and flexibility of the region being compared. Figure~\ref{fig:metric_rmsd} illustrates RMSD values for two predictions of varying quality.

For antibodies, we report \textbf{region-specific RMSD}: framework RMSD (typically $<$1~\AA\ for all methods) and CDR-H3 RMSD (the primary metric for antibody prediction quality, with state-of-the-art methods achieving 2--4~\AA).

\textbf{TM-score}~\citep{zhang2004scoring,zhang2005tm} is a length-normalized metric ranging from 0 to 1:
\begin{equation}
\text{TM-score} = \max \frac{1}{L} \sum_{i=1}^L \frac{1}{1 + (d_i / d_0)^2}
\end{equation}
where $d_i$ is the distance between aligned residues after superposition, $L$ is the target length, and $d_0$ is a length-dependent normalization factor. TM-score $>$0.5 indicates the same fold; $>$0.6 indicates high similarity. TM-score is less sensitive to local errors than RMSD.

\textbf{lDDT} (Local Distance Difference Test) measures local geometry preservation without requiring global superposition. For each residue, it computes the fraction of pairwise distances (to neighbors within 15~\AA) preserved within tolerance thresholds:
\begin{equation}
\text{lDDT} = \frac{1}{4} \sum_{t \in \{0.5, 1, 2, 4\}} \frac{\#\{\text{distances preserved within } t \text{ \AA}\}}{\#\{\text{total distances}\}}
\end{equation}
The formula above ranges from 0 to 1; lDDT is often reported on a 0--100 scale after multiplying by 100. Per-residue lDDT (pLDDT) follows this 0--100 reporting convention and is output by AlphaFold2 as a confidence estimate.

\subsection{Evaluation Metrics for Complexes}

For protein-protein complexes, we use interface-specific metrics that assess binding geometry.

\textbf{Interface RMSD (I-RMSD)} computes RMSD of interface residues (atoms within 10~\AA\ of the partner) after superimposing one chain. Thresholds: $<$2~\AA\ (high quality), 2--4~\AA\ (medium), $>$4~\AA\ (low).

\textbf{Ligand RMSD (L-RMSD)} superimposes the receptor and computes RMSD of the ligand, capturing both interface geometry and overall binding pose.

\textbf{F$_{\text{nat}}$ (Fraction of native contacts)} measures the fraction of residue-residue contacts in the reference structure that are preserved in the prediction. F$_{\text{nat}} > 0.5$ indicates acceptable contact recovery.

\textbf{DockQ}~\citep{basu2016dockq} integrates F$_{\text{nat}}$, L-RMSD, and I-RMSD into a single score:
\begin{equation}
\text{DockQ} = \frac{F_{\text{nat}} + \frac{1}{1 + (\text{L-RMSD}/8.5)^2} + \frac{1}{1 + (\text{I-RMSD}/1.5)^2}}{3}
\end{equation}
DockQ ranges from 0 to 1, with established quality thresholds:
\begin{itemize}
    \item \textbf{High quality:} DockQ $\geq$ 0.80 (accurate interface geometry)
    \item \textbf{Medium quality:} 0.49 $\leq$ DockQ $<$ 0.80 (correct binding mode)
    \item \textbf{Acceptable quality:} 0.23 $\leq$ DockQ $<$ 0.49 (approximate binding pose)
    \item \textbf{Incorrect:} DockQ $<$ 0.23 (wrong binding mode)
\end{itemize}

\begin{figure}[htbp]
\centering
\includegraphics[width=0.85\textwidth]{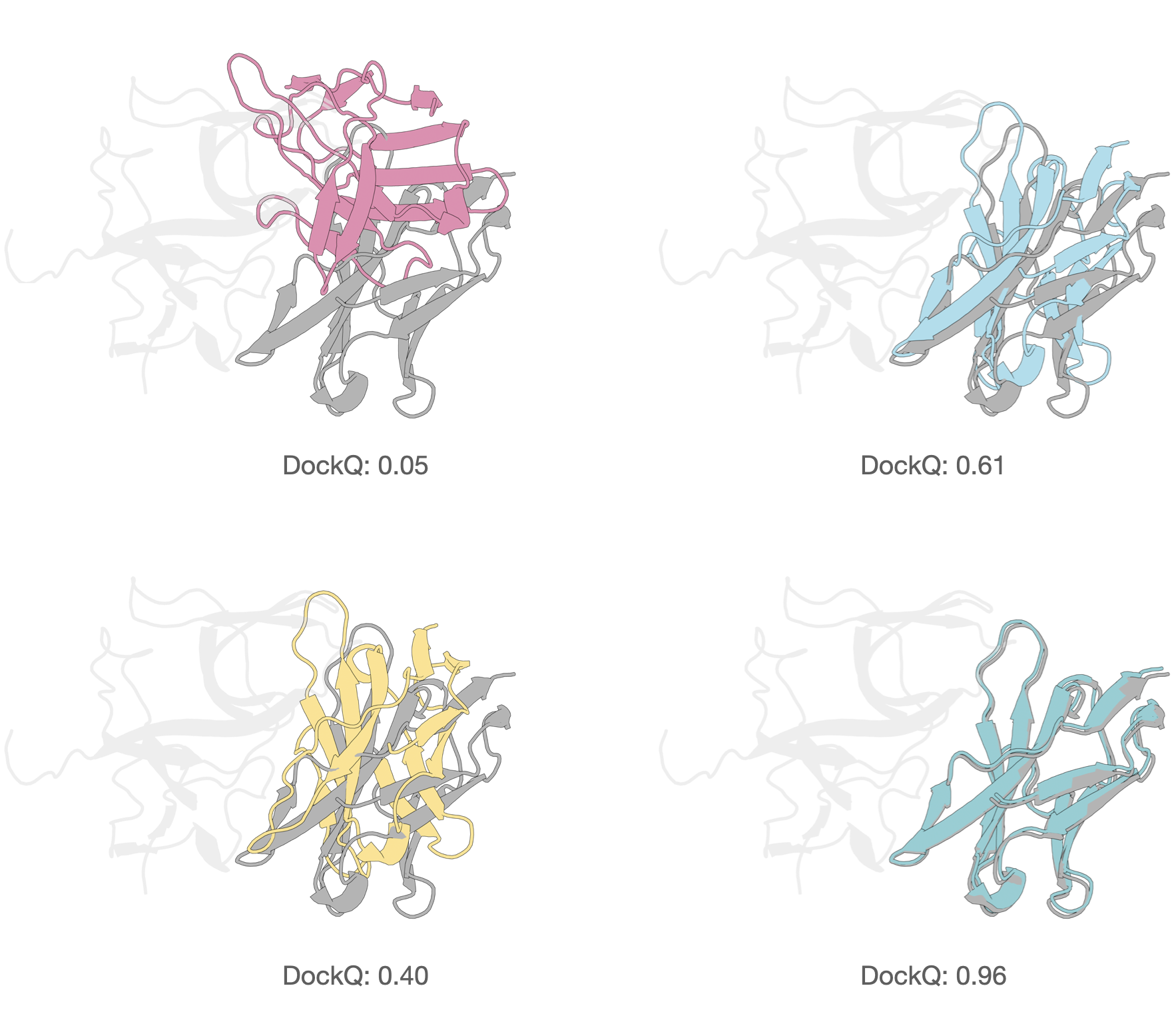}
\caption{DockQ quality thresholds illustrated with antibody-antigen complex predictions. Predicted antibodies (colored) are shown against reference structures (gray), with antigens in light gray background. Top-left: DockQ 0.05 (incorrect)---antibody binds wrong region. Top-right: DockQ 0.61 (medium)---correct binding site with minor interface errors. Bottom-left: DockQ 0.40 (acceptable)---approximate binding pose captured. Bottom-right: DockQ 0.96 (high)---near-perfect prediction matching reference geometry.}
\label{fig:metric_dockq}
\end{figure}

Figure~\ref{fig:metric_dockq} illustrates these quality thresholds with example antibody-antigen complex predictions.

DockQ is the primary metric for antibody-antigen complex prediction throughout this thesis.


\section{Summary}
\label{sec:bg_summary}

This chapter covered the background for computational antibody-antigen structure prediction. The key points for the rest of the thesis are: antibodies are generated through V(D)J recombination and somatic hypermutation rather than gradual evolution, which makes their MSAs sparse in CDR regions and removes the cooperative co-evolutionary signals that methods like AlphaFold2 rely on. CDR-H3, the most important region for binding specificity, shows extreme diversity in both length and sequence.

For structure prediction, there are two main approaches: MSA-based methods that use co-evolutionary signals from sequence alignments, and PLM-based methods that predict structure directly from single sequences using language model embeddings. Each has trade-offs for antibody modeling---MSA-based methods struggle with CDR hypervariability, while PLM-based methods are faster but may sacrifice some accuracy.

We evaluate monomer quality primarily by CDR-H3 RMSD and complex quality by DockQ score. The subsequent chapters build on these concepts: Section~\ref{sec:monomer_prediction} develops PLM-based methods for antibody monomers, Section~\ref{sec:antibody_antigen_complex} extends to complexes, and Chapter~\ref{ch:pipeline_adaptation} develops MSA-based methods for antibody-antigen prediction.

\chapter{PLM-Based Antibody Modeling: From Monomers to Complexes}
\label{ch:plm_modeling}

Protein language models (PLMs) provide a sequence-only alternative to MSA-based structure prediction by encoding evolutionary and structural regularities in model parameters rather than relying on explicit alignments. This chapter tests whether such representations support antibody modeling at two levels. First, we ask whether PLM embeddings provide sufficient intra-chain structural priors for antibody monomer prediction, especially in the hypervariable CDR loops. Second, we ask whether the same sequence-derived representations can support inter-chain interface identification in antibody-antigen complexes. The contrast between these two settings is the central observation of this chapter: PLMs are effective for antibody monomers, but they do not reliably identify antigen binding interfaces from sequence alone, except when external binding-site information is provided. This boundary motivates the MSA-based pipeline adaptation developed in Chapter~\ref{ch:pipeline_adaptation}.

\section{Antibody Monomer Prediction: Intra-Chain Structure}
\label{sec:monomer_prediction}

\subsection{Background and Motivation}

MSA-based structure prediction methods such as AlphaFold2~\citep{jumper2021highly} and RoseTTAFold~\citep{baek2021accurate} achieve high accuracy by extracting co-evolutionary signals from multiple sequence alignments. However, these methods face challenges for antibodies: (1) hypervariable CDR regions, especially CDR-H3, lack informative co-evolutionary signals in MSAs, and (2) MSA generation is computationally expensive, limiting high-throughput applications.

PLM-based methods offer an alternative by embedding evolutionary information implicitly in model parameters through self-supervised training on large sequence databases. Methods such as ESMFold~\citep{lin2023evolutionary}, OmegaFold~\citep{wu2022high}, and HelixFold-Single~\citep{fang2022helixfold} predict structures directly from single sequences, enabling faster inference without MSA construction. These characteristics make PLM-based approaches particularly suitable for antibody structure prediction, where the combination of conserved framework regions and hypervariable CDRs creates a unique modeling challenge that PLMs may address through learned structural patterns rather than evolutionary signals.

\subsection{Protein Language Models for Structure Prediction}

\subsubsection{Overview of Protein Language Models}

Protein language models (PLMs) adapt the concept of language models from natural language processing to protein sequences. The key insight is that amino acid sequences share similarities with natural language: large available corpora, limited supervised labels, and sequential, tokenized structure. By learning to predict masked amino acids from their context, PLMs implicitly capture statistical patterns in proteins from large-scale sequence databases.

PLM training typically uses masked language modeling, where the model learns to predict approximately 15\% of randomly masked amino acids from the surrounding context. This self-supervised objective requires the model to learn the conditional probability distribution of amino acids across the sequence. PLMs typically use Transformer architectures with multi-head self-attention mechanisms.

Two major PLM families are relevant to our work:
\begin{itemize}
\item \textbf{ESM models:} The Evolutionary Scale Modeling family includes ESM-1b~\cite{rives2021biological}, ESM-1v~\cite{meier2021language}, and ESM-2~\citep{lin2023evolutionary}. These BERT-style masked language models are trained on UniRef databases with different clustering parameters, with model sizes ranging from 650M to 15B parameters.
\item \textbf{ProtT5 models:} Based on the T5 encoder-decoder architecture, these models are pre-trained on the BFD database (2.5 billion sequences), providing exposure to a larger and more diverse sequence space.
\end{itemize}

Through self-supervised training, PLMs learn internal representations encoding statistical patterns from training sequences. The TAPE benchmark~\citep{rao2019evaluating} established that these embeddings transfer effectively to downstream tasks including secondary structure prediction, contact prediction, and remote homology detection. Notably, attention maps in PLMs often correlate with structural contacts, enabling structure prediction without explicit MSA inputs~\citep{rao2020transformer}.

\subsubsection{From Language Models to Structure Prediction}

Several methods have adapted PLM representations for structure prediction:

\begin{itemize}
    \item \textbf{RGN2:} Combines AminoBERT embeddings with recurrent geometric networks using Frenet-Serret frames to generate backbone structures from single sequences~\citep{chowdhury2022single}. This approach achieves up to $10^6$-fold speedup over MSA-based methods and outperforms AlphaFold2 on orphan proteins lacking homology information.

    \item \textbf{OmegaFold:} Combines PLM embeddings with a geometry-inspired transformer (Geoformer) to learn residue spatial relationships~\citep{wu2022high}. OmegaFold achieves accuracy comparable to AlphaFold2 on proteins with limited homology, including antibodies with noisy MSAs.

    \item \textbf{ESMFold:} Integrates the ESM-2 language model with an AlphaFold2-inspired structure decoder modified for single-sequence input. ESMFold achieves TM-scores of 0.83 on CAMEO and 0.68 on CASP14, with up to 60$\times$ faster inference than MSA-based methods.
\end{itemize}

\textbf{Advantages and limitations.} PLM-based methods offer two key advantages: (i) \textit{Inference speed}---eliminating MSA generation provides orders of magnitude speedup; (ii) \textit{Robustness to limited homology}---PLMs maintain performance when evolutionary information is sparse.

However, limitations remain: (i) \textit{Accuracy on homology-rich targets}---MSA-based methods outperform PLM-based methods when sufficient evolutionary information is available; (ii) \textit{Confidence estimation}---PLM-based methods lack the natural ensembling from MSA subsampling that MSA-based methods use for confidence estimation.

\subsection{Methods}
\label{sec:plm_method}

The reduced inference time and the performance on low-homology targets make PLM-based methods particularly suitable for antibody structure prediction, where hypervariable regions often lack meaningful evolutionary depth. To further improve the advantage of this type of methods, we develop a model that integrates multiple protein language models.

\subsubsection{Architecture Overview}

Our architecture builds on AlphaFold2's Evoformer and structure module~\citep{jumper2021highly}, adapted for single-sequence input by replacing MSA processing with PLM embeddings. The model feeds an individual protein sequence through different PLMs to generate rich sequence embeddings, then processes these through a modified Evoformer and structure module to predict atomic coordinates. Using multiple PLMs captures complementary information beyond a single model. On antibody benchmarks, this approach achieves competitive overall accuracy, improves CDR-H3 modeling, and offers much faster inference than MSA-based methods.

\begin{figure}[htbp]
\centering
\includegraphics[width=\textwidth]{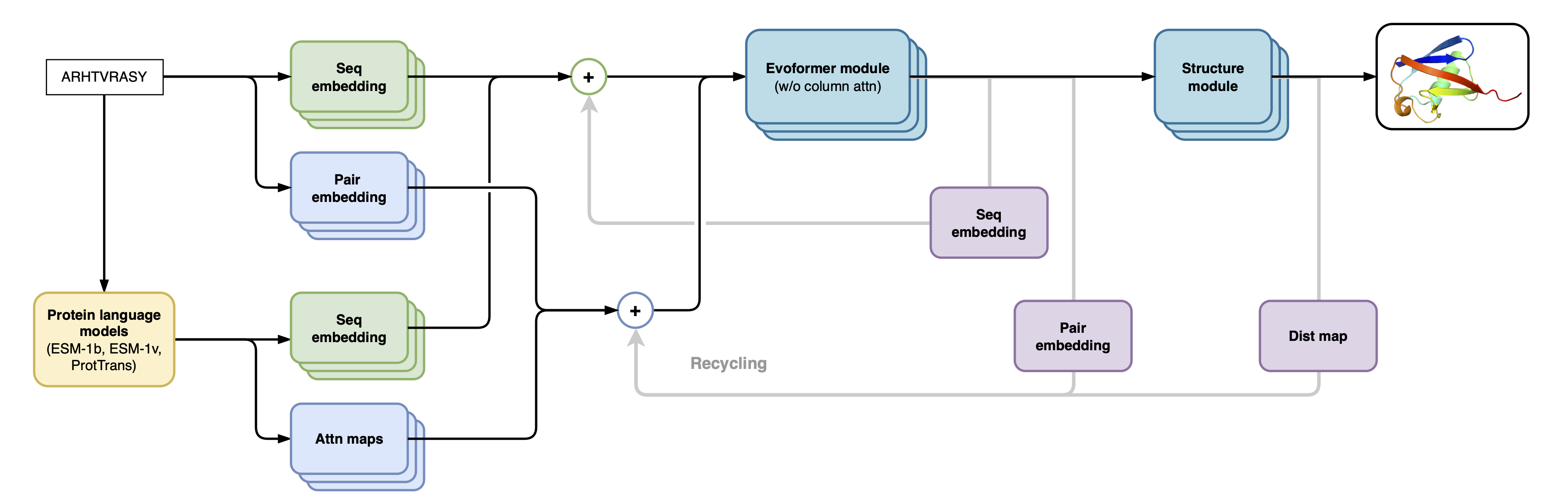}
\caption{Architecture of RaptorX-Single-Ab. The model takes a single amino acid sequence as input and processes it through multiple protein language models (ESM-1b, ESM-1v, ProtTrans) to generate sequence embeddings, pair embeddings, and attention maps. These are combined and fed into a modified Evoformer module (without MSA column attention) followed by a structure module that predicts the final 3D structure. The recycling loop feeds sequence embeddings, pair embeddings, and distance maps from the output back to the input for iterative refinement.}
\label{fig:architecture}
\end{figure}

Figure~\ref{fig:architecture} shows the network architecture of our model. The pipeline consists of three main components: (1) a multi-PLM embedding module, (2) a modified Evoformer, and (3) a structure module. Given a single sequence as input, the model first generates sequence and pair embeddings from multiple PLMs along with their attention maps, then feeds these representations into the modified Evoformer to refine them, and finally uses the structure module to generate the predicted 3D coordinates. A recycling mechanism feeds the output embeddings and distance map back into the input for iterative refinement.

\subsubsection{Protein Language Model Inputs}

The model integrates representations from ESM-1b, ESM-1v, and ProtT5-XL. These PLMs are trained on different sequence databases and clustering thresholds, so their embeddings and attention maps capture complementary sequence statistics. We fuse their residue embeddings, PLM attention maps, and recycled geometric features into the sequence and pair representations consumed by the modified Evoformer. Full model specifications and representation-fusion equations are provided in Appendix~\ref{sec:app_plm_fusion}.

\subsubsection{Modified Evoformer}

We adapt the Evoformer for single-sequence input, following design insights from OpenFold~\citep{ahdritz2024openfold}. Since we process only one sequence, MSA column attention is removed and the sequence trunk becomes a stack of 24 transformer blocks, each consuming the current pair features as attention bias. Given sequence tensor $\mathbf{s}$ and pair tensor $\mathbf{z}$, the $l$-th block computes:
\begin{equation}
\mathbf{s}^{(l+1)} = \text{TransformerLayer}\big(\mathbf{s}^{(l)}, \mathbf{z}^{(l)}, \text{mask}\big),
\end{equation}
where the bias terms are derived from $\mathbf{z}^{(l)}$ and a learned gating mechanism within the attention heads. Feed-forward transitions use GELU activations and residual connections.

Pair representations are updated through a dedicated layer using a triangle stack:
\begin{enumerate}
    \item \textbf{Seq-to-pair lifting}: Layer-normalizes $\mathbf{s}^{(l+1)}$, projects it into two 32-dimensional tensors, and forms an outer-product feature that is added to the recycled $\mathbf{z}^{(l)}$.
    \item \textbf{Triangle multiplication}: Two residual branches apply outgoing and incoming triangle multiplicative updates to propagate information around triplets of residues.
    \item \textbf{Triangle attention}: Starting and ending triangle self-attention layers operate on $\mathbf{z}$ and $\mathbf{z}^\top$, respectively, capturing pairwise relationships.
    \item \textbf{Pair feed-forward}: A residual feed-forward with hidden dimension $4 d_z$ completes the pair refinement.
\end{enumerate}
Residual connections wrap every sublayer, enabling stable gradient flow through the deep network.

\subsubsection{Structure Module}

The structure module uses invariant point attention (IPA). Sequence and pair activations are first projected to hidden sizes $(384,128)$, layer-normalized, and linearly remapped to initialize node/pair states. Each residue starts with an identity rotation and zero translation.

For $n_{\text{layer}} = 8$ iterations, we repeat:
\begin{enumerate}
    \item \textbf{Invariant Point Attention (IPA)}: Consumes the current sequence/pair embeddings alongside running rotations/translations to produce SE(3)-equivariant updates that mix scalar queries, learned 3D points, and pairwise biases.
    \item \textbf{Transition MLP}: A three-layer ReLU network with dropout~\citep{srivastava2014dropout} refines the node states using residual connections and LayerNorm.
    \item \textbf{Backbone update}: A linear head outputs quaternion-like parameters and translations, which are converted to rotation matrices and composed with running frames to yield updated C$_\alpha$ trajectories $\{\mathbf{r}_i\}$.
    \item \textbf{Side-chain torsions}: A dedicated head predicts normalized torsion angles from the current sequence features.
\end{enumerate}
Rotation tensors are detached before the next IPA call to stabilize gradients, following a stop-gradient strategy. The module outputs C$_\alpha$ coordinates, intermediate frames, torsion predictions, and per-residue pLDDT confidence scores.

\subsubsection{Recycling Mechanism}

We use iterative refinement through recycling. After each forward pass through the Evoformer and structure module, the model feeds outputs back as additional inputs to the next iteration:

\begin{enumerate}
    \item \textbf{Sequence recycling}: The previous query representation $\mathbf{s}^{(t-1)}$ is layer-normalized and added to the current sequence embedding $\mathbf{s}^{(t)}$.
    \item \textbf{Pair recycling}: The previous pair representation $\mathbf{r}_{ij}^{\text{pair}}$ is layer-normalized and added to the current pair embedding.
    \item \textbf{Distance map recycling}: Predicted C$_\alpha$ coordinates from the previous iteration are converted to pairwise distances, discretized into bins, and embedded as additional pair features $\mathbf{d}_{ij}$.
\end{enumerate}

PLM embeddings and attention maps are computed once and cached, so subsequent recycling iterations reuse them without recomputation. This design allows the model to iteratively refine predictions while maintaining computational efficiency.

\subsubsection{Training and Inference Summary}

Training follows a two-stage strategy: general protein pretraining using experimental and AlphaFold2-distilled structures, followed by antibody-specific fine-tuning on SAbDab variable-domain structures. At inference time, we generate multiple predictions and select the conformation with the highest overall pLDDT score (the AlphaFold2-style per-residue confidence averaged across the predicted structure). Full dataset construction, optimizer settings, loss terms, recycling schedule, and inference details are provided in Appendix~\ref{sec:app_plm_training}.

\subsection{Results}

\subsubsection{Evaluation Protocol}

\textbf{Datasets and splits.} The general RaptorX-Single model is trained on protein structures released before January 2020, using approximately 80k experimental PDB structures together with approximately 264k AlphaFold2-distilled structures. General-protein evaluation uses structures released between January 2020 and April 12, 2022, including an orphan-protein subset with no detectable homologs in the training databases. For antibody-specific fine-tuning, structures are drawn from SAbDab~\cite{dunbar2014sabdab}, with nanobody entries supplemented from SAbDab-nano~\citep{schneider2022sabdabnano}. RaptorX-Single-Ab is fine-tuned on SAbDab antibody chains released before 2021/03/31 (5,033 chains) and validated on structures released between 2021/04/01 and 2021/06/30 (178 structures). We evaluate antibody performance on three temporally separated benchmarks: IgFold-Ab structures released after 2021/07/01, SAbDab-Ab structures released in the first six months of 2022, and nanobody structures released in the first six months of 2022.

\textbf{Inference procedure.} For every target we generate five predictions per method using different random seeds/recycling schedules. Predictions are ranked by each model's internal confidence score: RaptorX-Single variants and AlphaFold2-style baselines are ranked by overall pLDDT (per-residue mean across the predicted structure), while other baselines use model-specific ranking heads where available. Unless explicitly noted, metrics are reported on the top-ranked prediction; we additionally include best-of-five (oracle) results for selected plots/tables.

\textbf{Metrics.} Backbone quality is assessed with C$_\alpha$ RMSD after Kabsch alignment~\citep{kabsch1976solution}. We report separate RMSDs for framework regions (FR), CDR-H1/H2/H3, and the light-chain analogs. CDR boundaries follow IMGT numbering applied via ANARCI~\citep{dunbar2016anarci}, and RMSDs are computed with PyRosetta~\citep{chaudhury2010pyrosetta}.

\textbf{Baselines.} We compare against four categories of methods: (1) MSA-based methods: AlphaFold2 (no templates) with MSAs from HHblits and Jackhmmer searches; (2) general PLM-based methods: ESMFold, OmegaFold~\citep{wu2022high}, and HelixFold-Single~\citep{fang2022helixfold}; (3) general coarse-grained structure prediction evaluated on antibodies: EquiFold~\citep{lee2022equifold}; and (4) antibody-specific methods: DeepAb~\citep{ruffolo2022antibody} and IgFold~\citep{ruffolo2023fast}. All methods follow the same inference protocol (five samples, top-1 ranking) for fair comparison.

\subsubsection{Overall Performance}

Table~\ref{tab:antibody_rmsd} presents antibody structure prediction performance across methods, broken down by heavy (H) and light (L) chains and subdivided into framework regions (Fr) and individual CDRs. RaptorX-Single-Ab achieves the best CDR-H3 RMSD (3.24~\AA) on this benchmark, surpassing the previous best EquiFold (3.37~\AA) and outperforming AlphaFold2 with MSAs (3.82~\AA) while eliminating the MSA generation step entirely.

\begin{table}[ht]
\centering
\caption{RMSD (\AA, $\downarrow$) values for different antibody structure prediction methods. Lower values indicate better prediction accuracy. Methods are grouped by category. Best results for each region are shown in \textbf{bold}.}
\begin{tabular}{l|cccc|cccc}
\toprule
& \multicolumn{4}{c|}{RMSD (H) $\downarrow$} & \multicolumn{4}{c}{RMSD (L) $\downarrow$} \\
\cmidrule{2-9}
& Fr & CDR-1 & CDR-2 & CDR-3 & Fr & CDR-1 & CDR-2 & CDR-3 \\
\midrule
\multicolumn{9}{l}{\textit{MSA-based}} \\
AlphaFold2 (MSA) & 0.63 & 1.08 & 0.89 & 3.82 & 0.59 & 0.89 & 0.69 & 1.39 \\
AlphaFold2 (Single) & 8.85 & 12.3 & 11.59 & 15.24 & 8.82 & 13.28 & 15.13 & 14.62 \\
\midrule
\multicolumn{9}{l}{\textit{General PLM-based}} \\
HelixFold-Single & 0.71 & 1.15 & 1.1 & 5.5 & 0.66 & 1.1 & 0.79 & 1.84 \\
OmegaFold & 0.63 & 1.05 & 0.86 & 4.11 & 0.58 & 0.9 & 0.69 & 1.42 \\
ESMFold & 0.64 & 1.11 & 1.02 & 4.56 & 0.6 & 1.16 & 0.72 & 1.74 \\
\midrule
\multicolumn{9}{l}{\textit{Antibody-specific}} \\
DeepAb & 0.62 & 1.08 & 0.9 & 3.83 & 0.66 & 0.96 & 0.75 & 1.43 \\
IgFold & 0.66 & 1.15 & 0.95 & 3.65 & 0.65 & 0.96 & 0.8 & 1.4 \\
EquiFold & 0.6 & 1.05 & 0.89 & 3.37 & 0.57 & 0.87 & 0.72 & 1.31 \\
\midrule
\multicolumn{9}{l}{\textit{Ours}} \\
RaptorX-Single & 0.64 & 1.17 & 1.06 & 4.66 & 0.64 & 1.12 & 0.77 & 2.14 \\
RaptorX-Single-Ab & \textbf{0.57} & \textbf{1.01} & \textbf{0.82} & \textbf{3.24} & \textbf{0.53} & \textbf{0.79} & \textbf{0.66} & \textbf{1.24} \\
\bottomrule
\end{tabular}
\label{tab:antibody_rmsd}
\end{table}

\subsubsection{Comparison with MSA-Based Methods}

Despite not using MSAs, RaptorX-Single-Ab achieves performance comparable to or better than AlphaFold2 with MSAs on several regions. On framework regions, RaptorX-Single-Ab achieves 0.57~\AA\ (heavy) and 0.53~\AA\ (light), slightly better than AlphaFold2's 0.63~\AA\ and 0.59~\AA. On CDR-H1 and CDR-H2, RaptorX-Single-Ab (1.01~\AA, 0.82~\AA) performs comparably to AlphaFold2 (1.08~\AA, 0.89~\AA). For the challenging CDR-H3, RaptorX-Single-Ab (3.24~\AA) improves over AlphaFold2's performance (3.82~\AA) on this benchmark.

In contrast, when AlphaFold2 is run without MSAs (single-sequence mode), performance drops sharply, with RMSDs exceeding 8~\AA\ on frameworks and 11--15~\AA\ on CDRs. This confirms that AlphaFold2's architecture relies strongly on evolutionary information from MSAs and cannot effectively use sequence information alone, highlighting the importance of PLM-based architectures for MSA-free structure prediction.

\subsubsection{Comparison with PLM-Based and Antibody-Specific Methods}

RaptorX-Single-Ab consistently outperforms other general PLM-based methods across all regions. Compared to ESMFold (CDR-H3: 4.56~\AA), OmegaFold~\citep{wu2022high} (4.11~\AA), and HelixFold-Single~\citep{fang2022helixfold} (5.5~\AA), RaptorX-Single-Ab achieves better CDR-H3 modeling (3.24~\AA). On light-chain CDR-L3, the improvement is similarly consistent: RaptorX-Single-Ab (1.24~\AA) outperforms ESMFold (1.74~\AA), HelixFold-Single (1.84~\AA), and the generic RaptorX-Single (2.14~\AA).

Among methods specifically designed for antibodies, RaptorX-Single-Ab also performs favorably. Compared to IgFold~\citep{ruffolo2023fast} (CDR-H3: 3.65~\AA), DeepAb~\citep{ruffolo2022antibody} (3.83~\AA), and EquiFold~\citep{lee2022equifold} (3.37~\AA), RaptorX-Single-Ab's 3.24~\AA\ is the best CDR-H3 accuracy on this benchmark. Across all methods, light-chain CDRs are easier to model than heavy-chain CDRs, with CDR-L3 RMSDs typically 1--3~\AA\ lower than CDR-H3.

\subsubsection{Effect of Antibody-Specific Fine-Tuning}
\label{sec:finetune_effect}

Comparing RaptorX-Single (general-purpose) with RaptorX-Single-Ab (antibody fine-tuned) isolates the effect of domain adaptation. Fine-tuning reduces CDR-H3 RMSD from 4.66~\AA\ to 3.24~\AA\ (30.5\% improvement) and CDR-L3 from 2.14~\AA\ to 1.24~\AA\ (42.1\% improvement). Framework regions also improve: heavy-chain FR drops from 0.64~\AA\ to 0.57~\AA\ and light-chain FR from 0.64~\AA\ to 0.53~\AA. These consistent gains across all regions show the importance of antibody-specific adaptation, even when the underlying architecture and PLM embeddings remain unchanged.

\subsubsection{Impact of Individual PLMs}
\label{sec:plm_ablation}

To understand the contribution of each protein language model, we evaluate three single-PLM variants: RaptorX-Single~(1b) using ESM-1b~\citep{rives2021biological}, RaptorX-Single~(1v) using ESM-1v~\citep{meier2021language}, and RaptorX-Single~(pt) using ProtT5~\citep{elnaggar2021prottrans}. These models differ in both scale and training data: ESM-1b and ESM-1v share the same architecture ($\sim$650M parameters) but are trained on UniRef50 and UniRef90 respectively, while ProtT5 is much larger ($\sim$3B parameters) and trained on BFD~\citep{steinegger2019protein}. The full ablation results on the SAbDab-Ab benchmark are presented in Table~\ref{tab:plm_ablation}.

\begin{table}[ht]
\centering
\caption{PLM ablation on the SAbDab-Ab dataset: RMSD (\AA, $\downarrow$) for single-PLM variants and their combinations, with and without antibody-specific fine-tuning. ESM-1b is trained on UniRef50 ($\sim$650M params), ESM-1v on UniRef90 ($\sim$650M params), and ProtT5 on BFD ($\sim$3B params). Best results per column are in \textbf{bold}.}
\resizebox{\textwidth}{!}{%
\begin{tabular}{l|cccc|cccc}
\toprule
& \multicolumn{4}{c|}{RMSD (H) $\downarrow$} & \multicolumn{4}{c}{RMSD (L) $\downarrow$} \\
\cmidrule{2-9}
& Fr & CDR-1 & CDR-2 & CDR-3 & Fr & CDR-1 & CDR-2 & CDR-3 \\
\midrule
\multicolumn{9}{l}{\textit{Single PLM (no fine-tuning)}} \\
RaptorX-Single (1b) & 0.72 & 1.65 & 1.41 & 5.06 & 0.65 & 1.33 & 0.82 & 2.37 \\
RaptorX-Single (1v) & 0.70 & 1.54 & 1.25 & 4.64 & 0.63 & 1.22 & 0.79 & 1.96 \\
RaptorX-Single (pt) & 0.68 & 1.40 & 1.27 & 4.89 & 0.64 & 1.23 & 0.84 & 2.25 \\
RaptorX-Single (combined) & 0.64 & 1.17 & 1.06 & 4.66 & 0.64 & 1.12 & 0.77 & 2.14 \\
\midrule
\multicolumn{9}{l}{\textit{Single PLM (antibody fine-tuned)}} \\
RaptorX-Single-Ab (1b) & \textbf{0.57} & 1.01 & 0.83 & 3.39 & \textbf{0.53} & 0.80 & \textbf{0.66} & \textbf{1.23} \\
RaptorX-Single-Ab (1v) & \textbf{0.57} & 1.01 & \textbf{0.82} & \textbf{3.17} & 0.54 & 0.80 & \textbf{0.66} & 1.25 \\
RaptorX-Single-Ab (pt) & \textbf{0.57} & \textbf{0.99} & 0.83 & 3.32 & \textbf{0.53} & 0.81 & \textbf{0.66} & 1.25 \\
RaptorX-Single-Ab (combined) & \textbf{0.57} & 1.01 & \textbf{0.82} & 3.24 & \textbf{0.53} & \textbf{0.79} & \textbf{0.66} & 1.24 \\
\bottomrule
\end{tabular}%
}
\label{tab:plm_ablation}
\end{table}

Despite being roughly four times smaller than ProtT5, RaptorX-Single~(1v) achieves the best CDR-H3 RMSD (4.64~\AA) among non-fine-tuned variants, outperforming both RaptorX-Single~(1b) (5.06~\AA), which shares the same $\sim$650M-parameter architecture as~(1v) but is trained on UniRef50, and the much larger RaptorX-Single~(pt) (4.89~\AA). After antibody-specific fine-tuning, this pattern persists: RaptorX-Single-Ab~(1v) achieves 3.17~\AA\ CDR-H3 RMSD, the best among all single-PLM fine-tuned variants, compared to RaptorX-Single-Ab~(1b) (3.39~\AA) and RaptorX-Single-Ab~(pt) (3.32~\AA)~\citep{jing2024single}.

This reveals an important insight: \emph{not only the scale but also the training data of protein language models impacts performance on domain-specific tasks}. ESM-1v, trained on the less-clustered UniRef90, retains more sequence diversity than ESM-1b (trained on UniRef50), making it more sensitive to the subtle sequence variations that distinguish antibody CDR conformations. ProtT5, despite its larger capacity, may underrepresent the specific sequence patterns characteristic of immunoglobulin hypervariable regions in its BFD training set.

The combined model RaptorX-Single integrates embeddings from all three PLMs and provides a balanced representation across regions, even though individual PLMs can be best for particular metrics (for example, ESM-1v gives the lowest CDR-H3 RMSD among the single-PLM fine-tuned variants). On SAbDab-Ab, the combined RaptorX-Single-Ab achieves 3.24~\AA\ CDR-H3 and 1.24~\AA\ CDR-L3, showing that the three language models capture complementary sequence statistics: ESM-1b and ESM-1v encode different granularities of evolutionary information due to their clustering thresholds, while ProtT5 contributes representations learned from a distinct and larger sequence database. Further performance gains are obtained by selecting predictions based on pLDDT confidence scores across the four model variants (three single-PLM plus the combined model), demonstrating that model diversity also benefits prediction ranking~\citep{jing2024single}.

\subsubsection{Validation on Nanobodies}
\label{sec:nanobody_results}

To assess generalization beyond traditional paired antibodies, we evaluated RaptorX-Single-Ab on a benchmark of 60 nanobodies (single-domain VHH antibodies) from SAbDab-nano~\citep{schneider2022sabdabnano}. Nanobodies present a distinct structural challenge: lacking light chains, they rely on elongated CDR3 loops and an adapted framework to achieve antigen binding, resulting in CDR conformations that differ from conventional antibodies.

\begin{table}[ht]
\centering
\caption{RMSD (\AA, $\downarrow$) on the Nanobody dataset (60 VHH single-domain antibodies from SAbDab-nano). Methods are grouped by category. Best results per column are in \textbf{bold}.}
\begin{tabular}{l|cccc}
\toprule
& Fr $\downarrow$ & CDR-1 $\downarrow$ & CDR-2 $\downarrow$ & CDR-3 $\downarrow$ \\
\midrule
\multicolumn{5}{l}{\textit{MSA-based}} \\
AlphaFold2 (MSA) & 0.73 & 2.05 & 1.15 & 4.01 \\
AlphaFold2 (Single) & 9.34 & 12.67 & 12.39 & 17.87 \\
\midrule
\multicolumn{5}{l}{\textit{General PLM-based}} \\
HelixFold-Single & 0.86 & 1.99 & 1.18 & 4.20 \\
OmegaFold & \textbf{0.71} & 2.02 & 1.12 & 3.77 \\
ESMFold & 0.80 & 2.06 & 1.12 & 4.23 \\
\midrule
\multicolumn{5}{l}{\textit{Antibody-specific}} \\
DeepAb & 0.92 & 2.38 & 1.34 & 8.76 \\
IgFold & 0.82 & 1.93 & 1.29 & 4.27 \\
EquiFold & 2.30 & 3.23 & 2.61 & 7.19 \\
\midrule
\multicolumn{5}{l}{\textit{Ours}} \\
RaptorX-Single & 0.83 & 2.19 & 1.14 & 4.06 \\
RaptorX-Single-Ab & 0.82 & \textbf{1.78} & \textbf{1.06} & \textbf{3.50} \\
\bottomrule
\end{tabular}
\label{tab:nanobody_rmsd}
\end{table}

As shown in Table~\ref{tab:nanobody_rmsd}, RaptorX-Single-Ab achieves a CDR3 RMSD of 3.50~\AA\ on this dataset, outperforming all baselines including IgFold (4.27~\AA), AlphaFold2 with MSAs (4.01~\AA), OmegaFold (3.77~\AA), and EquiFold (7.19~\AA)~\citep{jing2024single}. Notably, the performance gap between RaptorX-Single-Ab and EquiFold is much larger for nanobodies (3.69~\AA\ difference) than for traditional antibodies (0.13~\AA\ on SAbDab-Ab), suggesting that EquiFold's geometry-based approach---which does not utilize PLMs---struggles to generalize across antibody formats. DeepAb also degrades on nanobodies (CDR3: 8.76~\AA), likely because its architecture was designed specifically for paired VH-VL antibodies. These results highlight the value of PLM-derived representations for capturing structural diversity across different antibody modalities.

\subsubsection{Structural Case Studies}
\label{sec:ch3_case_studies}

\begin{figure}[htbp]
  \centering
  \begin{subfigure}[t]{0.48\textwidth}
    \centering
    \includegraphics[width=\textwidth]{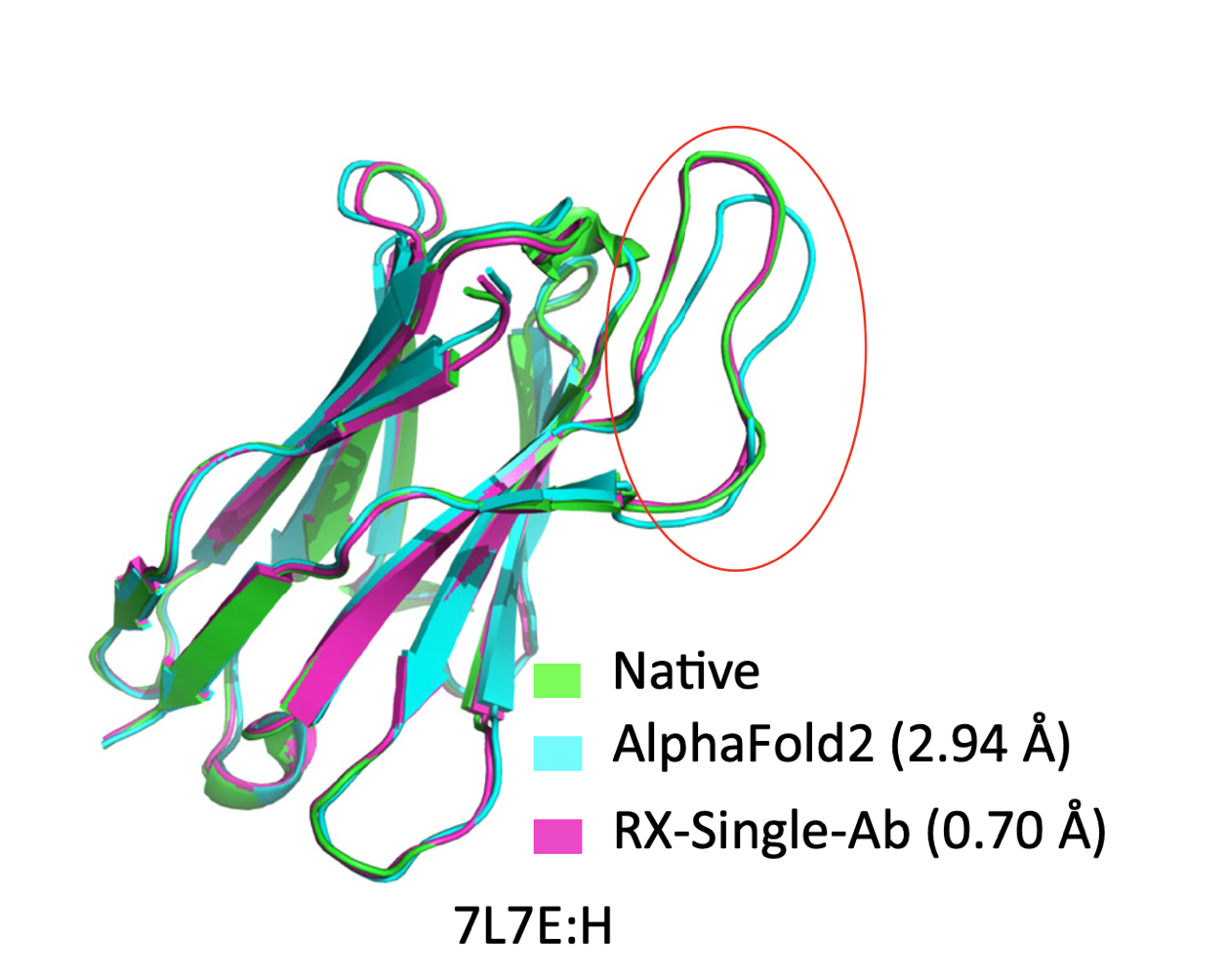}
    \caption{7L7E chain H}
    \label{fig:case_7L7E}
  \end{subfigure}
  \hfill
  \begin{subfigure}[t]{0.48\textwidth}
    \centering
    \includegraphics[width=\textwidth]{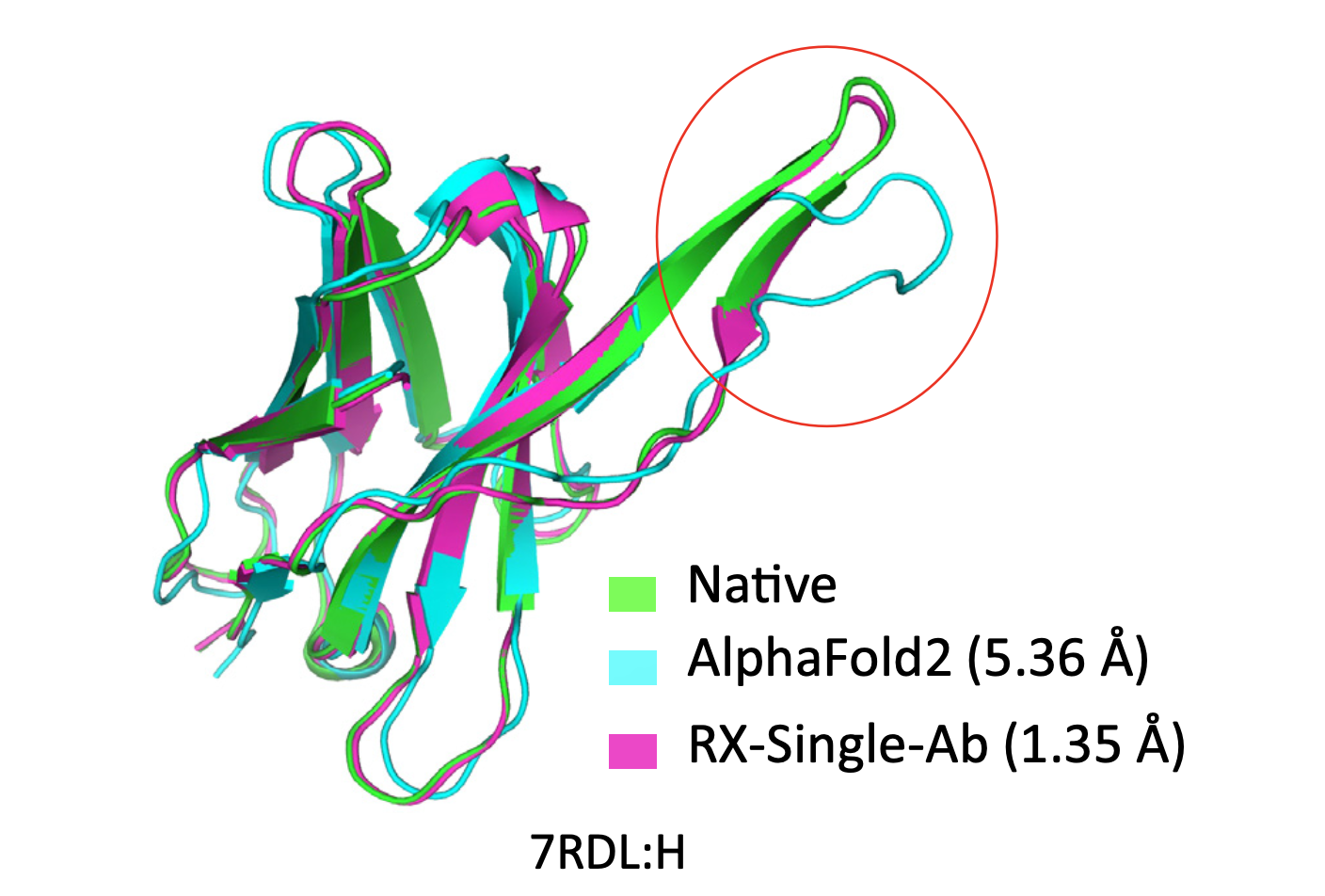}
    \caption{7RDL chain H}
    \label{fig:case_7RDL}
  \end{subfigure}
  \caption{Structural case studies comparing RaptorX-Single-Ab (magenta) and AlphaFold2 (cyan) against the native structure (green). RMSD values are shown in the legend. Red circles highlight CDR loop regions where AlphaFold2 deviates from the native conformation while RaptorX-Single-Ab maintains close agreement. (a)~7L7E:H --- AlphaFold2 produces a 2.94~\AA\ RMSD with visible CDR loop deviation, while RaptorX-Single-Ab achieves 0.70~\AA\ (4.2$\times$ improvement). (b)~7RDL:H --- a more difficult case, where AlphaFold2's CDR loop prediction (5.36~\AA) diverges in an entirely wrong direction, while RaptorX-Single-Ab (1.35~\AA) closely tracks the native (4.0$\times$ improvement).}
  \label{fig:ch3_case_studies}
\end{figure}

In addition to the quantitative evaluation, we present structural superpositions for two representative targets where RaptorX-Single-Ab outperforms AlphaFold2 (Figure~\ref{fig:ch3_case_studies}). In each panel, the native structure (green) is superimposed with predictions from AlphaFold2 (cyan) and RaptorX-Single-Ab (magenta), with the red circle highlighting the CDR loop region where predictions diverge most.

These cases illustrate a consistent pattern: the $\beta$-sandwich framework regions are well-predicted by both methods, confirming that the immunoglobulin fold is structurally conserved and straightforward to model. The critical differences emerge in the CDR loops, particularly CDR-H3, where AlphaFold2 without antibody-specific training frequently produces incorrect loop conformations. In contrast, RaptorX-Single-Ab's antibody-specific fine-tuning and PLM-derived representations capture the conformational preferences of hypervariable loops, producing predictions that closely follow the native backbone even for long and structurally diverse CDR-H3 loops.

The 7RDL:H case (Figure~\ref{fig:case_7RDL}) is especially revealing: AlphaFold2's predicted CDR loop extends in an entirely different direction from the native, suggesting that without antibody-specific priors, the model cannot correctly resolve the conformational ambiguity of CDR-H3 from MSA information alone. The 4$\times$ RMSD reduction achieved by RaptorX-Single-Ab (from 5.36~\AA\ to 1.35~\AA) on this target demonstrates the practical impact of domain adaptation for antibody structure prediction.

\subsection{Discussion}
\label{sec:ch3_discussion}

\subsubsection{Why PLM-Based Methods Excel at Antibody Monomers}

The competitive performance of PLM-based methods for antibodies comes from several factors:

\textbf{Limited utility of MSA information for antibodies.} Hypervariable CDRs lack meaningful co-evolutionary signals, reducing the advantage of MSA-based approaches. As shown in Table~\ref{tab:antibody_rmsd}, AlphaFold2 without MSAs produces RMSDs exceeding 8~\AA\ on frameworks and 11--15~\AA\ on CDRs, demonstrating that its architecture cannot extract structural information from single sequences. PLM-based methods, trained specifically on single-sequence inputs, avoid this dependency.

\textbf{PLMs capture general folding principles.} Through large-scale pre-training on hundreds of millions of protein sequences, PLMs learn sequence-to-structure mappings that generalize beyond specific evolutionary families. Our method achieves slightly better framework RMSD than AlphaFold2 with MSAs (Table~\ref{tab:antibody_rmsd}), suggesting that PLMs have learned the conserved $\beta$-sandwich fold of immunoglobulin frameworks without relying on co-evolutionary information.

\textbf{Complementary information from multi-PLM integration.} Our multi-PLM strategy uses models trained on different databases with different clustering thresholds: ESM-1b on UniRef50~\cite{rives2021biological}, ESM-1v on UniRef90~\cite{meier2021language}, and ProtT5 on BFD~\citep{steinegger2019protein}. We hypothesize that the diversity in training data sources allows these models to capture complementary sequence statistics, which benefits prediction on the highly variable CDR regions.

\textbf{Domain-specific fine-tuning.} Antibody-specific fine-tuning helps the model learn structural patterns of the immunoglobulin fold, including framework geometry and CDR loop conformations. As shown in Section~\ref{sec:finetune_effect}, this adaptation improves accuracy across all regions.

\subsubsection{When Do PLMs Outperform MSA-Based Methods?}
\label{sec:plm_vs_msa_depth}

The results above demonstrate that PLM-based methods excel at antibody monomer prediction, but this raises a broader question: under what conditions do single-sequence methods outperform MSA-based approaches? To investigate, we compare RaptorX-Single with MSA-based AlphaFold2 across proteins with varying MSA depths, using 353 targets from CASP14, CAMEO, and an additional set of proteins with limited homologs~\citep{jing2024single}.

\begin{figure}[t]
  \centering
  \includegraphics[width=0.85\linewidth]{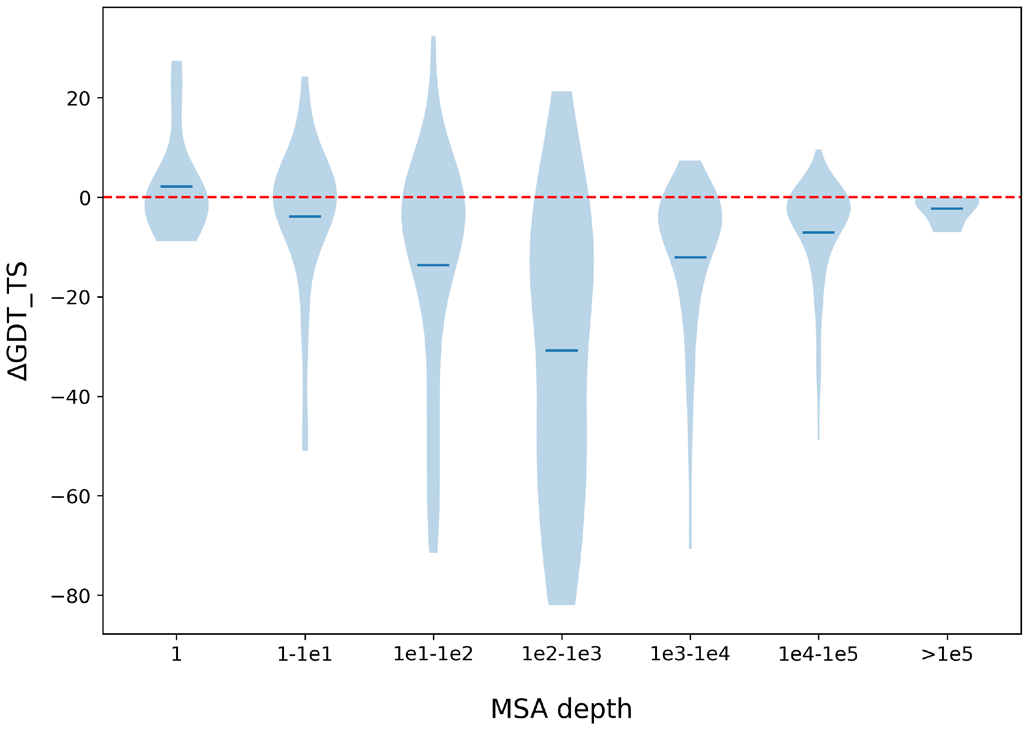}
  \caption{Performance comparison between RaptorX-Single and MSA-based AlphaFold2 with respect to MSA depth. The performance is measured by $\Delta$GDT\_TS (GDT\_TS of RaptorX-Single minus that of AlphaFold2). Positive values indicate RaptorX-Single outperforms AlphaFold2. Violin plots show the distribution at each MSA depth bin. Figure reproduced from~\citet{jing2024single}.}
  \label{fig:msa_depth_vs_performance}
\end{figure}

Figure~\ref{fig:msa_depth_vs_performance} shows a clear relationship between MSA depth and the relative advantage of each approach:

\begin{itemize}
  \item \textbf{No homologs (MSA depth $= 1$):} RaptorX-Single slightly outperforms AlphaFold2 (GDT\_TS: 0.48 vs.\ 0.46), confirming that PLMs provide useful structural priors even when no evolutionary information is available.
  \item \textbf{Shallow MSAs (depth $10$--$10^3$):} AlphaFold2 gains a clear advantage as co-evolutionary signals become available. The performance gap is largest in the $10^2$--$10^3$ range, where MSAs contain enough sequences to extract meaningful co-variation patterns but PLMs cannot access this target-specific information.
  \item \textbf{Deep MSAs (depth $> 10^4$):} The gap narrows. With depth $> 10^5$, RaptorX-Single (GDT\_TS: 0.87) approaches AlphaFold2 (GDT\_TS: 0.89). This convergence suggests that PLMs have implicitly learned co-evolutionary patterns of large protein families during pre-training, effectively internalizing the information that deep MSAs provide explicitly.
\end{itemize}

This analysis connects to antibody-antigen complex prediction. Antibody CDR regions are hypervariable by design, limiting co-evolutionary signal, but belong to a structurally conserved superfamily that provides some MSA coverage through framework regions. For antibody \emph{monomers}, CDRs fall in the shallow-MSA regime where PLMs can make up for weak co-evolutionary signals, which explains the results in this chapter. However, for antibody-antigen \emph{complexes}, predicting the binding interface requires inter-chain contacts---information that single-sequence PLM embeddings do not provide. In Section~\ref{sec:antibody_antigen_complex}, we find that current single-sequence PLM representations do not reliably infer interface contacts from sequence alone, which leads us back to MSA-based methods for complex prediction.

\subsubsection{Future Directions}

\textbf{Integration of antibody-specific PLMs.} Antibody-focused language models such as AbLang, AntiBERTy, and BioPhi/Sapiens~\citep{olsen2022ablang,ruffolo2021deciphering,prihoda2022biophi} encode repertoire-specific statistics. Incorporating these specialized models could further improve CDR modeling by capturing somatic hypermutation patterns that general PLMs may underrepresent.

\textbf{Confidence calibration.} Current pLDDT scores may not be optimally calibrated for antibody CDRs. Developing antibody-specific confidence metrics could improve prediction selection from multiple samples.

\subsection{Summary}

This chapter presented PLM-based antibody structure prediction. We developed RaptorX-Single-Ab, which combines multiple PLMs (ESM-1b, ESM-1v, ProtT5) with antibody-specific fine-tuning to predict antibody structures without MSAs. The model uses a modified Evoformer without MSA column attention, operating directly on PLM embeddings from a single sequence.

RaptorX-Single-Ab achieves CDR-H3 RMSD of 3.24~\AA, outperforming all other PLM-based methods including ESMFold (4.56~\AA), OmegaFold~\citep{wu2022high} (4.11~\AA), and HelixFold-Single~\citep{fang2022helixfold} (5.5~\AA). It also outperforms AlphaFold2 with MSAs on this benchmark for both CDR-H3 (3.24~\AA\ vs. 3.82~\AA) and framework regions (0.57~\AA\ vs. 0.63~\AA\ for heavy chains; 0.53~\AA\ vs. 0.59~\AA\ for light chains). The method is also much faster than MSA-based approaches, which is useful for high-throughput antibody design.

These results show that PLM representations capture useful intra-chain structural priors for antibody variable domains. The next question is whether the same sequence-only representations also contain enough information for inter-chain recognition: identifying where an antibody binds on an antigen and assembling the complex. Section~\ref{sec:antibody_antigen_complex} tests this directly. Complex prediction requires identifying the binding interface and assembling the correct geometry, without co-evolutionary signals between antibody and antigen.

\section{Antibody-Antigen Complex Prediction: Inter-Chain Interfaces}
\label{sec:antibody_antigen_complex}

\subsection{Motivation and Challenges}

Section~\ref{sec:monomer_prediction} demonstrated that PLM-based methods can predict antibody monomer structures with accuracy approaching MSA-based approaches, achieving a CDR-H3 RMSD of 3.24~\AA{} without requiring evolutionary information. A natural next question is: can these PLM-derived representations extend beyond individual chains to predict the bound configuration of an antibody-antigen complex? This question is important because therapeutic antibody development depends on understanding how an antibody recognizes its target, not just the antibody structure alone. This chapter investigates whether the structural information captured by PLMs is enough for complex prediction, or whether the missing co-evolutionary signals between antibodies and antigens pose a clear limitation.

\subsubsection{Why Antibody-Antigen Complex Prediction Matters}

Complex structures show how the antibody contacts the antigen and which residues are involved in binding. This information is useful for both designing new antibodies and improving existing ones---for example, by guiding targeted mutations at interface positions or by identifying which framework positions can be changed without affecting binding. Experimental structure determination by X-ray crystallography or cryo-EM takes months per complex and often fails for flexible interactions. Computational prediction could allow rapid testing of thousands of antibody variants, helping select the best candidates for experiments.

\subsubsection{Challenges in Complex Prediction}

Predicting antibody-antigen complexes is harder than predicting antibody monomers. The main difficulty is not the larger size, but identifying where the antibody binds on the antigen.

The central challenge is \textbf{interface identification}: finding which region of the antigen surface the antibody binds. A typical antigen has a large surface area, with many regions that could serve as epitopes. Individual epitopes typically occupy only a small portion of the antigen surface~\cite{conte1999atomic,ramaraj2012antigen,reis2022antibody,madsen2024structural}. The antibody binding site (paratope) is localized to the CDR loops, but epitopes can occur anywhere on the antigen and vary widely in size, shape, and chemistry.

Structural flexibility adds another difficulty. CDR loops, especially CDR-H3, often change shape when binding the antigen~\citep{liu2024antibody,wang2013conformational,keskin2007binding}. These changes can be large, so an accurate unbound antibody structure may not show the bound conformation. The antigen may also change shape at the binding site.

For PLM-based methods, the question is whether sequence information alone can identify the correct binding interface. We test this in two scenarios: blind docking where the epitope is unknown, and epitope-guided docking where we provide approximate binding region information.

\subsubsection{Our Approach}

Building on the monomer prediction pipeline from Section~\ref{sec:monomer_prediction}, we test whether protein language models combined with antibody-specific training can predict complex structures from sequences alone. Our method uses structural information learned by multiple PLMs, combined with predicted monomer structures, to predict bound complexes.

This approach differs from physics-based docking methods that search all possible orientations and rank them by energy. By training on experimental complex structures, our model can learn binding patterns specific to antibody-antigen interactions. The key question is whether these learned patterns can make up for the missing co-evolutionary information between antibodies and antigens.

The next section reviews existing docking methods.

\subsection{Related Work}
\label{sec:ch4_related_work}

Protein-protein docking methods predict how two proteins bind given their individual structures. Traditional approaches search possible binding poses and rank them using scoring functions. This section reviews prior work on antibody-antigen docking, organized by methodology, and positions our approach within this landscape.

\subsubsection{Physics-Based Docking}

Physics-based methods sample binding orientations and score them using energy functions derived from physical principles. FFT-based approaches such as ZDOCK~\cite{chen2003zdock,pierce2014zdock} and ClusPro~\cite{kozakov2017cluspro} efficiently evaluate millions of rigid-body orientations through shape complementarity and electrostatic matching. These methods are general-purpose and require no training data, but assume rigid protein backbones and rely on scoring functions that may not capture the specificity of antibody-antigen recognition.

SnugDock~\cite{sircar2010snugdock} is an early antibody-specific docking method, extending RosettaDock~\cite{gray2003protein} with CDR loop flexibility by stochastically mixing antibody-antigen rigid-body perturbations, VL--VH rigid-body reorientation, and CDR loop remodeling under a Monte Carlo sampler. HADDOCK~\cite{dominguez2003haddock,van2016haddock} takes a data-driven approach, incorporating experimental restraints from sources such as mutagenesis, cross-linking, or NMR to guide the docking search, which can help when partial epitope information is available.

While physics-based methods produce interpretable energy scores, they typically require significant computation and careful parameter tuning. Large-scale benchmarks continue to show both their strengths and weaknesses for antibody-antigen complexes~\citep{guest2021,hwang2010protein}, especially for cases involving large conformational changes upon binding.

\subsubsection{Template-Based Methods}

Template-based pipelines combine antibody monomer modeling with subsequent docking against the antigen. RosettaAntibody~\cite{weitzner2017modeling} pairs Rosetta-based antibody modeling with Rosetta docking refinement, while AbAdapt~\cite{davila2022abadapt} takes antibody and antigen sequences as input, homology-models both chains, predicts paratope and epitope using machine-learning components, and docks the modeled structures with Piper and Hex. Such pipelines can be useful when suitable templates or homology models are reliable, but their accuracy tends to drop on novel binding modes that are not well represented in existing databases.

\subsubsection{Deep Learning Approaches}

Deep learning methods have explored several strategies for protein-protein docking. EquiDock~\cite{ganea2021independent} uses SE(3)-equivariant graph neural networks to predict rigid-body transformations directly from input structures, avoiding exhaustive search over poses. Surface-based methods such as MaSIF~\citep{gainza2020} learn interaction fingerprints from local molecular surface geometry and chemistry, and related geometric models such as ScanNet~\citep{tubiana2022} predict binding-site residues from structural features; both lines of work have inspired antibody-specific variants.

For antibody-antigen prediction specifically, DyMean~\cite{kong2023end} is an end-to-end antibody design framework based on a dynamic multi-channel equivariant graph network that, given an epitope, jointly docks the antibody and refines full-atom CDR loops through iterative message passing. HERN~\cite{jin2023iterative} adopts a hierarchical equivariant-refinement procedure that iteratively updates interface geometry. DockGPT~\cite{mcpartlon2023deep} is a transformer-based generative model that formulates flexible, site-aware docking as completion of inter-chain geometry, optionally conditioned on interface residues. These learning-based methods are typically faster at inference than physics-based approaches but often rely on epitope information or structural templates, and their reported accuracy on blind antibody-antigen docking is still limited.

\subsubsection{Our Approach in Context}

Our approach tests whether protein language models can provide enough information for antibody-antigen complex prediction without evolutionary signals or epitope constraints, extending PLM-based methods from monomer prediction (Section~\ref{sec:monomer_prediction}) to complexes. Unlike physics-based methods that search over orientations, or template-based methods that rely on known complexes, our model directly predicts bound structures from PLM embeddings and predicted monomer structures. In our experiments, we compare against representative methods from each category---HDOCK for physics-based docking, DyMean for antibody-specific deep learning, and ESMFold and AlphaFold2-Multimer for general-purpose single-sequence prediction---to test how our PLM-based approach performs relative to different docking strategies. We evaluate both blind docking and epitope-guided scenarios to see how experimental information improves results.

\subsection{Architecture Design}
\label{sec:ch4_architecture}

\subsubsection{Overall Pipeline}

\begin{figure}[htbp]
\centering
\includegraphics[width=\textwidth]{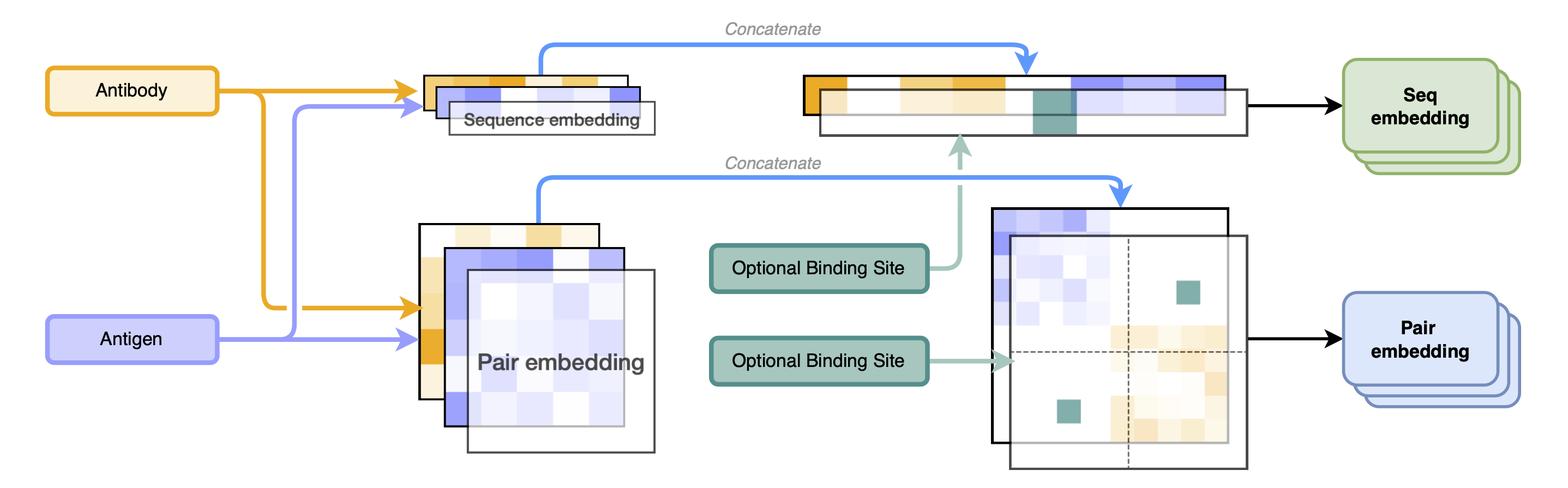}
\caption{Feature extraction pipeline for antibody-antigen complex prediction. Multi-PLM sequence embeddings and block-structured pair embeddings encode both intra-chain and cross-chain relationships, with optional binding-site guidance for epitope-guided docking.}
\label{fig:complex_architecture_part1}
\end{figure}

\begin{figure}[htbp]
\centering
\includegraphics[width=\textwidth]{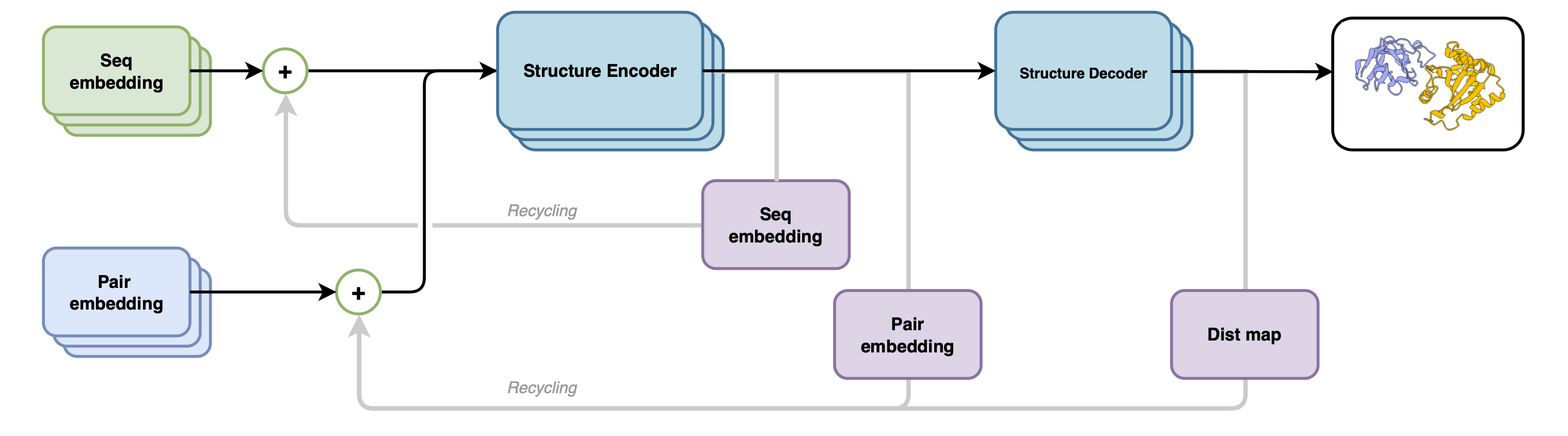}
\caption{Structure prediction module. A 6-block transformer encoder refines representations through triangular operations, and an IPA-based decoder generates 3D coordinates with iterative recycling for interface refinement.}
\label{fig:complex_architecture_part2}
\end{figure}

The complex predictor extends the monomer architecture from Section~\ref{sec:monomer_prediction} through a four-stage pipeline (Figures~\ref{fig:complex_architecture_part1} and~\ref{fig:complex_architecture_part2}):

\begin{enumerate}
    \item \textbf{Feature extraction}: Sequence representations from multiple PLMs are extracted for each chain and combined with geometric features from predicted monomer structures.
    \item \textbf{Pair representation construction}: A block-structured pair matrix captures intra-chain structure (diagonal blocks from monomer predictions) and initializes cross-chain blocks with basic sequence-derived features such as amino acid identity, chain type, and relative position. In the epitope-guided setting, optional binding-site information is also added to the cross-chain blocks.
    \item \textbf{Structure encoder}: A 6-block transformer refines sequence and pair embeddings, sharing information across chains through triangular attention and multiplication operations.
    \item \textbf{Structure decoder with recycling}: An IPA-based decoder generates 3D coordinates from the refined representations, with three recycling iterations to progressively correct interface geometry.
\end{enumerate}

This design uses pre-trained monomer representations while focusing new computation on the cross-chain interface prediction problem.

\subsubsection{Design Principles and Key Decisions}

Our architecture builds on the antibody-specific single-sequence model from Section~\ref{sec:monomer_prediction}, extending it to handle antibody-antigen complexes. We make several key design decisions:

\textbf{Predicted vs. ground-truth monomer structures.} We use predicted monomer structures (from Section~\ref{sec:monomer_prediction}) rather than ground-truth unbound structures for two reasons. First, this matches the practical setting where only sequences are available. Second, using predicted structures during training keeps training and inference consistent. The model learns to work with realistic predictions that may have local errors, making it more robust.

\textbf{Minimal architectural changes.} We reuse the antibody-finetuned Evoformer weights for antibody chains and general PLM weights for antigen chains. The only new trainable parameters are the feature projection layers (Eq.~\ref{eq:seq_proj}, \ref{eq:pair_proj}). This design uses the structural patterns learned during antibody training while adding cross-chain modeling with few new parameters.

\textbf{Interface-focused pair features.} We fill only the off-diagonal blocks of the pair matrix with complex-specific features. The diagonal blocks keep the single-sequence Evoformer outputs, preserving intra-chain structure while focusing new computation on inter-chain interfaces where antibody meets antigen.

\textbf{Lighter encoder for efficiency.} Complex prediction involves longer sequences (heavy + light + antigen, typically 400--800 residues) than monomer prediction. We use 6 encoder blocks instead of the 24 Evoformer layers in Section~\ref{sec:monomer_prediction} to balance speed and accuracy, since much structural information is already in the predicted monomers and PLM embeddings. The encoder mainly needs to model cross-chain interactions rather than rebuild full structure from scratch.

\subsubsection{Sequence Features}

The feature extraction pipeline transforms raw sequences and predicted monomer structures into the representations consumed by the encoder and decoder. We describe the sequence-derived features here and the pair (structure-derived) features in the next subsection. Every chain uses the same multi-PLM embeddings described in Section~\ref{sec:monomer_prediction}. For antibodies we reuse the weights from the antibody-finetuned model, while antigen embeddings come from the general pre-trained model to keep broad structural knowledge. The PLM outputs concatenate the hidden states from three models:
\begin{equation}
\mathbf{h}_{i,c}^{\text{LM}} = \mathbf{e}_{i,c}^{\text{ESM1b}} \oplus \mathbf{e}_{i,c}^{\text{ESM1v}} \oplus \mathbf{e}_{i,c}^{\text{ProtT5}}.
\end{equation}
We denote the Evoformer sequence activations after the final recycling pass from the single-sequence model as $\mathbf{h}_{i,c}^{\text{Evo}}$. Sequence-derived annotations include several features:
\begin{equation}
\mathbf{h}_{i,c}^{\text{seq}} = \mathbf{o}_{i,c} \oplus \mathbf{c}_{i,c}^{\text{type}} \oplus m_{i,c}^{\text{CDR}} \oplus \mathbf{s}_{i,c}^{\text{sym}},
\end{equation}
where $\mathbf{o}_{i,c}$ is the one-hot amino acid identity, $\mathbf{c}_{i,c}^{\text{type}}$ indicates chain type (heavy, light, or antigen), $m_{i,c}^{\text{CDR}}$ is a binary CDR marker (1 for CDR residues, 0 otherwise), and $\mathbf{s}_{i,c}^{\text{sym}}$ provides symmetry identifiers. Optional external guidance (e.g., binding-site hints from experiments) provides an additional signal $\mathbf{h}_{i,c}^{\text{guide}}$, which is zero when not available.

Because all chains are processed together, we concatenate them along the sequence dimension before projection. All four feature groups are then projected to the common embedding dimension and added:
\begin{equation}
\label{eq:seq_proj}
\hat{\mathbf{s}}_{i,c} = U_{\text{LM}}^{(c)} \mathbf{h}_{i,c}^{\text{LM}} + U_{\text{Evo}}^{(c)} \mathbf{h}_{i,c}^{\text{Evo}} + U_{\text{seq}}^{(c)} \mathbf{h}_{i,c}^{\text{seq}} + U_{\text{guide}}^{(c)} \mathbf{h}_{i,c}^{\text{guide}} + \mathbf{b}_{\text{seq}}^{(c)}.
\end{equation}
This projection step is the only place new weights are added for sequence features, keeping the rest of the encoder unchanged.

\subsubsection{Structure-Derived Pair Features}

Pair features reuse both the general pre-trained antigen encoder and the antibody-finetuned encoder from Section~\ref{sec:monomer_prediction}. The pair activations are arranged in block matrices: diagonal blocks contain the single-sequence Evoformer outputs (intra-chain structure), and off-diagonal blocks describe cross-chain interactions.

Sequence-derived pair features extend the single-sequence inputs with symmetry information:
\begin{equation}
\mathbf{h}_{ij}^{\text{seq-pair}} = \mathbf{p}_{ij}^{(0)} \oplus \mathbf{c}_{ij}^{\text{type}} \oplus \pi_{ij}^{\text{sym}} \oplus \mathbf{s}_{ij}^{\text{entity}},
\end{equation}
where $\mathbf{p}_{ij}^{(0)} = W_L \mathbf{o}_i + W_R \mathbf{o}_j + W_{\text{pos}} \mathbf{p}_{ij}^{\text{pos}}$ encodes amino-acid identities and relative positions. The other components are: $\mathbf{c}_{ij}^{\text{type}}$ for the pair of chain types (e.g., heavy-antigen), $\pi_{ij}^{\text{sym}}$ as a symmetry marker, and $\mathbf{s}_{ij}^{\text{entity}}$ for entity identifiers. The Evoformer pair activations from the single-sequence model are denoted $\mathbf{h}_{ij}^{\text{Evo}} = \tilde{\mathbf{p}}_{ij}$.

Structure-derived signals come from the monomeric structures predicted in Section~\ref{sec:monomer_prediction}:
\begin{equation}
\mathbf{h}_{ij}^{\text{struct}} = d_{ij} \oplus \phi_{ij} \oplus \psi_{ij} \oplus \omega_{ij} \oplus \mathbf{s}_i^{\text{sec}} \oplus \mathbf{s}_j^{\text{sec}} \oplus \mathbf{d}_{ij}^{\text{recycle}},
\end{equation}
where $d_{ij}$ is the C$_\alpha$ distance, $(\phi_{ij}, \psi_{ij}, \omega_{ij})$ encode relative orientations, $\mathbf{s}^{\text{sec}}$ are secondary-structure labels, and $\mathbf{d}_{ij}^{\text{recycle}}$ is the recycled distance map from the structure module. Optional contact guidance $\mathbf{h}_{ij}^{\text{guide}}$ can bias the model toward specific interfaces when prior knowledge is available, and is zero otherwise.

As with sequence features, we project each feature group with its own linear layer before adding:
\begin{equation}
\label{eq:pair_proj}
\mathbf{z}_{ij} = U_{\text{seq-pair}} \mathbf{h}_{ij}^{\text{seq-pair}} + U_{\text{Evo-pair}} \mathbf{h}_{ij}^{\text{Evo}} + U_{\text{struct}} \mathbf{h}_{ij}^{\text{struct}} + U_{\text{guide-pair}} \mathbf{h}_{ij}^{\text{guide}} + \mathbf{b}_{\text{pair}}.
\end{equation}
These projection layers are the only new trainable weights for pair features. The off-diagonal blocks of $\mathbf{z}_{ij}$ use the complex-specific features above, while the diagonal blocks keep the single-sequence Evoformer outputs.

\subsubsection{Structure Encoder}

The structure encoder refines sequence and pair representations through multiple transformer blocks. Unlike the 24-layer Evoformer in Section~\ref{sec:monomer_prediction}, we use a 6-block encoder here for two reasons. First, complex prediction uses longer sequences (heavy + light + antigen, typically 400--800 residues vs. 200--400 for antibodies alone), needing more memory per layer. Second, much structural information is already in the predicted monomers and PLM embeddings, so the encoder mainly needs to model cross-chain interactions rather than rebuild full structure from scratch. The encoder follows triangular attention and multiplication operations to combine monomer geometry with cross-chain pair features; pseudocode is provided in Appendix~\ref{sec:app_plm_complex_details} (Algorithm~\ref{alg:structure_encoder_app}).

\subsubsection{Structure Decoder}

The structure decoder uses Invariant Point Attention (IPA) to generate 3D coordinates from the refined representations. Following Section~\ref{sec:monomer_prediction}, IPA layers work in SE(3) frame space so the output does not depend on input orientation.

We initialize residue frames from the predicted monomer structures rather than identity transformations. Each antibody residue starts with the frame predicted by the antibody-specific model (Section~\ref{sec:monomer_prediction}), and each antigen residue starts with the frame from a general structure predictor. These initial frames give strong geometric guidance that helps the IPA layers find reasonable bound structures, greatly narrowing the search compared to random starting frames.

The frames are then updated through IPA layers:
\begin{equation}
\mathbf{T}_i \gets \text{IPA}(\mathbf{T}_i, \mathbf{s}_i, \{(\mathbf{T}_j, \mathbf{s}_j, \mathbf{z}_{ij})\}_j),
\end{equation}
where $\mathbf{T}_i \in \text{SE}(3)$ is the frame for residue $i$. The IPA mechanism attends to neighboring residues based on both sequence similarity (through $\mathbf{s}_j$) and spatial proximity (through $\mathbf{T}_j$), letting interface residues adjust their positions based on cross-chain contacts in $\mathbf{z}_{ij}$. The pair features in the off-diagonal blocks guide the model to find and refine the antibody-antigen interface.

The final frames produce backbone and side-chain coordinates. We apply recycling three times so later passes can refine earlier predictions using feedback from the full complex geometry. Each recycling iteration feeds the predicted structure back through the encoder and decoder, letting the model fix interface errors and improve accuracy.

\subsubsection{Training Procedure}

We follow a two-stage training strategy similar to Section~\ref{sec:monomer_prediction}: pre-training on general protein-protein complexes followed by fine-tuning on antibody-antigen complexes. The complex predictor reuses antibody-finetuned weights from the monomer model, uses predicted monomer structures during training to match inference, and adds interface/contact supervision for cross-chain interaction learning. Detailed training data, augmentation, and loss settings are provided in Appendix~\ref{sec:app_plm_complex_details}.

With the architecture and training procedure defined, the next section describes the evaluation protocol and baseline methods used to assess complex prediction performance.

\subsection{Evaluation}
\label{sec:ch4_evaluation}

\subsubsection{Evaluation Metrics}

We evaluate complex prediction quality using DockQ~\cite{basu2016dockq}, a composite metric introduced in Chapter~\ref{ch:background}. DockQ combines fraction of native contacts, ligand RMSD, and interface RMSD into a single score from 0 to 1, with standard quality thresholds: High quality (DockQ $\geq 0.8$), Medium ($0.49 \leq$ DockQ $< 0.8$), Acceptable ($0.23 \leq$ DockQ $< 0.49$), and Incorrect (DockQ $< 0.23$). We report \textbf{success rate} as the percentage of predictions with at least Acceptable quality (DockQ $\geq 0.23$).

\subsubsection{Baseline Methods}

The baselines cover several docking settings. \textbf{ESMFold}~\cite{lin2023evolutionary} is a general PLM-based structure predictor that also supports multi-chain complex prediction, testing whether a general single-sequence model can handle antibody-antigen docking without task-specific training. \textbf{AlphaFold2-Multimer}~\cite{jumper2021highly,evans2021protein} is run in single-sequence mode (without MSAs), testing whether its multi-chain architecture can handle complex prediction from sequence alone. \textbf{DyMean}~\cite{kong2023end} is an end-to-end antibody design framework based on a dynamic multi-channel equivariant graph network that jointly docks the antibody to a given epitope and refines CDR loops through iterative message passing; it requires epitope information as input. \textbf{HDOCK}~\cite{yan2017hdock} uses template-free rigid-body docking with FFT-based shape matching and knowledge-based scoring, representing fast physics-based docking without antibody-specific features. All baselines use recommended settings from published papers or official servers.

To fit long antibody-antigen sequences and multi-PLM embeddings within GPU memory, we enable gradient checkpointing throughout the Evoformer and structure module, trading additional compute for lower activation storage~\citep{chen2016training}.

\subsection{Experimental Results}
\label{sec:ch4_results}

\subsubsection{Evaluation Settings}

The test set consists of 40 antibody-antigen complexes from SAbDab: 21 antibodies (heavy + light chains) and 19 nanobodies (single-domain). All structures were deposited after January 2023, ensuring no temporal overlap with the training data. Complexes are filtered for crystallographic resolution $\leq 3.0$ \AA, complete CDR regions and epitopes, and sequence identity $\leq 40\%$ to any training complex.

The two subsets differ in several structural characteristics relevant to docking difficulty. Antibodies present a larger paratope surface spanning six CDR loops across two chains, requiring the model to correctly orient both heavy and light chains relative to the antigen. Nanobodies, by contrast, use only three CDR loops on a single domain, resulting in a smaller and more constrained interface. The test set covers diverse antigen targets to evaluate generalization across different binding contexts and antigen sizes.

Two docking scenarios test different levels of prior information:

\begin{enumerate}
    \item \textbf{Blind docking (no epitope prior):} The model receives only sequence inputs and predicted monomer structures. It must identify the correct binding interface without any epitope information, searching the entire antigen surface. This scenario tests whether sequence information alone suffices for complex prediction.

    \item \textbf{Epitope-guided docking (with epitope prior):} Binding-site information is provided by marking native epitope residues (antigen residues within $10$ \AA\ of any antibody residue in the native complex). This is an upper-bound setting for binding-site guidance. It simulates the type of information that experiments such as alanine scanning, cross-linking mass spectrometry, or hydrogen-deuterium exchange try to provide, but without experimental noise.
\end{enumerate}

For the epitope-guided docking setting, 12 epitope site initializations are sampled per prediction, and 15 predictions per target are generated with different random seeds, providing diversity in both binding-site hypothesis and structural sampling. Each prediction uses 3 recycling iterations, with predicted monomer structures from Section~\ref{sec:monomer_prediction} for antibody chains and ESMFold for antigen chains, and all three PLMs (ESM-1b, ESM-1v, ProtT5-XL). Predictions are ranked by a confidence score that combines interface-residue pLDDT with predicted aligned error (PAE), weighting interface confidence more heavily than global alignment quality. This ranking scheme prioritizes predictions where the model is confident about the interface geometry, which is the most challenging aspect of antibody-antigen complex prediction.

\subsubsection{Blind Docking Results}

\begin{figure}[htbp]
\centering
\includegraphics[width=0.6\textwidth]{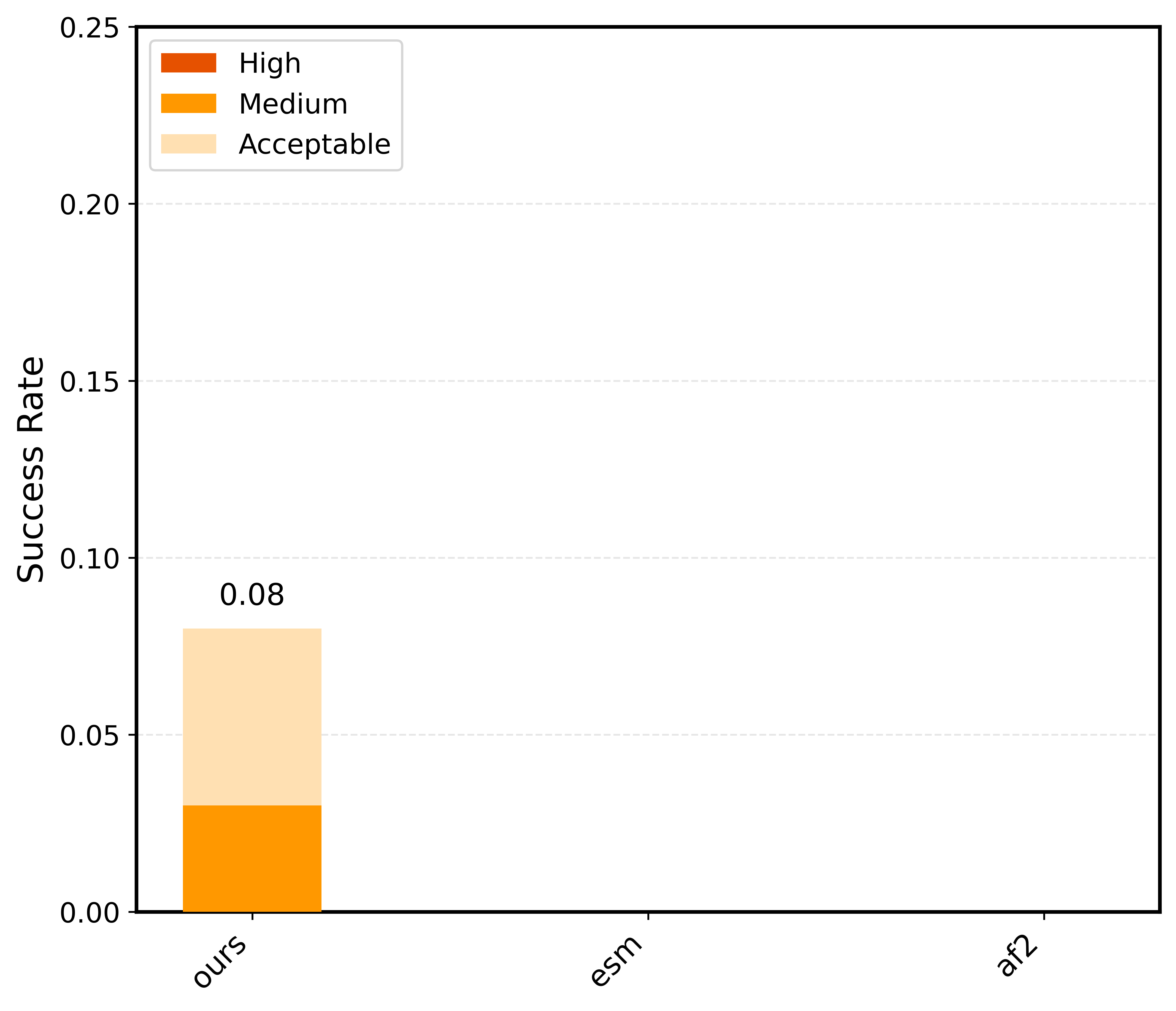}
\caption{Blind docking results without epitope information. Our method achieves 8\% success rate (DockQ $\geq 0.23$), while both ESMFold and AlphaFold2-Multimer achieve 0\% success rate on antibody-antigen complex prediction in this setting.}
\label{fig:blind_docking}
\end{figure}

Figure~\ref{fig:blind_docking} presents results for blind docking without any epitope prior information.

\begin{figure}[htbp]
\centering
\includegraphics[width=0.5\textwidth]{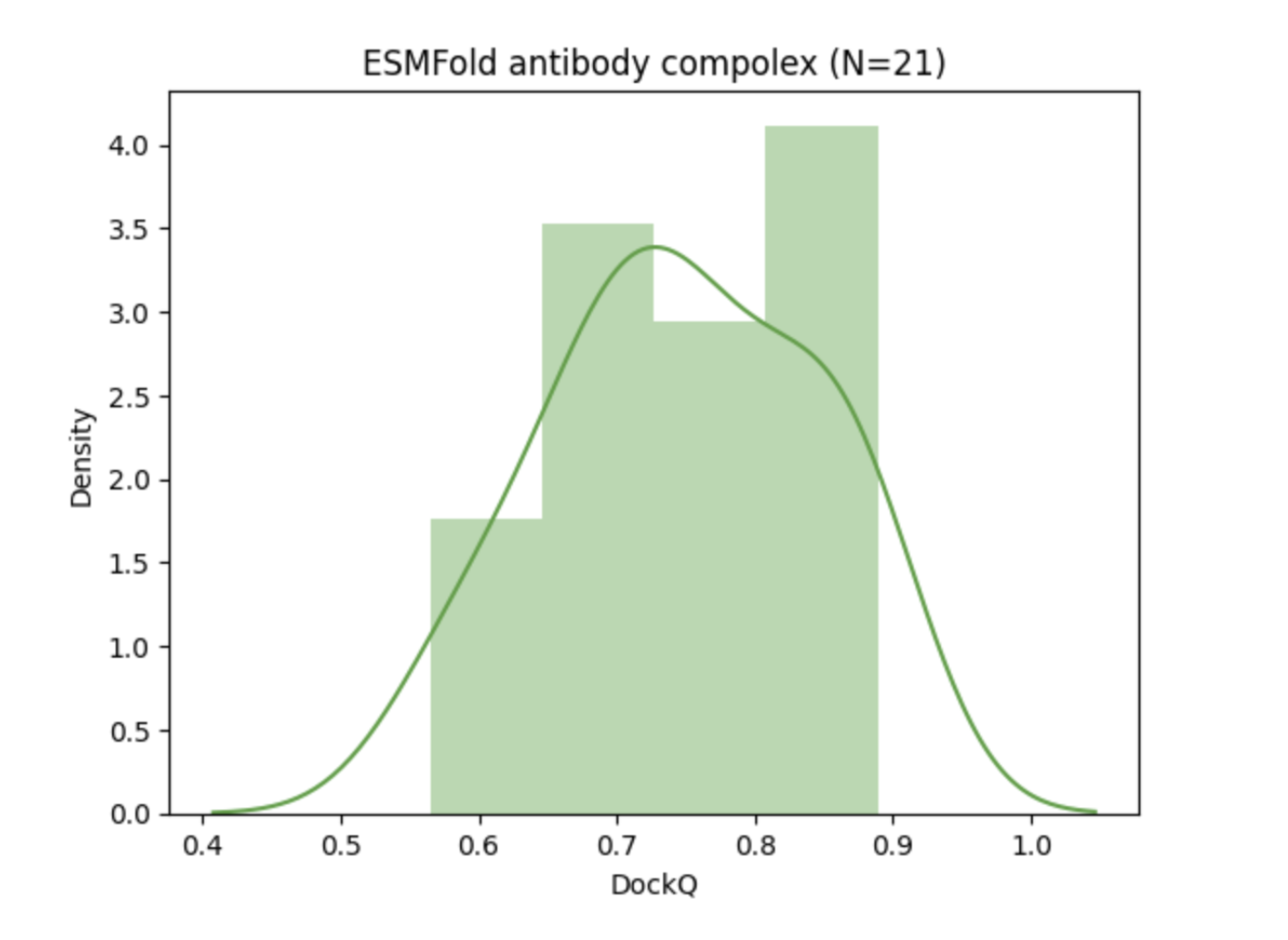}
\caption{DockQ distribution for ESMFold on antibody-antigen complexes with heavy and light chains (N=21), evaluated separately for the heavy-light interface and the antibody-antigen interface. ESMFold preserves intra-antibody geometry (DockQ 0.6--0.9) but all inter-protein interface predictions remain below the acceptable threshold (DockQ $< 0.23$).}
\label{fig:esmfold_analysis}
\end{figure}

\textbf{General-purpose sequence-based baselines achieve 0\% success.} Both ESMFold and AlphaFold2-Multimer achieve 0\% success rate on antibody-antigen complex prediction (Figure~\ref{fig:blind_docking}), despite their general-purpose strength on monomer benchmarks (Section~\ref{sec:monomer_prediction}). To understand the source of failure, we analyze ESMFold's predictions on the 21 antibody-antigen complexes (heavy + light chains), separately evaluating the intra-antibody interface (heavy-light chain positioning) and the inter-protein interface (antibody-antigen binding) (Figure~\ref{fig:esmfold_analysis}).

The key findings from blind docking are:

\begin{enumerate}
    \item \textbf{Modest success with antibody-specific training}: Our method achieves 8\% success rate (3\% medium + 5\% acceptable quality), demonstrating that some complex prediction is possible from sequence and predicted monomer structure alone, but the success rate remains low.

    \item \textbf{General methods achieve 0\% success}: Both ESMFold and AlphaFold2-Multimer achieve 0\% success rate. ESMFold correctly models the intra-antibody structure---heavy-light chain pairing is accurate in all 21 cases (Figure~\ref{fig:esmfold_analysis})---but places the antigen essentially at random relative to the antibody. The model cannot distinguish binding surfaces from non-binding surfaces without evolutionary or supervised interface signals.

    \item \textbf{Interface identification as the bottleneck}: The dominant failure mode is incorrect interface identification rather than poor local geometry. Antibody-specific training provides a weak signal for binding site localization, but it is insufficient for reliable blind docking.
\end{enumerate}

\subsubsection{Epitope-Guided Docking Results}

\begin{figure}[htbp]
\centering
\includegraphics[width=\textwidth]{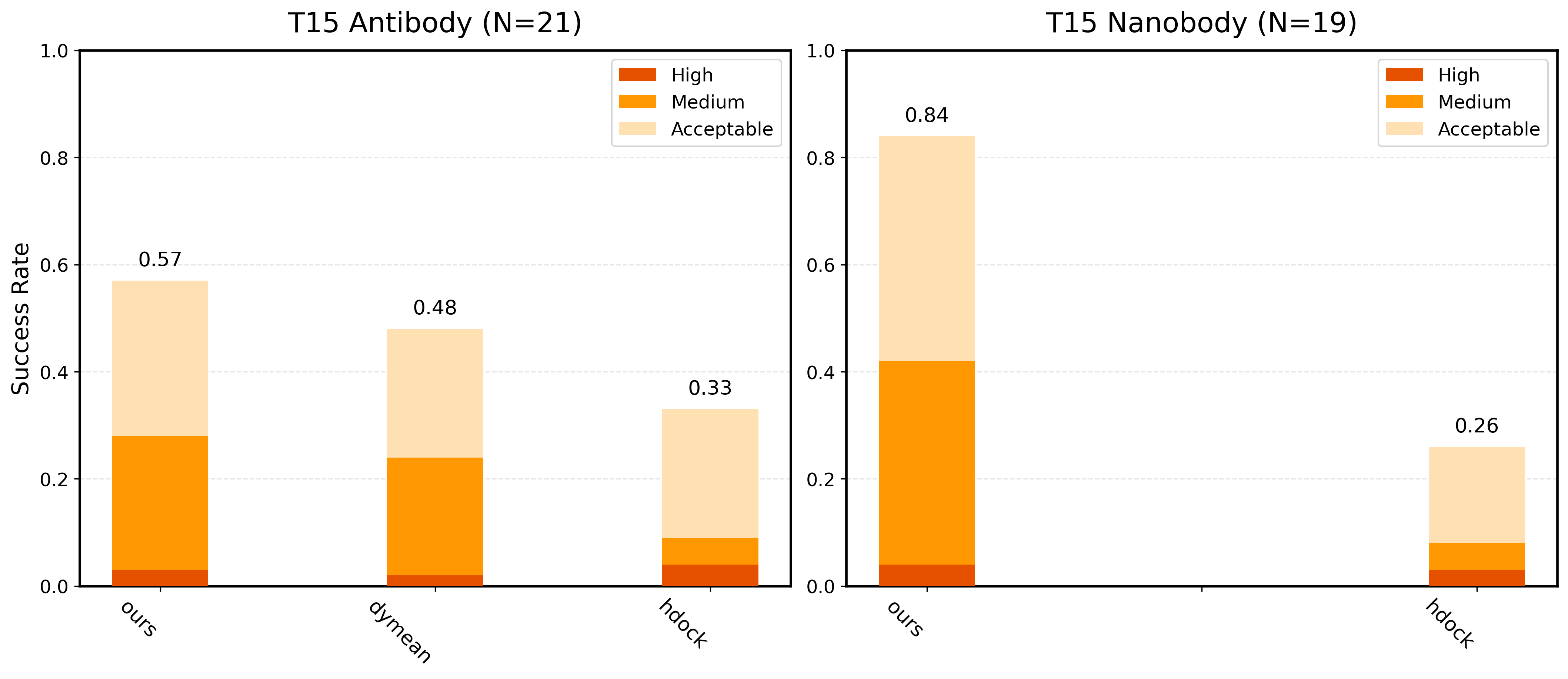}
\caption{Epitope-guided docking results with 15 predictions per target. Left: Antibodies with heavy and light chains (N=21). Right: Nanobodies (N=19). Our method achieves 57\% and 84\% success rates respectively, outperforming both DyMean and HDOCK.}
\label{fig:epitope_guided}
\end{figure}

Figure~\ref{fig:epitope_guided} presents results when binding-site information guides interface prediction.

The key findings from epitope-guided docking are:

\begin{enumerate}
    \item \textbf{Large improvement with epitope information}: Native epitope guidance increases the success rate from 8\% (blind) to 57\% on antibodies, a $7.1\times$ improvement. This gap shows that interface identification, not geometric assembly, is the primary bottleneck for PLM-based complex prediction.

    \item \textbf{Outperforms existing methods}: With epitope guidance, our method (57\%) outperforms both HDOCK (33\%), a physics-based docking approach, and DyMean (48\%), an antibody-specific end-to-end framework that jointly docks the antibody to a given epitope and refines CDR loops.

    \item \textbf{Nanobodies achieve higher success}: Nanobody predictions reach 84\% success rate (4\% high + 38\% medium + 42\% acceptable quality), much higher than the 57\% for antibodies. The simpler single-domain architecture, smaller interface area, and reduced conformational flexibility of nanobodies all contribute to this advantage. HDOCK achieves only 26\% on nanobodies, indicating that learning-based methods benefit particularly from the structural regularity of single-domain antibodies.

    \item \textbf{Most successes are acceptable quality}: On antibodies, the quality breakdown is 3\% high, 25\% medium, and 29\% acceptable. Most successful predictions identify the correct binding mode but do not reach near-native accuracy, suggesting that local interface refinement remains a challenge.

    \item \textbf{High-quality predictions remain rare}: Across both subsets, high-quality predictions (DockQ $\geq 0.8$) make up only 3--4\% of cases. Achieving atomic-level accuracy likely requires better handling of conformational flexibility and fine-grained interface packing.
\end{enumerate}

\subsection{Discussion}
\label{sec:ch4_discussion}

\subsubsection{Why PLMs Fail at Interface Identification}

The blind docking results (Section~\ref{sec:ch4_results}) show that PLM-derived representations, despite encoding useful structural information for monomers, cannot reliably identify antibody-antigen interfaces. The problem is not model size or training data---these PLM representations do not provide enough target-specific information for interface prediction. Several properties of antibody-antigen binding make it different from typical protein-protein interactions:

\begin{enumerate}
\item \textbf{Adversarial co-evolution:} Antibodies and antigens evolve in competition, where each partner changes to counter the other. This removes the cooperative evolutionary patterns that structure predictors normally use. Unlike stable protein complexes where paired MSAs show correlated mutations at interface positions, antibody-antigen MSAs have no useful pairing signals.

\item \textbf{Large search space:} The antibody-antigen interface occupies only a small portion of the antigen surface, while the rest of the antigen exposes many candidate patches of comparable size and chemistry (hydrophobic cores, charged regions, protruding loops). Without evolutionary signals to narrow the search, the model must tell the true epitope from many incorrect sites using geometric and sequence information alone.

\item \textbf{Many-to-many sequence-structure mapping:} CDR-H3 hypervariability means that different antibodies binding the same epitope have almost no sequence similarity in their paratopes. This breaks the connection between sequence and structure that PLMs depend on---similar binding modes come from different sequences, and similar sequences can bind different epitopes.
\end{enumerate}

These factors explain the low 8\% success rate in blind docking. PLM embeddings help with monomer folding (Section~\ref{sec:monomer_prediction}) but do not extend to inter-molecular recognition. Chapter~\ref{ch:pipeline_adaptation} develops training-free MSA-based refinements for antibody-antigen targets.

\subsubsection{Interface Identification as the Main Bottleneck}

The $7.1\times$ improvement from blind docking (8\%) to epitope-guided docking with native epitope labels (57\%) shows that interface identification is the main source of error. Once the binding region is known, the model can assemble complexes with reasonable accuracy, showing that the geometric assembly parts of the model---the structure encoder, IPA decoder, and recycling mechanism---work well for their intended purpose. The bottleneck is upstream: finding \emph{where} on the antigen the antibody binds, not \emph{how} it binds once the site is known.

This also suggests that noisy or partial binding-site information---from low-resolution experiments, computational epitope predictors, or other data sources---may help by narrowing the search space, although the results here use native epitope labels.

\subsubsection{Implications for Experimental Integration}

The performance of epitope-guided docking with native epitope labels (57\% antibodies, 84\% nanobodies) shows that PLM-based methods remain valuable for complex prediction when binding-site information is available. This suggests an appealing direction: combining experimental or computational epitope information with PLM-based prediction. Several types of experiments can provide binding-site information at various levels of detail:

\begin{itemize}
    \item \textbf{Cross-linking mass spectrometry} can identify residue pairs that are close in the bound state, providing distance restraints that help narrow the interface search.
    \item \textbf{Hydrogen-deuterium exchange (HDX)} shows regions whose solvent accessibility changes upon binding, giving coarse epitope localization.
    \item \textbf{Alanine scanning} finds residues important for binding affinity, directly showing key interface contacts.
    \item \textbf{Computational binding-site prediction.} Tools can provide priors even before experiments are performed: antigen-side epitope predictors such as DiscoTope, ElliPro, and EpiScan~\citep{sanchez2017fundamentals,kringelum2012reliable,ponomarenko2008ellipro,wang2024episcan} predict candidate epitope regions on the antigen; antibody-side paratope predictors such as Parapred~\citep{liberis2018parapred} predict CDR residues likely to contact the antigen; and recent AlphaFold2-based adaptations such as PAbFold~\citep{deroo2024} predict linear antibody epitopes.
\end{itemize}

Even partial or noisy data from these sources may narrow the search and improve accuracy. A practical workflow would combine computational epitope prediction with experimental validation, using the resulting binding-site priors to guide complex prediction.

\subsubsection{Comparison with Baseline Methods}

Our method with native epitope guidance (57\% success on antibodies, 84\% on nanobodies) outperforms physics-based HDOCK (33\% on antibodies, 26\% on nanobodies) and antibody-specific DyMean (48\%). This advantage comes from several factors: (i) antibody-specific training that teaches typical binding patterns and CDR shapes; (ii) good use of epitope information through binding-site features in the pair representation; and (iii) learned structural patterns from the encoder that find matching surfaces better than rigid-body search.

The 0\% blind docking success of general methods (ESMFold and AlphaFold2-Multimer) shows why task-specific training matters. These methods work well for monomer prediction but cannot tell binding from non-binding interfaces without evolutionary signals or interface supervision during training. Specialized antibody structure predictors such as NanoNet, ABodyBuilder3, and tFold-Ab~\citep{cohen2022nanonet,kenlay2024abodybuilder3,wu2024tfold} show similar benefits from domain-specific architecture and training; our results show that the same principle extends to complex prediction, though interface guidance remains necessary for high success rates.

\subsubsection{Limitations and Future Directions}

Several limitations remain in the current approach:

\paragraph{Conformational flexibility.} The model treats monomer structures as rigid inputs, so it cannot handle shape changes upon binding where CDR loops or antigen surface regions rearrange. Future work could add learned conformational sampling---for example, generating an ensemble of CDR-H3 conformations and selecting the best-fitting pose during docking---to better handle cases where the bound structure differs from the unbound prediction.

\paragraph{Confidence estimation.} The current confidence score, which combines interface-residue pLDDT with predicted aligned error, provides a rough ranking but does not reliably distinguish correct from incorrect poses. Developing interface-aware confidence metrics, potentially trained on a set with known DockQ scores, could improve model selection and bring the top-ranked performance closer to the oracle (best-of-$N$) performance.

\subsection{Summary}
\label{sec:ch4_summary}

This chapter tested PLM-based methods for antibody-antigen complex prediction, extending the monomer architecture from Section~\ref{sec:monomer_prediction} to multi-chain complexes. We combined PLM embeddings with predicted monomer structures through a pipeline with block-structured pair representations and an IPA-based decoder.

The main finding is that interface identification is the bottleneck. General methods (ESMFold, AlphaFold2-Multimer) achieve 0\% success rate in blind docking, and even antibody-specific training achieves only 8\% without epitope information. However, providing native epitope labels raises the success rate to 57\% on antibodies and 84\% on nanobodies, outperforming HDOCK (33\%, 26\%) and DyMean (48\%) in this setting. This shows that geometric assembly works well once the binding region is known.

Nanobodies consistently perform better than conventional antibodies in our benchmark, likely because of their simpler single-domain architecture~\citep{vanbockstaele2009nanobody}. These results show the limits of what PLM-based methods can achieve for complex prediction, and motivate the MSA-based methods developed in Chapter~\ref{ch:pipeline_adaptation}.

\section{Synthesis: Where PLM Representations Help and Fail}
\label{sec:plm_synthesis}

The two studies in this chapter show a clear boundary for PLM-based antibody modeling. For antibody monomers, PLM representations provide useful intra-chain structural priors. The immunoglobulin framework has a conserved fold, and even the hypervariable CDR loops follow structural constraints that can be learned from large-scale sequence pretraining and antibody-specific fine-tuning. This explains why RaptorX-Single-Ab improves CDR-H3 modeling without requiring target-specific MSAs.

For antibody-antigen complexes, the limiting factor changes. The main challenge is no longer whether the antibody chain can fold, but whether the model can identify the correct binding interface on a large antigen surface. This requires inter-chain information: which antigen residues form the epitope, which antibody residues form the paratope, and how the two surfaces should be oriented. Generic PLM embeddings are learned from single-chain sequence statistics and do not provide reliable target-specific evidence for these contacts. As a result, blind docking remains poor even when monomer structures and local geometry are modeled well.

The epitope-guided results clarify this distinction. When approximate binding-site information is provided, the same PLM-based complex model performs much better, indicating that the geometric assembly problem is tractable once the search space is constrained. The failure mode is therefore interface identification rather than complete inability to model complexes.

The limitation of PLM-based complex prediction is therefore not a failure of antibody structure modeling in general. Rather, it is a failure to recover target-specific inter-chain binding signal from sequence-only representations. This points to MSA-based predictors as the next step, since their evolutionary alignments encode structural constraints that single-sequence embeddings do not. Chapter~\ref{ch:pipeline_adaptation} develops training-free MSA-based refinements for antibody-antigen prediction along this line.

\chapter{MSA-Based Methods for Antibody-Antigen Complex Prediction}
\label{ch:pipeline_adaptation}

\section{Motivation}
\label{sec:ch4_motivation}

\subsection{From PLM Limitations to MSA-Based Complex Prediction}

Chapter~\ref{ch:plm_modeling} showed that PLM-based models can learn intra-chain antibody structure but cannot reliably identify inter-chain antibody-antigen interfaces from sequence alone. This motivates a return to MSA-based methods, where evolutionary alignments provide target-specific structural constraints that single-sequence embeddings do not encode. Among current MSA-based predictors, AlphaFold3~\citep{abramson2024accurate} is a leading recent general structure predictor; however, use and modification of its parameters are restricted, so unrestricted model-level fine-tuning on antibody-antigen data is not available in the usual open-source sense. We therefore focus on training-free refinements to MSA construction and inference behavior, which can be applied without changing model weights whenever these stages are configurable.

Two characteristics of antibody-antigen prediction shape the failure modes that motivate our interventions. First, antibody MSAs are dominated by conserved framework matches that align easily across many antibody sequences, while the hypervariable CDR loops---the regions most relevant for binding---tend to be poorly covered. Second, antibodies and antigens evolve in a non-cooperative manner, so reliable cross-chain co-evolutionary signal is sparse, which can make AlphaFold3's iterative recycling less stable on these targets.

This chapter develops two training-free MSA-based interventions:

\begin{itemize}
  \item \textbf{MSA refinement}: improve the binding-relevant content and effective depth of antibody MSAs through CDR-focused filtering and depth recovery.
  \item \textbf{Convergence-aware recycling}: select a stable intermediate recycle state for final diffusion sampling, based on a per-recycle convergence signal.
\end{itemize}

Both interventions adjust MSA construction or inference behavior without changing model parameters, which has several practical consequences:
\begin{itemize}
  \item They can be applied without retraining or model-weight access when MSA construction and inference stages are configurable, as in AF3-style MSA-based predictors.
  \item They require no retraining and can in principle be composed with model-level adaptation strategies, since the two directions modify different parts of the prediction stack.
  \item Changes to inputs or inference behavior are easier to inspect and reason about than learned weight changes.
\end{itemize}

\section{MSA Refinement for Antibody-Antigen Prediction}
\label{sec:ch6_msa_refinement}

\subsection{Framework-Dominated Antibody MSAs}

Antibodies combine highly conserved framework regions with hypervariable CDR loops, creating an important challenge for MSA construction. The conserved frameworks align easily across different antibody sequences, while the highly variable CDRs align poorly due to variable length and diverse sequences.

\begin{figure}[htbp]
\centering
\includegraphics[width=\textwidth]{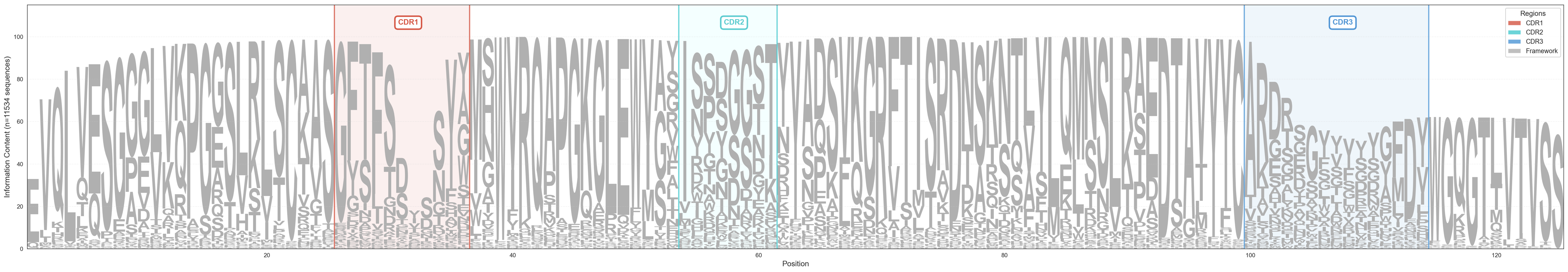}
\caption{Sequence logo plot for a representative antibody heavy chain MSA. Framework regions (left and right) show strong conservation with tall amino acid preferences, while CDR regions (boxes) show reduced information content with shorter letters and more uniform distributions. CDR-H3 shows the most severe degradation, reflecting its hypervariable nature and alignment difficulty.}
\label{fig:antibody_msa_logo}
\end{figure}

This results in MSAs with abundant high-quality framework alignments but extensive gaps in CDR regions (Figure~\ref{fig:antibody_msa_logo}). Since CDRs determine binding specificity, these framework-dominated alignments provide little useful CDR-informative signal for the regions that matter most.

To quantify this effect, we analyzed the gap frequency across IMGT-numbered regions for all antibody chains in our validation set (Table~\ref{tab:gap_profile}). The contrast is stark: framework regions (FR1--FR3) exhibit gap means below 0.12, while CDR3 and FR4---which borders the hypervariable loop---show gap means of 0.42 and 0.48 for heavy chains, respectively. Light chains follow the same pattern, with CDR3 (0.33) and FR4 (0.65) as the most gap-heavy regions. The high gap frequency in FR4 is likely a cascading alignment artifact: insertions and deletions in CDR3 disrupt the positional alignment of downstream residues, causing FR4 to be systematically misaligned. These numbers confirm that standard MSA construction yields alignments where CDR3 coverage is severely degraded compared to framework regions, with gap frequencies 5--30$\times$ higher in binding-critical positions.

\begin{table}[htbp]
\centering
\caption{Mean gap frequency per IMGT region in unpaired MSAs (validation set, AF3 pipeline). CDR3 and FR4 have much higher gap frequencies than other regions, showing that standard MSA construction produces framework-dominated alignments with poor coverage of binding-critical positions.}
\label{tab:gap_profile}
\begin{tabular}{lccccccc}
\toprule
\textbf{Chain} & \textbf{FR1} & \textbf{CDR1} & \textbf{FR2} & \textbf{CDR2} & \textbf{FR3} & \textbf{CDR3} & \textbf{FR4} \\
\midrule
Heavy & 0.103 & 0.043 & 0.016 & 0.038 & 0.018 & \textbf{0.418} & \textbf{0.477} \\
Light & 0.116 & 0.092 & 0.012 & 0.009 & 0.018 & \textbf{0.328} & \textbf{0.646} \\
\bottomrule
\end{tabular}
\end{table}

\subsection{CDR-Focused Filtering}
\label{sec:ch6_cdr_filter}

We hypothesize that filtering out MSA entries with excessive CDR gaps while retaining those with meaningful CDR alignments will improve prediction accuracy. Removing framework-only sequences increases the relative weight of CDR-informative sequences, shifting the model's attention toward binding-relevant positions, and improves the signal-to-noise ratio at functionally relevant positions by focusing on sequences that contain CDR-informative content rather than diluting it with framework-only matches.

For each antibody chain (heavy or light), we first identify CDR regions using the IMGT numbering scheme, which provides standardized numbering for immunoglobulin variable domains. We focus on CDR3 as the most variable and functionally critical region that dominates binding specificity, then define a CDR3 window by expanding the CDR3 boundaries by a small offset to capture the immediate structural context surrounding the hypervariable loop.

For each MSA row (sequence), we apply a simple but effective filtering criterion based on gap content in the CDR3 window. We check the CDR3 window for gap characters and keep only sequences with at least one non-gap residue in the CDR3 window, thereby discarding sequences where the entire CDR3 window consists of gaps. This preserves CDR-informative content specific to the binding loops while discarding framework-only matches that provide no information about CDR structure or binding interactions. Non-antibody chains and antibody chains without valid IMGT numbering are passed through unchanged. The complete pseudocode is provided in Appendix~\ref{sec:appB_cdr3_filter_algo} (Algorithm~\ref{alg:cdr3-filter}).

\paragraph{Unpaired MSA setting.}
Recent work by~\citet{feldman2025af3complex} shows that removing paired MSAs and using only unpaired MSAs can improve AlphaFold3's prediction performance across several complex categories, including antibody-antigen complexes. This finding aligns with our understanding of antibody-antigen co-evolution: unlike protein complexes that co-evolve cooperatively over long evolutionary timescales, antibodies and antigens have adversarial co-evolutionary dynamics where the immune system generates novel binding solutions rather than conserved interaction patterns. Consequently, paired MSAs constructed through standard cross-species pairing heuristics may introduce misleading co-evolutionary signals that confuse rather than inform the model, consistent with computational structural epitope-profiling studies on coronavirus antibodies~\citep{robinson2021}, which build on resources such as the CoV-AbDab repertoire~\citep{raybould2021}. Following this insight, we adopt the unpaired-MSA-only setting throughout this chapter and apply CDR-focused filtering to the individual chain MSAs. For antibody heavy and light chains, filtering is applied independently based on their respective CDR3 regions; antigen chain MSAs are used without CDR-based filtering, since standard evolutionary information remains valuable for antigen structure prediction.

\subsection{MSA Depth Recovery}
\label{sec:ch6_msa_depth}

CDR-focused filtering improves the relevance of antibody MSAs, but it also reduces raw MSA depth: many sequences that were framework-only matches are removed. The second part of the MSA refinement therefore aims to restore effective sequence diversity after filtering. We use BFD, which is much larger than AlphaFold3's default Small BFD, as a practical implementation.

\paragraph{Why depth matters here.}
MSA depth---the number of homologous sequences available for a query protein---strongly correlates with prediction accuracy~\citep{jumper2021highly,jing2024single}. Shallow MSAs provide limited signal, forcing models to rely more heavily on learned priors. For antibody-antigen complexes this relationship is particularly pronounced: antibodies often yield shallow effective MSAs because of the framework-dominated alignment problem described above, and after CDR-focused filtering removes framework-only sequences, the remaining MSA may be much smaller in raw depth. The optimal balance depends on the initial MSA depth; if the starting MSA is already shallow, aggressive filtering may remove too much information.

\paragraph{BFD as the implementation.}
The choice of sequence database directly determines the available evolutionary depth. AlphaFold3's default pipeline uses Small BFD, a clustered subset containing approximately 65 million representative sequences designed to reduce computational cost while maintaining adequate coverage for most proteins. In our implementation, we use BFD~\citep{steinegger2019protein} (approximately 2.5 billion sequences) for the depth-recovery step. For well-studied protein families with abundant homologs, this difference would be negligible; for antibody-antigen complexes with sparse evolutionary coverage, the larger search space increases the probability of finding distant homologs that share structural features with the query antibody, particularly in the hypervariable CDR regions where close homologs are rare by definition. Together, CDR-focused filtering and depth recovery form the MSA refinement strategy used throughout the rest of this chapter: filtering improves relevance at CDR positions, while larger-database search restores effective diversity after filtering.

\subsection{Validation: CDR Coverage and Effective Diversity}
\label{sec:ch6_msa_validation}

We now examine how the two components of MSA refinement affect MSA composition on the validation set, before turning to prediction-quality experiments.

\begin{figure}[htbp]
\centering
\includegraphics[width=0.8\textwidth]{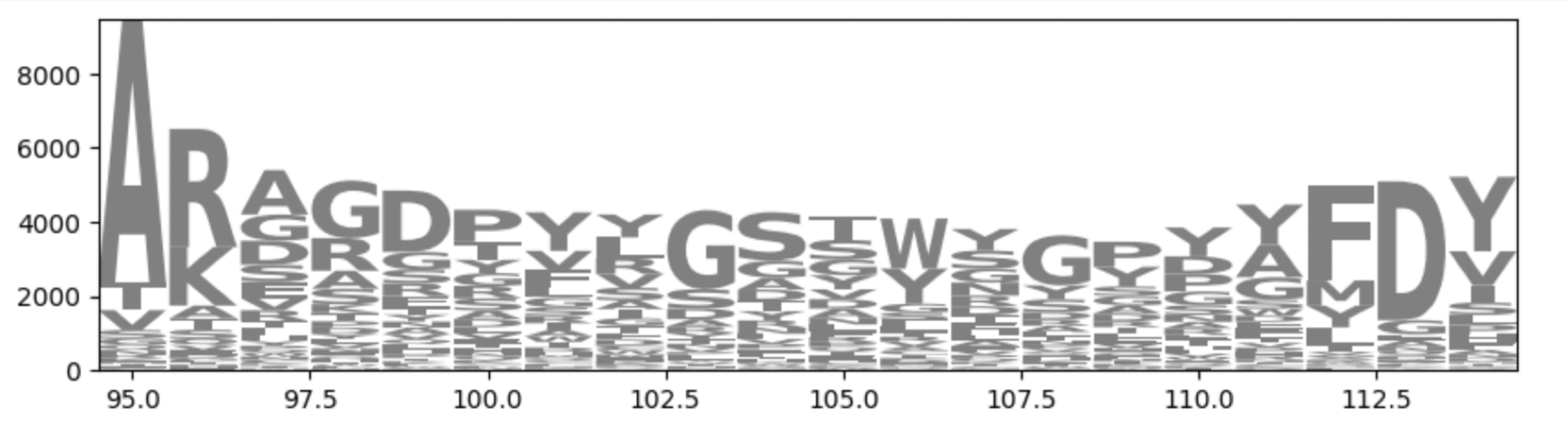}
\caption{Sequence logo plot for CDR-H3 region before filtering. The MSA contains many sequences with gaps in positions 95--114, resulting in low information content. Most of these sequences align well to framework regions but have complete gaps across the hypervariable loop.}
\label{fig:cdr3_before_filter}
\end{figure}

\begin{figure}[htbp]
\centering
\includegraphics[width=0.8\textwidth]{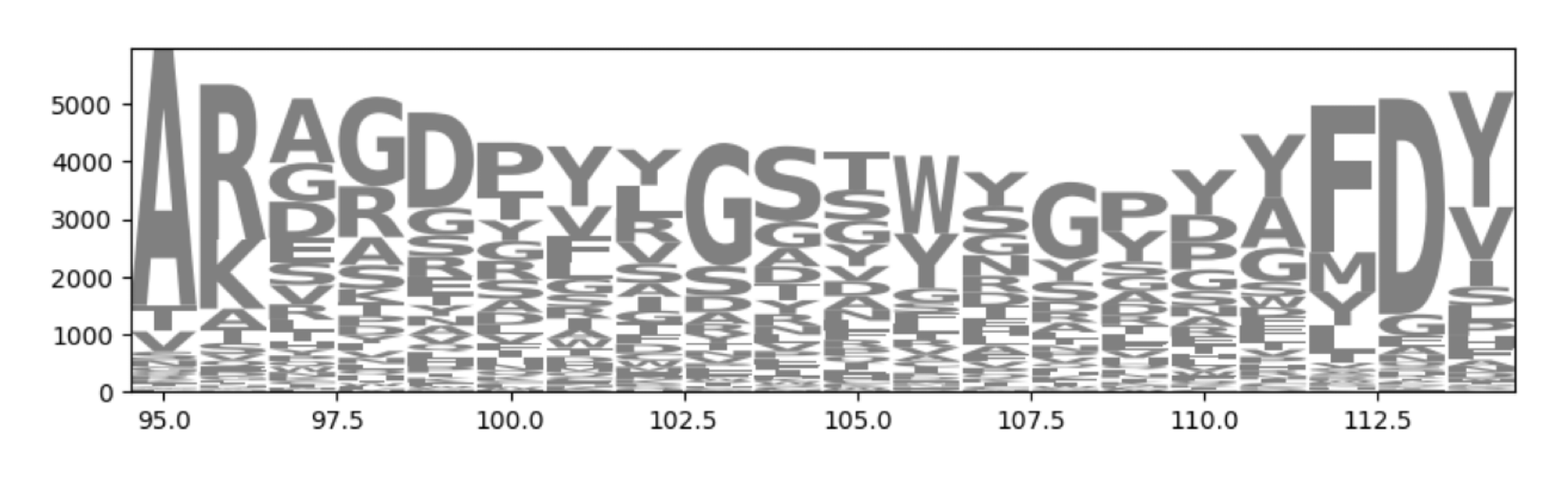}
\caption{Sequence logo plot for CDR-H3 region after CDR-focused filtering. Removing sequences with complete CDR3 gaps improves information content, with taller and more defined amino acid preferences. The remaining sequences contain actual CDR3 alignments, providing stronger CDR-informative signal for the binding region.}
\label{fig:cdr3_after_filter}
\end{figure}

\begin{figure}[htbp]
\centering
\begin{minipage}{0.48\textwidth}
\centering
\includegraphics[width=\textwidth]{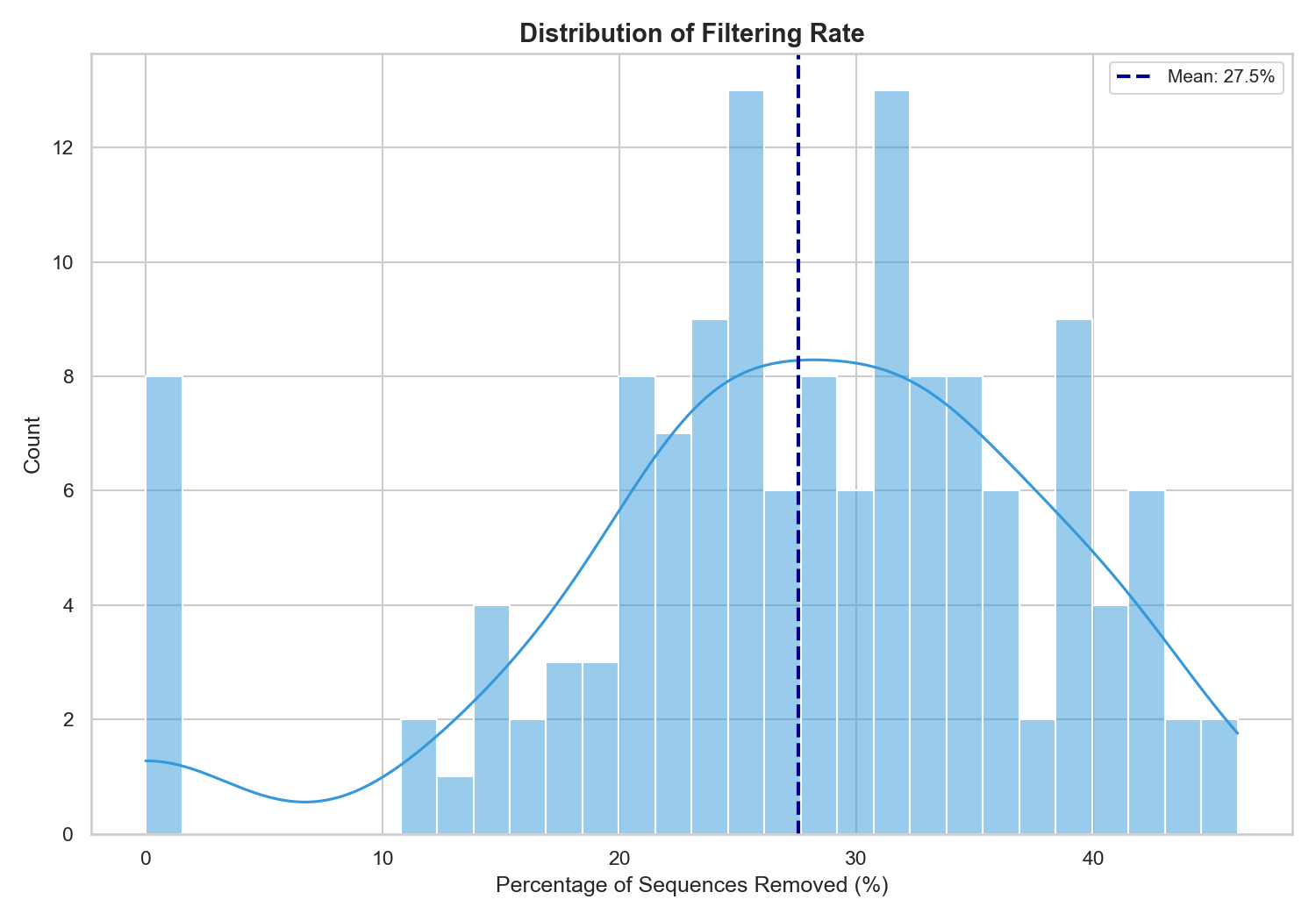}
\end{minipage}
\hfill
\begin{minipage}{0.48\textwidth}
\centering
\includegraphics[width=\textwidth]{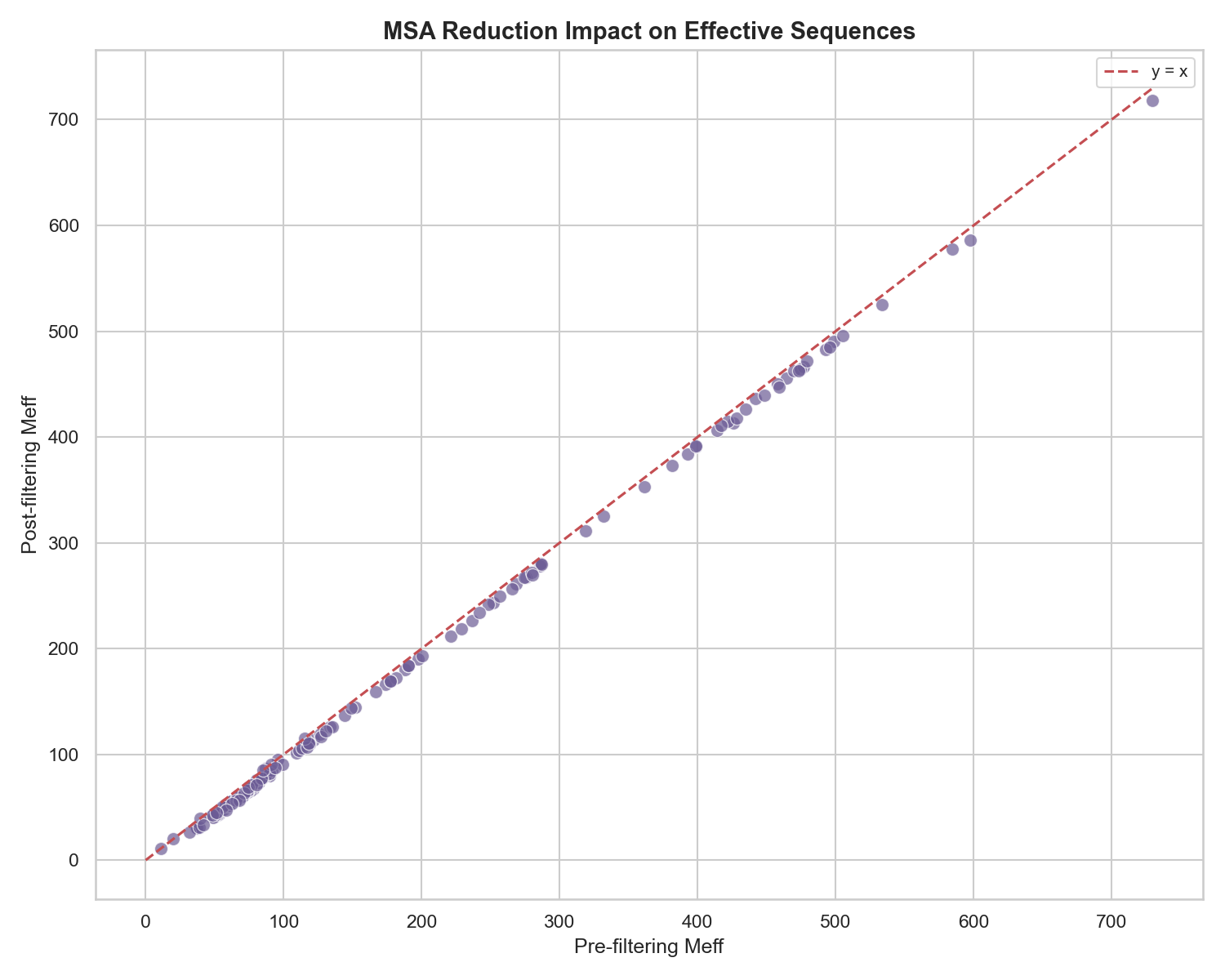}
\end{minipage}
\caption{Impact of CDR-focused filtering on MSA composition. \textbf{Left:} Distribution of filtering rates across the validation set (mean 27.5\% removed). \textbf{Right:} Effective sequence diversity (Meff) before and after filtering. Points cluster along the $y=x$ line, showing that filtering preserves nearly all effective diversity despite removing many raw sequences.}
\label{fig:filtering_impact}
\end{figure}

\paragraph{Effect of CDR-focused filtering on CDR3 coverage.}
Applying CDR3-focused filtering reduces MSA depth by 20--45\% (mean 27.5\%; Figure~\ref{fig:filtering_impact}, left), removing framework-only sequences that contribute minimal information about binding. Despite this reduction, the filtered MSA increases average CDR3 coverage from approximately 40\% to 75\%, nearly doubling the availability of CDR-informative alignment content for the functionally critical binding region, and the effective diversity (Meff) is preserved (Figure~\ref{fig:filtering_impact}, right). The before/after sequence logo plots in Figure~\ref{fig:cdr3_before_filter} and Figure~\ref{fig:cdr3_after_filter} make this concrete for a representative target.

The gap reduction achieved by filtering is targeted (Table~\ref{tab:gap_reduction}). Heavy-chain CDR3 gap frequency drops from 0.418 to 0.211 (a 49.5\% reduction), and FR4 drops from 0.477 to 0.241 (also 49.5\%). Light-chain CDR3 shows a similar pattern, dropping from 0.328 to 0.181 (44.8\% reduction). Framework regions that are already well-aligned show only modest reductions, consistent with filtering selectively improving the regions that need it most.

\begin{table}[htbp]
\centering
\caption{Effect of CDR3-focused filtering on gap frequency per IMGT region (validation set). Filtering reduces CDR3 and FR4 gap frequencies by 40--50\% while framework regions remain largely unchanged, demonstrating targeted improvement of binding-critical positions.}
\label{tab:gap_reduction}
\begin{tabular}{llccccccc}
\toprule
\textbf{Chain} & \textbf{MSA Type} & \textbf{FR1} & \textbf{CDR1} & \textbf{FR2} & \textbf{CDR2} & \textbf{FR3} & \textbf{CDR3} & \textbf{FR4} \\
\midrule
\multirow{2}{*}{Heavy} & Unpaired & 0.103 & 0.043 & 0.016 & 0.038 & 0.018 & 0.418 & 0.477 \\
 & Filtered & 0.075 & 0.032 & 0.009 & 0.026 & 0.003 & \textbf{0.211} & \textbf{0.241} \\
\midrule
\multirow{2}{*}{Light} & Unpaired & 0.116 & 0.092 & 0.012 & 0.009 & 0.018 & 0.328 & 0.646 \\
 & Filtered & 0.099 & 0.086 & 0.009 & 0.005 & 0.004 & \textbf{0.181} & \textbf{0.543} \\
\bottomrule
\end{tabular}
\end{table}

\begin{figure}[htbp]
\centering
\includegraphics[width=0.8\textwidth]{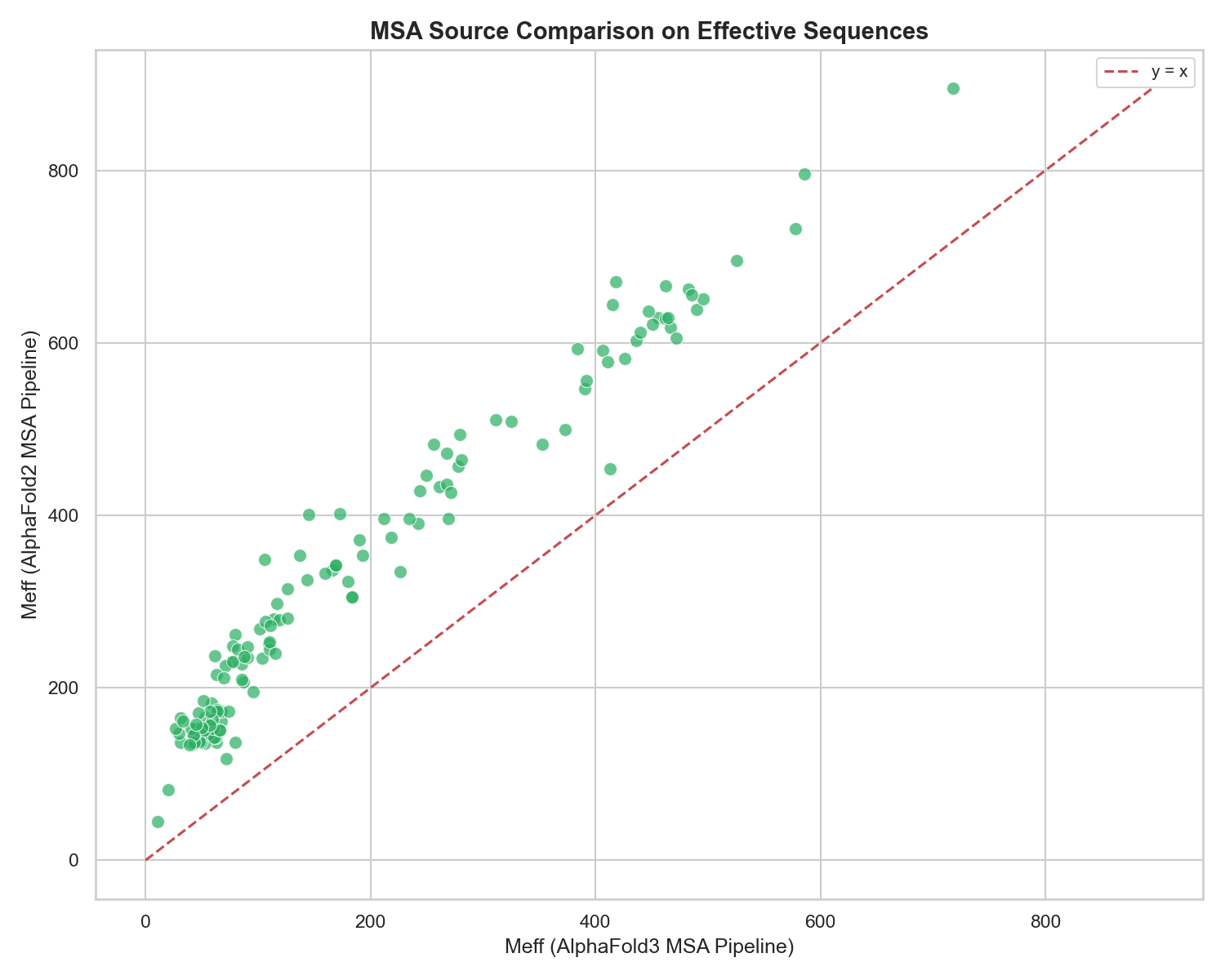}
\caption{Comparison of effective sequence diversity (Meff) between AlphaFold3's default MSA pipeline (Small BFD, x-axis) and the AlphaFold2 MSA pipeline with BFD (y-axis). Nearly all points lie above the $y=x$ diagonal, showing that the BFD retrieval setup consistently yields higher Meff values. The improvement is particularly pronounced for targets with lower initial diversity (left side of plot), where BFD often doubles or triples the effective sequence count.}
\label{fig:msa_source_comparison}
\end{figure}

\paragraph{Effect of depth recovery on effective diversity.}
We compute effective sequence diversity (Meff)~\citep{morcos2011direct} over the CDR3 region using a 40\% sequence identity threshold matching our dataset filtering criteria. Figure~\ref{fig:msa_source_comparison} compares the AlphaFold2 MSA pipeline with BFD against AlphaFold3's default Small BFD pipeline across the validation set: nearly all points lie above the $y=x$ diagonal, with many targets showing 1.5--2$\times$ higher Meff values. Targets that benefit most from the larger-search MSA setup are those with low initial Meff values---precisely the challenging cases where additional evolutionary information is most needed. The average source ratio (BFD Meff / Small BFD Meff) is 2.29$\times$ at CDR3 positions.

\paragraph{Interaction between filtering and depth recovery.}
Across all targets, the median filter ratio (filtered Meff / unfiltered Meff) is 0.945, even though filtering removes a median of 27\% of raw sequences (median nseq ratio = 0.729). This 5\% Meff reduction versus 27\% sequence reduction confirms that the removed sequences are overwhelmingly redundant framework-only matches that contribute minimally to effective diversity at CDR positions. The two interventions are thus complementary: BFD increases the pool of informative sequences, and CDR-focused filtering removes uninformative ones.

\subsection{Comparison with Alternative Prediction Strategies}
\label{sec:ch4_comparison}

\paragraph{Dataset and protocols.}\label{sec:ch6_evaluation}
The test-set analyses in this chapter use a 76-target subset (restricted to antigen length $\leq 600$ residues for computational tractability) of a temporally held-out SAbDab test set. The subset is drawn from train/validation/test splits defined by PDB release date (training $<$ 2021-09-30, validation 2021-09-30 to 2023-01-13, test 2023-01-13 to 2024-01-01). Non-redundancy with the training data is enforced by clustering SAbDab structures on antigen sequence and IMGT-numbered heavy-chain CDR3 sequence using MMseqs2~\citep{steinegger2017mmseqs2}, with targets assigned to held-out clusters separated from training clusters at the 40\% sequence-identity level; targets with incomplete CDR or epitope regions are excluded. In this section we use a \emph{five-sample comparison protocol} (1~seed~$\times$~5~samples per target), in which the top-1 prediction is selected by the AF3Complex interface confidence score (pIS)~\citep{feldman2025af3complex} for AF3-based methods or each method's native ranking score otherwise. DockQ is the primary metric, with thresholds Medium (DockQ~$\geq 0.49$) and Acceptable (DockQ~$\geq 0.23$). For DockQ calculation, antibody chain(s) (heavy and light when present) are treated as a single receptor group and evaluated against the antigen chain(s). Full protocol details are in Appendix~\ref{app:pipeline_details}.

\paragraph{Setup.}
To evaluate whether MSA refinement improves prediction in a common five-sample setting, Table~\ref{tab:main_comparison} compares methods using one seed and five samples per target on the 76-target subset. The MSA-refinement-only configuration does not use convergence-aware recycling; it uses unpaired MSAs, CDR-focused filtering, and depth recovery via BFD. We include both alternative architectures (AlphaFold2-Multimer, Chai-1~\citep{chai2024chai}, Protenix~\citep{bytedance2025protenix}, Boltz-1~\citep{wohlwend2024boltz1}) and a model-level adaptation baseline---LoRA~\citep{hu2022lora} fine-tuning of AlphaFold2-Multimer, detailed in Appendix~\ref{app:lora}. The AlphaFold2-Multimer baseline uses model~4 of the AF2-Multimer ensemble (the same model that LoRA fine-tunes) with five sample variations, matching the five-sample setting used by the other methods.

\begin{table}[htbp]
\centering
\caption{Five-sample comparison on the 76-target subset of the temporal test set. Each method reports five prediction candidates per target; the top-1 prediction is selected by the method's confidence score (AF3Complex pIS for AF3-based methods, ranking\_score for AlphaFold2-Multimer-based methods). SR@$x$ denotes the success rate at DockQ thresholds for acceptable ($\geq 0.23$) and medium ($\geq 0.49$) quality. The AlphaFold2-Multimer baseline uses model~4 from the AF2-Multimer ensemble with five sample variations; the LoRA row applies parameter-efficient fine-tuning to this same model.}
\label{tab:main_comparison}
\begin{tabular}{lcc}
\toprule
Method & SR@$0.23$ (\%) & SR@$0.49$ (\%) \\
\midrule
\multicolumn{3}{l}{\textit{AlphaFold2-Multimer based}} \\
\quad Baseline (model~4)                   & 25.0          & 18.4          \\
\quad + LoRA fine-tuning                   & 32.9          & 22.4          \\
\midrule
\multicolumn{3}{l}{\textit{Diffusion-based architectures}} \\
\quad Chai-1                                   & 22.4          & 14.5          \\
\quad Protenix                                 & 22.4          & 14.5          \\
\quad Boltz-1                                  & 28.9          & 14.5          \\
\midrule
\multicolumn{3}{l}{\textit{AlphaFold3 based}} \\
\quad Baseline                                 & 31.6          & 21.1          \\
\quad \textbf{Ours}                            & \textbf{36.8} & \textbf{30.3} \\
\bottomrule
\end{tabular}
\end{table}

\paragraph{Top-ranked performance.} Under the single-seed, five-sample setting, MSA refinement achieves the highest top-1 success rate at both the acceptable ($36.8\%$) and medium ($30.3\%$) thresholds, improving over the AlphaFold3 baseline by $5.2$ and $9.2$ percentage points respectively. The medium-threshold improvement ($21.1\% \rightarrow 30.3\%$) is the largest single gain over the AlphaFold3 baseline among the methods compared. LoRA fine-tuning of AlphaFold2-Multimer improves over the single-model AF2-Multimer baseline at acceptable ($25.0\% \rightarrow 32.9\%$) and medium ($18.4\% \rightarrow 22.4\%$) quality, providing a useful model-level adaptation reference under the same five-sample setting. The three diffusion-based architectures (Chai-1, Protenix, Boltz-1) all perform below the AlphaFold3 baseline on antibody-antigen complexes in this five-sample setting, indicating that architecture alone does not explain performance in this domain; MSA-construction choices also matter, which is consistent with the improvement reported here.

\section{Convergence-Aware Recycling}
\label{sec:ch6_adaptive_recycling}

The MSA refinement of Section~\ref{sec:ch6_msa_refinement} improves prediction performance over the AlphaFold3 baseline at the acceptable and medium quality thresholds (Section~\ref{sec:ch4_comparison}). Examining per-target behavior, we observed that even with refined MSAs the AlphaFold3 recycling trajectory remains unstable for many antibody-antigen targets: per-recycle structural changes do not decay monotonically, and the final recycle state---which the default pipeline uses as input to diffusion sampling---can be further from a previously stable intermediate state than earlier iterations. This points to a second source of antibody-antigen prediction error, complementary to MSA quality, that originates in recycling behavior rather than MSA composition.

The rest of this section develops a convergence-aware recycling strategy that monitors per-recycle stability and uses the saved most-stable recycle state for final diffusion sampling. The component is then combined with the MSA refinement of Section~\ref{sec:ch6_msa_refinement} and evaluated under the test-time-scaling protocol in Section~\ref{sec:ch6_results}.

\subsection{Recycling Instability under Sparse MSA Signal}

Many structure prediction models use recycling~\citep{jumper2021highly}: feeding predicted structures back into the network for iterative refinement, gradually correcting errors over multiple passes. Models typically use a fixed number of iterations determined empirically on general protein benchmarks---AlphaFold3~\citep{abramson2024accurate} uses 10 recycles by default---without adaptive stopping mechanisms, even though earlier extensions such as AF2Complex~\citep{gao2022af2complex} have demonstrated that monitoring convergence can improve prediction quality on multimeric targets.

For antibody-antigen complexes, where cross-chain co-evolutionary signal is sparse and antibody MSAs themselves are noisier than general protein MSAs (Section~\ref{sec:ch6_msa_refinement}), recycling behavior differs from general proteins. The model is more sensitive to the randomly sampled MSA rows used in each recycle iteration; sparse signals typically require more iterations to converge compared to well-aligned co-evolving families; and too many recycles may cause the model to jump away from correct predictions when confidence is low, producing oscillation between conformations rather than smooth convergence.

Fixed recycling is therefore suboptimal for antibodies: different targets have vastly different convergence characteristics. Some predictions converge quickly and do not need many recycles, wasting computation and risking late-cycle degradation. Others need more iterations to reach stable predictions but may diverge if recycled too much with weak MSA guidance. Most critically, fixed recycling provides no mechanism to detect convergence or divergence, so it cannot stop at the optimal point for each individual prediction.

Convergence-based early stopping for recycling was introduced by AF2Complex~\citep{gao2022af2complex}, which monitors the backbone C$_\alpha$ distogram across recycles and stops once consecutive distograms have converged. We extend this convergence-based view to antibody-antigen complexes, where early stopping and intermediate-state selection are particularly important due to the oscillation behavior described above.

\subsection{$D_{\min}$ (min\_diff) as a Convergence Signal}
\label{sec:ch6_dmin}

We define convergence based on inter-residue distance changes between consecutive recycles. At recycle iteration $t$, let $d_{ij}^{(t)}$ denote the C$_\alpha$ distance between residues $i$ and $j$ in the recycle's structural state, and let $\mathcal{R}$ denote a representative set of residue pairs (e.g., interface residues or CDR residues, where the largest conformational changes are expected). The per-iteration change is
\begin{equation}
D^{(t)} = \text{RMS}\left(\{d_{ij}^{(t)} - d_{ij}^{(t-1)}\}_{(i,j) \in \mathcal{R}}\right),
\end{equation}
where RMS denotes the root mean square over the selected residue pairs. We track the running minimum across all recycles completed so far,
\begin{equation}
D^{\min}_t = \min_{1 \le t' \le t} D^{(t')},
\end{equation}
and refer to this running minimum as $\mathrm{min\_diff}$ in figures and pseudocode. We stop recycling as soon as the running minimum falls below a fixed threshold,
\begin{equation}
D^{\min}_t < \theta,
\end{equation}
with $\theta = 0.5$\,\AA\ in our experiments. Because $D^{\min}_t$ is non-increasing in $t$, the iteration at which this condition first holds is the one that produced the new minimum, so the stored recycle state at that iteration is what is used for final diffusion sampling and confidence estimation. This anchors the final prediction to the most stable intermediate state observed, rather than to whatever state the model happens to be in at the last iteration.

\subsection{Convergence-Aware Recycling with Mini Rollout}

For diffusion-based models that only produce structures after full diffusion sampling (e.g., AlphaFold3), intermediate structures are not directly available across recycles, so $D^{(t)}$ cannot be computed from the recycle state alone. We address this by inserting a \emph{mini rollout} between recycling steps: a small fraction (e.g., 10\%) of diffusion steps is run to obtain a fast, low-cost structure proxy used only for convergence assessment. This idea is related to AlphaFold3's training procedure, which uses mini rollouts to generate structure samples for training the confidence head~\citep{abramson2024accurate}; we repurpose a similar mechanism at inference time for convergence monitoring.

In brief, after each recycling iteration we run a lightweight diffusion rollout to obtain provisional 3D coordinates. The first rollout initializes the previous-distance reference; subsequent rollouts compute $D^{(t)}$ as the distance change from the preceding recycle and update the running minimum $D^{\min}_t$ along with the recycle state that achieved it. If $D^{\min}_t$ falls below the convergence threshold we stop early; otherwise we continue until the maximum recycle count is reached. In either case, the saved minimum-difference state is used for final diffusion sampling and confidence estimation, rather than the last iteration's state. The complete pseudocode is provided in Appendix~\ref{sec:appB_adaptive_recycle_algo} (Algorithm~\ref{alg:early-stop}).

The procedure is designed to address several antibody-specific failure modes simultaneously: it avoids late-cycle oscillations by stopping when the structure has converged, rather than continuing when predictions have begun to degrade; it saves computation for predictions that converge quickly, while allowing more iterations for harder targets; and it selects the most stable intermediate structure rather than always using the final recycle, which may not be optimal for targets with oscillatory behavior.

\subsection{Validation: min\_diff Correlates with Prediction Quality}
\label{sec:ch6_recycling_validation}

\begin{figure}[htbp]
\centering
\includegraphics[width=0.8\textwidth]{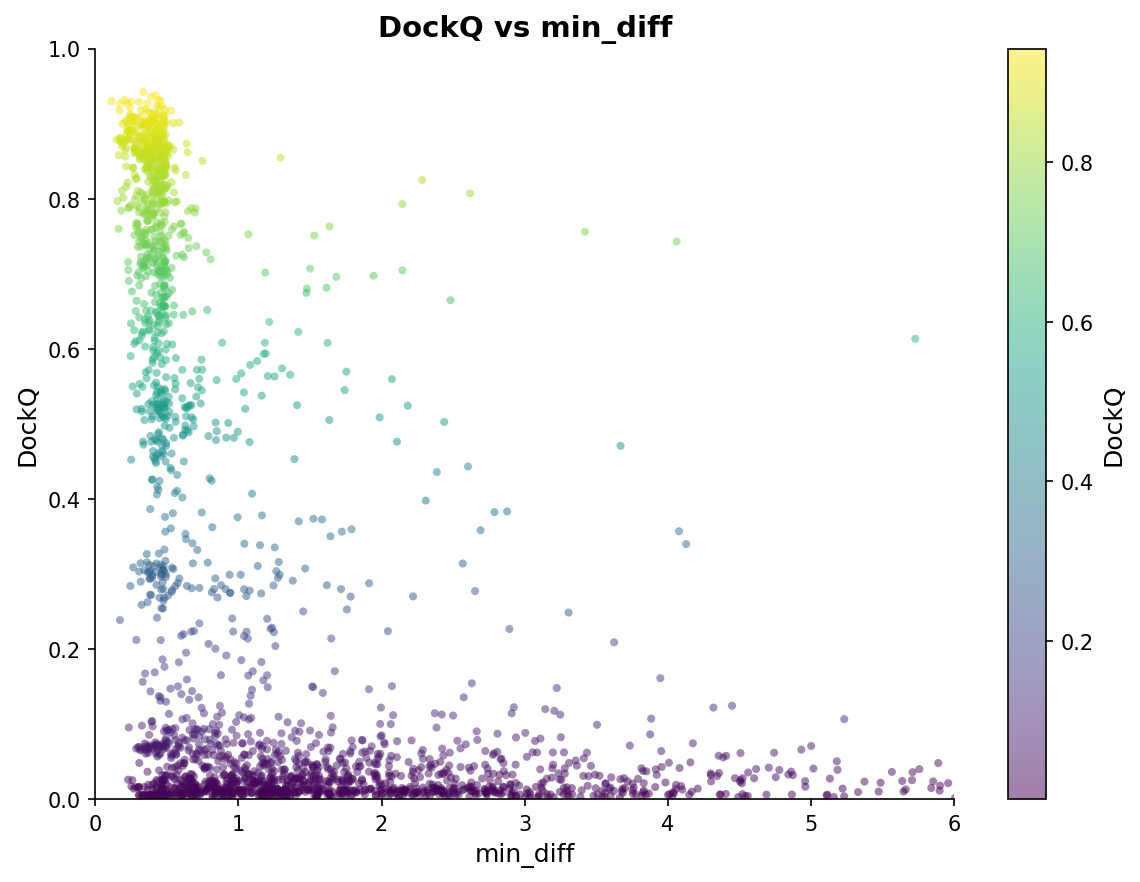}
\caption{Convergence stability ($\mathrm{min\_diff}$) versus prediction quality (DockQ) across the validation set. Each point is one prediction seed, colored by DockQ. Predictions with low $\mathrm{min\_diff}$ (stable convergence) tend to achieve high DockQ scores, while those with high $\mathrm{min\_diff}$ (continued oscillation) mostly fail.}
\label{fig:dockq_vs_mindiff}
\end{figure}

Figure~\ref{fig:dockq_vs_mindiff} shows a strong relationship between convergence behavior and prediction quality across the validation set. Each point represents one prediction (5 seeds per target), with the x-axis showing $\mathrm{min\_diff}$ ($D^{\min}_t$ at the end of recycling) and the y-axis showing the final DockQ score. The pattern is clear: predictions with low $\mathrm{min\_diff}$ (stable convergence) cluster in the high-DockQ region, while predictions with high $\mathrm{min\_diff}$ (continued oscillation) predominantly yield poor results.

This correlation supports the use of $\mathrm{min\_diff}$ as a proxy for selecting a stable recycle state for final sampling. We emphasize that $\mathrm{min\_diff}$ is a convergence signal used for stable-state selection within the recycling loop, not a replacement for the AF3Complex interface confidence score (pIS) used to rank top-1 predictions across multiple samples; pIS-based ranking remains unchanged.

\subsection{Evaluation with Test-Time Scaling}
\label{sec:ch6_results}

Recent work has shown that aggregating multiple inference runs per target---test-time scaling---improves AlphaFold prediction quality, particularly on harder tasks like antibody-antigen complexes~\citep{wallner2023afsample,abramson2024accurate}. To evaluate the complete method (MSA refinement combined with convergence-aware recycling), we adopt a \emph{test-time-scaling protocol} in which each target is predicted with 20~seeds and 5~samples per seed, and the top-1 prediction is selected by the AF3Complex interface confidence score (pIS) across all 100 candidates. Protocol details are in Appendix~\ref{app:pipeline_details}.

\begin{table}[htbp]
\centering
\caption{Multi-seed test set results on the 76-target subset under the test-time-scaling protocol (20~seeds~$\times$~5~samples per target, top-1 by AF3Complex pIS). Success rates are reported for medium (DockQ $\geq 0.49$) and acceptable (DockQ $\geq 0.23$) quality predictions. The table compares the AlphaFold3 baseline, AF3Complex, and our final method, matching the methods shown in Figure~\ref{fig:seed_ranking_comparison}.}
\label{tab:ch6_results}
\begin{tabular}{lcc}
\toprule
\textbf{Method} & \textbf{Medium SR@0.49} & \textbf{Acceptable SR@0.23} \\
\midrule
AlphaFold3 (baseline) & 36.8\% & 43.4\% \\
AF3Complex & 36.8\% & 44.7\% \\
Ours & \textbf{44.7\%} & \textbf{48.7\%} \\
\bottomrule
\end{tabular}

\end{table}

Under this test-time-scaling protocol, our method achieves $48.7\%$ acceptable and $44.7\%$ medium success rates, improving over the AlphaFold3 baseline by $5.3$ and $7.9$ percentage points respectively; under the same protocol, the AlphaFold3 baseline reaches $43.4\%$ acceptable and $36.8\%$ medium success rates.

\paragraph{Relation to AF3Complex.}
The AF3Complex configuration~\citep{feldman2025af3complex} (unpaired MSAs only, no CDR-focused filtering or depth recovery) reaches $44.7\%$ acceptable and $36.8\%$ medium success rates. Our method improves these to $48.7\%$ acceptable and $44.7\%$ medium quality, indicating that the gain is not explained by the unpaired-MSA-only setting alone. Component-level validation on the validation set is summarized in Appendix~\ref{sec:appB_ablations}.

\begin{figure}[htbp]
\centering
\includegraphics[width=0.95\textwidth]{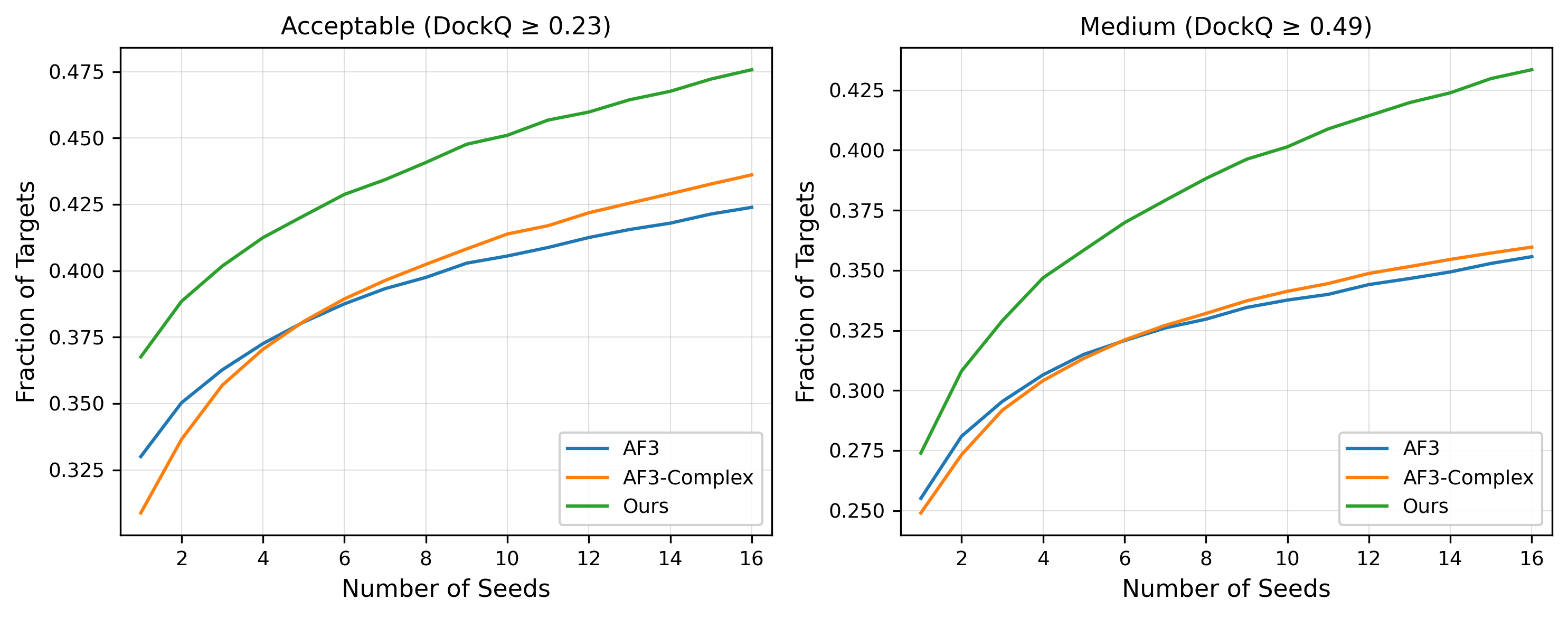}
\caption{Performance scaling with multiple seeds. Success rates for acceptable (DockQ $\geq 0.23$, left) and medium (DockQ $\geq 0.49$, right) quality predictions as a function of the number of seeds used for ranking. Each point shows the mean over 1{,}000 random samples of $n$ seeds from 20 available seeds. Predictions are selected by the highest AF3Complex pIS among sampled seeds. Our method consistently outperforms both AF3 and AF3Complex baselines across the sampled seed budgets shown ($n = 1$ to $16$).}
\label{fig:seed_ranking_comparison}
\end{figure}

Figure~\ref{fig:seed_ranking_comparison} shows how prediction quality scales with the number of seeds used for ranking. Our method maintains a consistent advantage over both baselines across the sampled seed budgets shown. At $n=16$ seeds, our method achieves $47.6\%$ acceptable and $43.4\%$ medium success rates, compared to $42.4\%/35.6\%$ for AF3 and $43.6\%/36.0\%$ for AF3Complex. The full 20-seed result is reported in Table~\ref{tab:ch6_results}. The performance gap remains stable across the sampled seed range, suggesting that the improvement is not driven by a single favorable seed choice.

\subsubsection{Structural Case Studies}
\label{sec:ch6_case_studies}

\begin{figure}[htbp]
  \centering
  \begin{subfigure}[t]{0.48\textwidth}
    \centering
    \includegraphics[width=\textwidth]{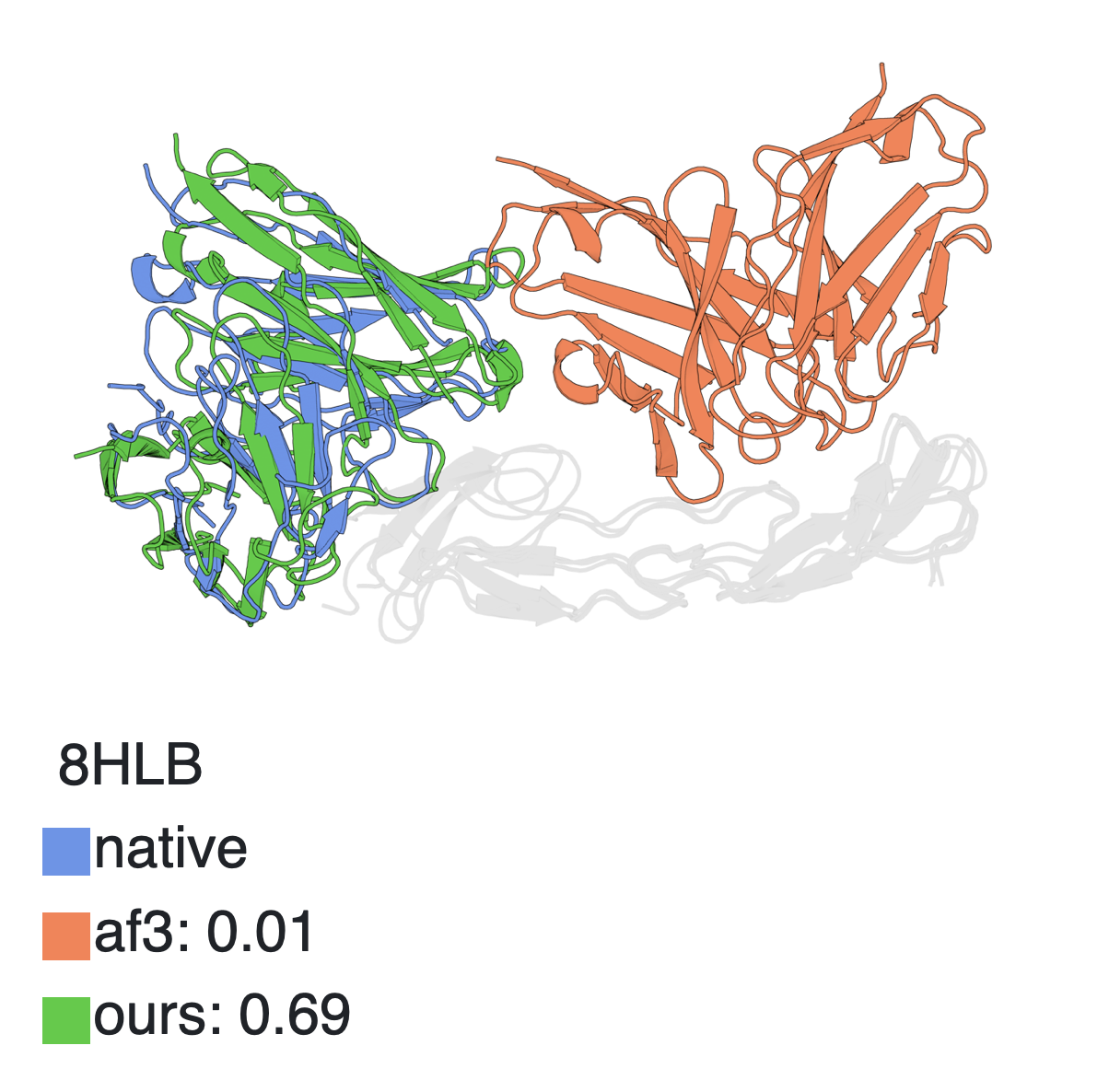}
    \caption{8HLB}
    \label{fig:case_8HLB}
  \end{subfigure}
  \hfill
  \begin{subfigure}[t]{0.48\textwidth}
    \centering
    \includegraphics[width=\textwidth]{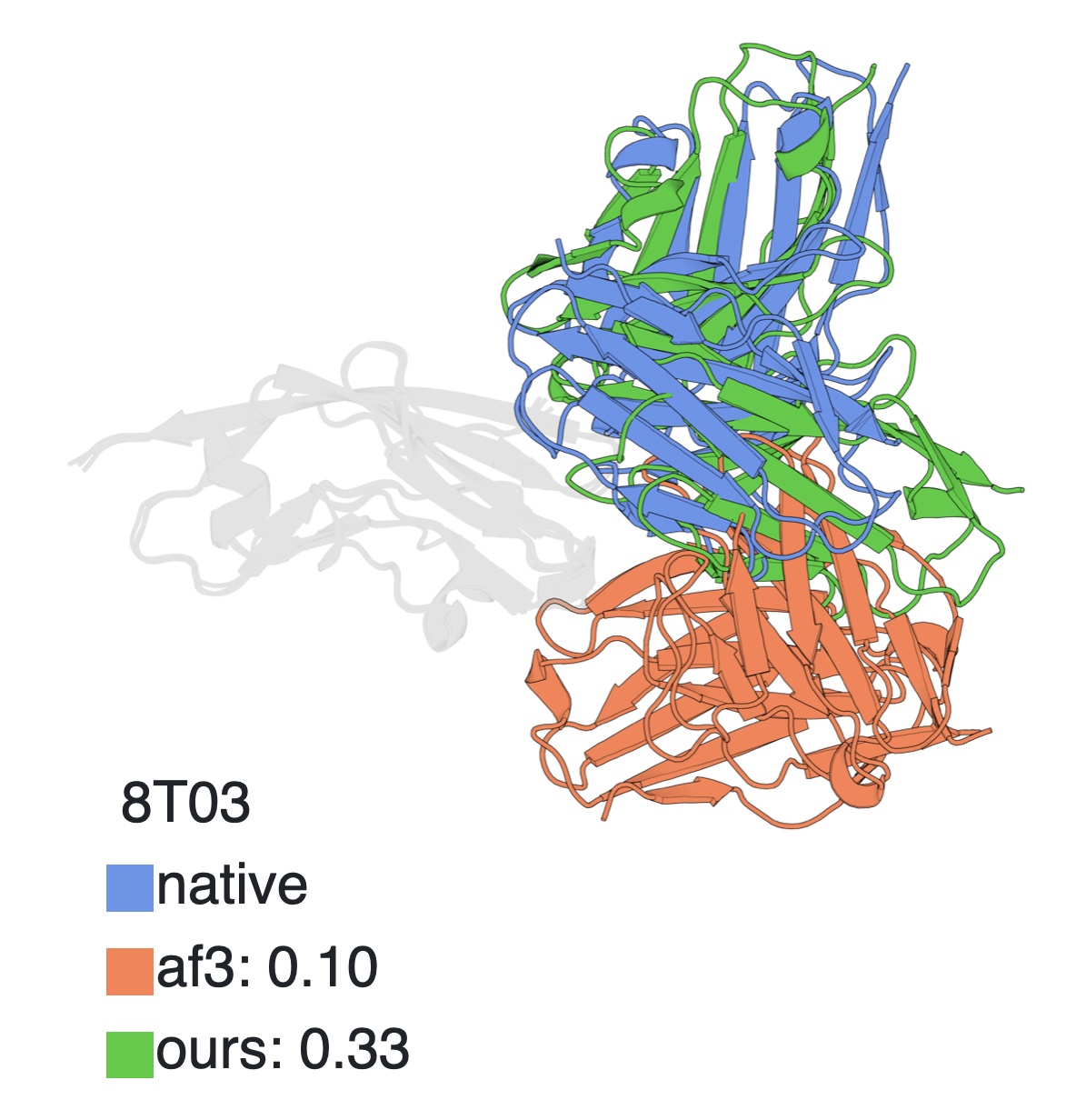}
    \caption{8T03}
    \label{fig:case_8T03}
  \end{subfigure}
  \caption{Structural case studies comparing our method (green) against AlphaFold3 (orange) and the native complex (blue). The antigen is shown in gray. DockQ scores are indicated in the legend. (a)~8HLB --- AlphaFold3 places the antibody in an entirely incorrect orientation (DockQ~$= 0.01$), while our method correctly predicts the binding interface with medium quality (DockQ~$= 0.69$). (b)~8T03 --- AlphaFold3 produces a poor docking pose (DockQ~$= 0.10$), whereas our method achieves acceptable quality (DockQ~$= 0.33$) with the antibody positioned near the correct binding site.}
  \label{fig:ch6_case_studies}
\end{figure}

To illustrate the practical impact of our method, we present structural visualizations of two test targets where AlphaFold3 fails but our method succeeds (Figure~\ref{fig:ch6_case_studies}). In each panel, the native complex is shown in blue, the AlphaFold3 baseline prediction in orange, and our method's prediction in green. The antigen is rendered in gray for spatial context.

The 8HLB case (Figure~\ref{fig:case_8HLB}) is particularly striking: AlphaFold3's prediction achieves a DockQ of only 0.01, indicating an essentially random binding pose with no correct interface contacts. In contrast, our method produces a DockQ of 0.69 (medium quality), with the antibody correctly positioned at the native binding interface. The green structure closely overlaps the native blue structure, demonstrating that CDR-filtered MSAs combined with convergence-aware recycling enable the model to identify the correct binding mode.

The 8T03 case (Figure~\ref{fig:case_8T03}) represents a more challenging target where our method achieves acceptable but not medium quality (DockQ~$= 0.33$). While our prediction captures the general binding orientation, some interface contacts remain imprecise. Nevertheless, the improvement over AlphaFold3's failed prediction (DockQ~$= 0.10$) is large---the baseline places the antibody in an incorrect region of the antigen surface, whereas our method identifies the correct binding site.

These cases illustrate that on individual targets where the method helps, the gains can be substantial; aggregate behavior across the full test set is reported in Table~\ref{tab:ch6_results}.

\section{Discussion}
\label{sec:ch6_discussion}

\subsection{Why Does MSA Refinement Help?}

The two MSA-side components---CDR-focused filtering and depth recovery---are most informative when considered together rather than in isolation. CDR-focused filtering improves the relevance of the MSA by removing framework-only sequences that contribute little to the binding-relevant region. By itself, however, filtering also reduces MSA depth, which can remove weak but useful signal. Depth recovery (implemented via the BFD database in this thesis) compensates by restoring evolutionary coverage with a larger pool of candidate sequences. The validation set ablations summarized in Appendix~\ref{sec:appB_ablations} are consistent with this view: filtering alone is not reliably a standalone improvement, but filtering combined with depth recovery improves practical top-ranked performance. The effective MSA-side intervention is therefore filtering plus depth recovery, with the two components addressing complementary aspects of MSA signal quality.

\subsection{Why Does Convergence-Aware Recycling Help?}

Convergence-aware recycling does not change what the model sees, but how the model processes and selects across recycling iterations. For antibody-antigen targets with weak or noisy MSA signal, continuing to recycle does not always improve the prediction, and late iterations can move away from a previously stable state. Saving the recycle state with the smallest $D^{(t)}$ and using it as the input to final diffusion sampling reduces the risk of relying on such transient late-cycle states. We emphasize again that $\mathrm{min\_diff}$ is a per-recycle stability proxy used inside the recycling loop, not a new ranker; AF3Complex pIS still selects the top-1 prediction across diffusion samples, exactly as in the AlphaFold3 baseline.

\subsection{Complementarity of MSA Refinement and Recycling}

The two interventions act on different parts of the prediction workflow: MSA refinement modifies the evolutionary signal that the model receives, while convergence-aware recycling modifies how the model traverses its recycling trajectory. The validation ablations suggest that each contributes, but not uniformly across metrics: depth recovery, filtering, and recycling affect partly different aspects of success, with recycling contributing most visibly to medium-quality gains in the test-time-scaling comparison. We do not claim that the two components are strictly independent, since both ultimately affect the same downstream confidence-ranking step; we treat them as complementary interventions whose effects can be combined within the same setting.

The two components also differ in how their effects appear across evaluation protocols. MSA refinement directly changes the information available to the model and accounts for most of the five-sample comparison improvement observed in Section~\ref{sec:ch4_comparison}. Convergence-aware recycling instead changes which recycle state is used for final diffusion sampling; its contribution is clearest under the test-time-scaling protocol (Section~\ref{sec:ch6_results}), where it accounts for most of the additional medium-quality improvement beyond MSA refinement alone. We therefore view the two interventions as complementary rather than uniformly additive across every metric.

\subsection{Limitations and Future Work}

The improvements reported here are modest in absolute terms and depend on the choice of filtering window, gap threshold, and convergence threshold; optimal values may vary across antibody types or MSA generation protocols. The approach only applies to MSA-based predictors and does not help PLM-based methods such as those in Sections~\ref{sec:monomer_prediction}--\ref{sec:antibody_antigen_complex}.

Future work could explore more selective MSA refinement strategies, such as quality-weighted filtering or epitope-aware filtering that incorporates binding-site predictions. On the recycling side, alternative convergence metrics or adaptive sampling for diffusion steps could be explored. Combining MSA refinement with experimental data from mutagenesis or epitope mapping is another natural direction. The two interventions introduced here---MSA refinement and convergence-aware recycling---may also extend to other immune proteins such as T-cell receptors and MHC complexes, where similar challenges with sparse cross-chain co-evolutionary signal arise.

\section{Summary}
\label{sec:ch6_summary}

This chapter presented two training-free MSA-based refinements for antibody-antigen complex prediction, implemented and evaluated with AlphaFold3. First, MSA refinement combines CDR-focused filtering with depth recovery, implemented here using the BFD database; we also adopt the unpaired-MSA-only setting following recent findings that paired MSAs can hurt antibody-antigen prediction. Second, convergence-aware recycling uses a per-recycle stability signal ($D_{\min}$, referred to as $\mathrm{min\_diff}$ in figures and pseudocode) computed from mini rollouts to select a stable intermediate recycle state for final diffusion sampling, avoiding late-cycle oscillations on noisy MSAs.

On the 76-target subset of the temporal test set, our method achieves $48.7\%$ acceptable-quality success rate, improving over the $43.4\%$ AlphaFold3 baseline by $5.3$ percentage points, and improves medium-quality predictions from $36.8\%$ to $44.7\%$ under the test-time-scaling protocol (20~seeds~$\times$~5~samples per target). Because the methods modify MSA construction and recycling behavior rather than model parameters, they do not require retraining or weight access; they are most directly applicable to AF3-style MSA-based predictors where these stages can be controlled.

\chapter{Conclusion and Future Work}
\label{ch:conclusion}

\section{Summary of Contributions}

This thesis explored computational methods for antibody-antigen structure prediction along two complementary directions. The work is organized around a central observation: protein language models predict antibody monomers effectively, but they cannot identify binding interfaces in antibody-antigen complexes without co-evolutionary signals (Chapter~\ref{ch:plm_modeling}). MSA-based methods can leverage such signals, but require specific adaptation for the sparse and adversarial evolutionary patterns of antibody-antigen interactions. We address this through training-free MSA-based refinements evaluated with AlphaFold3, without requiring access to model weights (Chapter~\ref{ch:pipeline_adaptation}).

Our contributions are:

\begin{enumerate}
  \item \textbf{Protein language model-based antibody modeling.} We developed a multi-PLM approach with antibody-specific fine-tuning for antibody monomer prediction, achieving the best CDR-H3 accuracy among the compared PLM-based methods (Section~\ref{sec:monomer_prediction}). Extending this approach to antibody-antigen complex prediction (Section~\ref{sec:antibody_antigen_complex}), we identified interface identification as the main bottleneck for sequence-only methods: PLM-derived representations achieve only 8\% blind docking success but reach 57\% (antibodies) and 84\% (nanobodies) when native epitope labels are provided. This negative finding clarifies the boundary of what PLM-based methods can achieve and motivates a return to MSA-based approaches for complex prediction.

  \item \textbf{MSA-based methods for antibody-antigen complex prediction.} We developed training-free improvements for AF3-style MSA-based predictors whose MSA construction and inference stages can be controlled (Chapter~\ref{ch:pipeline_adaptation}). The improvements consist of two interventions on the MSA-based workflow. \emph{MSA refinement} combines CDR-focused MSA filtering, which removes framework-only sequences (\textasciitilde 28\% reduction) while preserving CDR-informative alignments and nearly doubling CDR3 coverage, with depth recovery (implemented in this thesis using the BFD database) that restores evolutionary diversity after filtering for antibody-antigen targets with sparse coverage. \emph{Convergence-aware recycling} monitors per-recycle structural change and uses the most stable intermediate recycle state for final diffusion sampling, avoiding late-cycle oscillations on noisy MSAs. Combined, these methods improve AlphaFold3's success rate from 43.4\% to 48.7\% at the acceptable threshold and from 36.8\% to 44.7\% at the medium threshold on the held-out antibody-antigen test set.
\end{enumerate}

The comparison between adaptation strategies (Section~\ref{sec:ch4_comparison}) shows a useful trade-off: model-level adaptation requires access to model weights and can use domain-specific training directly, while training-free MSA-based adaptation is model-weight-agnostic and can fit with model-level approaches. For systems where model weights are unavailable, these methods provide a practical route to antibody-specific improvement when input generation and inference settings can be controlled.

\section{Future Directions}

Several directions could extend this work:

\begin{enumerate}
  \item \textbf{Using experimental data}: Incorporating binding-site information from cross-linking mass spectrometry or alanine scanning could improve complex prediction accuracy, as our epitope-guided results suggest.

  \item \textbf{Better confidence estimation}: A consistent gap between top-ranked and oracle (best-of-$n$) performance across both adaptation strategies suggests that better ranking could provide immediate improvement without model retraining. Antibody-specific confidence metrics that account for CDR flexibility may help close this gap.

  \item \textbf{Conformational flexibility}: Better handling of induced-fit binding and CDR flexibility remains an important challenge for accurate complex modeling.

  \item \textbf{Combining model-level and training-free adaptation}: Since the two adaptation strategies modify different parts of the prediction pipeline, applying MSA refinement and convergence-aware recycling on top of fine-tuned models could provide additive gains. We did not pursue this combination, but the empirical comparison in Section~\ref{sec:ch4_comparison} suggests it as a useful future direction.

  \item \textbf{Other immune proteins}: The adaptation strategies developed here could be applied to T-cell receptors and MHC complexes, where similar challenges with co-evolutionary signal sparsity arise.

  \item \textbf{Generative antibody design}: The methods developed here---particularly complex prediction and confidence estimation---can support generative antibody design~\citep{watson2023novo,dauparas2022robust,cutting2025novo} by providing rapid structural validation of designed candidates.
\end{enumerate}

\appendix
\chapter{Supplementary Material for Chapter 3: PLM Implementation Details}
\label{app:plm_details}

This appendix provides implementation details for the PLM-based monomer and complex prediction methods in Chapter~\ref{ch:plm_modeling}.

\section{Multi-PLM Representation Fusion}
\label{sec:app_plm_fusion}

The monomer predictor integrates three protein language models that provide complementary structural information through their different training data and architectures.

\textbf{ESM-1b} (650M parameters) is a BERT-style~\citep{devlin2019bert} masked language model trained on UniRef50 (45M sequences clustered at 50\% identity)~\citep{suzek2015uniref}. It learns global structural patterns and fold topologies from diverse, non-redundant protein sequences, outputting 1280-dimensional embeddings per residue.

\textbf{ESM-1v} (650M parameters) shares the same architecture as ESM-1b but is trained on UniRef90 (98M sequences at 90\% identity)~\citep{suzek2015uniref}. The higher sequence redundancy exposes the model to evolutionary variation within protein families, teaching it which positions tolerate mutations versus which are conserved.

\textbf{ProtT5-XL} (3B parameters) uses a T5 encoder-decoder architecture pre-trained on BFD (2.5 billion sequences)~\citep{steinegger2019protein}. The massive metagenomic training data captures rare structural motifs and orphan proteins absent from curated databases. We use only the encoder representations (1024-dimensional embeddings) for structure prediction.

Each residue $i$ starts from a learned amino-acid embedding and adds contributions from every PLM through model-specific linear projections:
\begin{equation}
\mathbf{s}_i = W_{\text{aa}} \mathbf{o}_i + \sum_{m \in \mathcal{M}} W_m \mathbf{e}_i^{(m)} + \mathbf{b}_s,
\end{equation}
where $\mathbf{o}_i$ is the one-hot representation of the residue type, $\mathcal{M}=\{\text{ESM1b}, \text{ESM1v}, \text{ProtT5}\}$ indexes the PLMs whose token embeddings form $\mathbf{e}_i^{(m)}$, and $\mathbf{b}_s$ is a learned bias vector. The coefficient matrices $W_{\text{aa}}$ and $W_m$ project each representation to a common hidden dimension before fusion.

Pair embeddings are built symmetrically from amino-acid identities and relative positions:
\begin{equation}
\mathbf{z}_{ij}^{(0)} = W_L \mathbf{o}_i + W_R \mathbf{o}_j + W_{\text{pos}} \mathbf{p}_{ij}^{\text{pos}},
\end{equation}
where $\mathbf{p}_{ij}^{\text{pos}}$ encodes relative position:
\begin{equation}
\mathbf{p}_{ij}^{\text{pos}} = \text{OneHot}(\text{clip}(j - i, -32, 32)).
\end{equation}
The final pair representation fuses recycled geometric features and PLM attention maps:
\begin{equation}
\mathbf{z}_{ij} = \mathbf{z}_{ij}^{(0)} + W_{\text{dist}} \mathbf{d}_{ij} + \mathbf{r}_{ij}^{\text{pair}} + \sum_{m \in \mathcal{M}} U_m \mathbf{a}_{ij}^{(m)},
\end{equation}
where $\mathbf{d}_{ij} = \text{OneHot}(\text{bucketize}(d_{ij}, \text{bins}))$ embeds the recycled distance map, $\mathbf{r}_{ij}^{\text{pair}}$ is the layer-normalized pair feature from the previous iteration, and $\mathbf{a}_{ij}^{(m)}$ concatenates attention heads from selected layers of PLM $m$ before projection with $U_m$. A final LayerNorm maintains consistent scale across the fused representations.

\section{PLM Monomer Training and Inference Configuration}
\label{sec:app_plm_training}

The monomer model is trained on approximately 340,000 proteins from two sources. The first is 80,852 proteins with experimental structures released before January 2020 in PDB (denoted BC100), clustered at 40\% sequence identity (BC100By40). The second is 264,000 AlphaFold2-distilled proteins from Uniclust30\_2018\_08~\cite{mirdita2017uniclust} with maximum 30\% mutual identity. Distillation~\cite{jumper2021highly} expands training coverage beyond PDB to improve generalization. Per epoch, we sample one sequence per BC100By40 cluster with length-dependent acceptance rates: 0.5 for sequences shorter than 256 residues, linearly increasing from 0.5 to 1.0 for sequences of 256--512 residues, and 1.0 for longer sequences. We mix BC100By40 and distillation data at a 1:3 ratio.

For antibody-specific fine-tuning, we use SAbDab~\cite{dunbar2014sabdab} structures released before 2021/03/31. We split complexes into individual chains, yielding 5,033 heavy and light chains for training and 178 antibody structures (released 2021/04/01--2021/06/30) for validation.

We implement the model in PyTorch~\cite{paszke2019pytorch} with distributed training via PyTorch Lightning~\cite{falcon2019pytorch}. We use the AdamW optimizer~\cite{loshchilov2017decoupled} with $\beta_1 = 0.9$, $\beta_2 = 0.999$, $\epsilon = 10^{-8}$, and weight decay $10^{-4}$.

The learning rate follows a warmup-constant-decay schedule: it warms from $1\times10^{-6}$ to $5\times10^{-4}$ over approximately 1,000 steps, remains at $5\times10^{-4}$ for the first third of training, then linearly decays to $1\times10^{-6}$ for the remainder. Sequence crop length increases during training: we use crops of 256 residues for the first two-thirds of training, then 384 residues for the final third.

The loss function combines distogram-based and structure-based terms. Distogram losses include distance and orientation prediction~\cite{yang2020trrosetta}, while structure losses include Frame Aligned Point Error (FAPE, clamped at 20 \AA) and per-residue confidence (pLDDT)~\cite{jumper2021highly}.

For antibody-specific fine-tuning, we train for an additional 50 epochs with learning rate linearly decaying from $2\times10^{-4}$ to $1\times10^{-5}$. This domain adaptation uses the same loss function and architecture, allowing the model to learn structural patterns specific to the immunoglobulin fold. During inference, we generate predictions with different random seeds and select the conformation with the highest overall pLDDT score (per-residue mean confidence across the predicted structure), following the same selection convention as the original RaptorX-Single ensemble procedure.

During training, the number of recycling iterations is sampled uniformly from 0 to 3. During inference, we use 3 recycling iterations by default. PLM embeddings and attention maps are computed once and cached, so subsequent recycling iterations reuse them without recomputation.

\section{PLM Complex Predictor Implementation Details}
\label{sec:app_plm_complex_details}

The structure encoder in Section~\ref{sec:ch4_architecture} refines sequence and pair representations through triangular attention and multiplication operations. Algorithm~\ref{alg:structure_encoder_app} summarizes the encoder procedure used by the PLM-based complex predictor.

\begin{algorithm}[h]
\caption{Structure Encoder for PLM-Based Complex Prediction}
\label{alg:structure_encoder_app}
\begin{algorithmic}[1]
\State \textbf{Input:} Sequence embeddings $\{\mathbf{s}_i\}$, pair embeddings $\{\mathbf{z}_{ij}\}$
\For{$\ell = 1$ to $N_{\text{enc}}$}
    \State // Sequence self-attention with pair bias
    \State $\mathbf{s}_i \gets \text{Attention}(\mathbf{s}_i, \{\mathbf{s}_j\}_j, \text{bias}=\mathbf{z}_{ij})$
    \State $\mathbf{s}_i \gets \text{FFN}(\mathbf{s}_i)$
    \State
    \State // Pair updates with triangular attention
    \State $\mathbf{z}_{ij} \gets \mathbf{z}_{ij} + \text{TriangleAttentionStart}(\mathbf{z}_{ij})$
    \State $\mathbf{z}_{ij} \gets \mathbf{z}_{ij} + \text{TriangleAttentionEnd}(\mathbf{z}_{ij})$
    \State $\mathbf{z}_{ij} \gets \mathbf{z}_{ij} + \text{TriangleMultiplicationOut}(\mathbf{z}_{ij})$
    \State $\mathbf{z}_{ij} \gets \mathbf{z}_{ij} + \text{TriangleMultiplicationIn}(\mathbf{z}_{ij})$
    \State $\mathbf{z}_{ij} \gets \text{FFN}(\mathbf{z}_{ij})$
\EndFor
\State \textbf{Output:} Refined $\{\mathbf{s}_i\}$, $\{\mathbf{z}_{ij}\}$
\end{algorithmic}

\end{algorithm}

Each encoder block uses the same triangular attention and multiplication operations as AlphaFold2's Evoformer (Chapter~\ref{ch:background}). These operations allow information to flow between residue pairs: triangular multiplication lets the model reason about triplets of residues (e.g., if residue $i$ is near $j$ and $j$ is near $k$, then $i$ and $k$ should have consistent geometry), while triangular attention gathers information along rows and columns of the pair matrix to capture distant relationships. For complex prediction, these operations help combine geometric information from predicted monomers with sequence features to find matching binding surfaces.

The PLM complex predictor follows a two-stage training strategy similar to Section~\ref{sec:monomer_prediction}. The model is first pre-trained on general protein-protein complexes from the DIPS dataset~\cite{townshend2019end}, which covers obligate, transient, enzyme-inhibitor, and other interaction types. This stage allows the network to acquire general knowledge of protein binding interfaces. We then fine-tune on antibody-antigen complexes from SAbDab using a temporal split (pre-2021 structures for training). During fine-tuning, we apply light data augmentation through random rotations and small perturbations to the predicted monomer structures to encourage robustness.

The loss function combines multiple terms:
\begin{equation}
\mathcal{L} = \mathcal{L}_{\text{FAPE}} + \lambda_{\text{interface}} \mathcal{L}_{\text{interface}} + \lambda_{\text{contact}} \mathcal{L}_{\text{contact}},
\end{equation}
where $\mathcal{L}_{\text{FAPE}}$ is the Frame Aligned Point Error objective from AlphaFold2, $\mathcal{L}_{\text{interface}}$ enforces accurate placement of interface residues, and $\mathcal{L}_{\text{contact}}$ encourages correct inter-chain contact prediction. We set $\lambda_{\text{interface}} = 2.0$ and $\lambda_{\text{contact}} = 1.0$, giving the interface term the highest weight since interface residues are a small fraction of all residues and their gradient signal can be easily dominated by the bulk FAPE loss without upweighting. Unlike the monomer prediction task in Section~\ref{sec:monomer_prediction} which used only FAPE and CDR-focused losses, the interface and contact losses are introduced here to specifically supervise cross-chain interaction learning.

We initialize encoder weights from the antibody-finetuned model in Section~\ref{sec:monomer_prediction}, transferring learned CDR structural priors to the complex prediction task. Predicted monomer structures serve as input during training to match the inference setting, whereas Section~\ref{sec:monomer_prediction} trained directly from sequences to coordinates.

\chapter{Supplementary Material for Chapter 4: LoRA Fine-Tuning Baseline}
\label{app:lora}

This appendix documents the LoRA fine-tuning baseline included in Section~\ref{sec:ch4_comparison}. The baseline represents a model-level adaptation strategy for AlphaFold2-Multimer, included as a reference point against the training-free MSA-based methods developed in Chapter~\ref{ch:pipeline_adaptation}. The main text reports the 76-target comparison result; this appendix specifies the adapter placement, training configuration, and evaluation protocol.

\section{Low-Rank Adaptation (LoRA)}
\label{sec:ch5_lora}

\subsection{LoRA Methodology}

Low-Rank Adaptation (LoRA)~\cite{hu2022lora} reduces the number of trainable parameters by adding low-rank matrices to existing weight matrices. For a weight matrix $W \in \mathbb{R}^{d \times k}$, LoRA learns low-rank matrices $B \in \mathbb{R}^{d \times r}$ and $A \in \mathbb{R}^{r \times k}$ with $r \ll \min(d, k)$, so that the adapted weight becomes:
\begin{equation}
W' = W + \Delta W = W + BA
\end{equation}
The original weight matrix $W$ stays frozen during training, and only $B$ and $A$ are updated. Matrix $A$ is initialized from a Gaussian distribution while $B$ is initialized to zero, so training starts from the original pre-trained weights ($\Delta W = 0$ at initialization). The number of trainable parameters is $r(d + k)$, which is much smaller than $dk$ when $r \ll \min(d, k)$.

In this baseline, LoRA provides a lightweight form of model-level adaptation: the original AlphaFold2-Multimer weights remain frozen, while a small set of low-rank adapter parameters is trained on antibody-antigen structures. This isolates the effect of parameter-efficient fine-tuning from the training-free MSA-based refinements evaluated in Chapter~\ref{ch:pipeline_adaptation}.

Related parameter-efficient adaptation studies in protein modeling have explored protein-language-model tuning, immune-protein structure prediction, and AlphaFold2-based immune-complex modeling~\citep{sledzieski2024democratizing,zhu2024accurate,zhang2024seqproft,wu2024fafe}.

\subsection{LoRA Target Layers in AlphaFold2-Multimer}

We apply LoRA to the attention layers in the Evoformer, which is the main module of AlphaFold2. The Evoformer has 48 stacked blocks, each with two processing tracks: the \textbf{MSA track} processes the MSA representation $\mathbf{M} \in \mathbb{R}^{N_{\text{seq}} \times N_{\text{res}} \times d_{\text{msa}}}$, and the \textbf{pair track} processes the pair representation $\mathbf{Z} \in \mathbb{R}^{N_{\text{res}} \times N_{\text{res}} \times d_{\text{pair}}}$. We target three attention types:

\paragraph{MSA Row-Wise Gated Self-Attention with Pair Bias.}
This attention processes each MSA row (a single sequence across all residue positions) with attention biased by the pair representation:
\begin{equation}
\mathbf{M}_i' = \text{LayerNorm}(\mathbf{M}_i) \cdot \sigma(\mathbf{g}_i) \odot \text{Attention}(\mathbf{Q}_i, \mathbf{K}_i, \mathbf{V}_i, \mathbf{b}_{\text{pair}})
\end{equation}
The pair bias $\mathbf{b}_{\text{pair}}$ adds geometric constraints from the pair representation, linking evolutionary patterns with geometric information.

\paragraph{MSA Column-Wise Gated Self-Attention.}
This attention processes each MSA column (a single residue position across all sequences) to combine evolutionary information:
\begin{equation}
\mathbf{M}_{\cdot j}' = \text{LayerNorm}(\mathbf{M}_{\cdot j}) \cdot \sigma(\mathbf{g}_j) \odot \text{Attention}(\mathbf{Q}_{\cdot j}, \mathbf{K}_{\cdot j}, \mathbf{V}_{\cdot j})
\end{equation}
Column-wise attention separates conserved positions from variable positions. For antibodies, this helps the model recognize that CDR variability is meaningful rather than noise.

\paragraph{Triangle Self-Attention.}
The pair representation uses two triangle attention operations---starting node and ending node attention---that combine information along rows and columns of the pair matrix:
\begin{equation}
\mathbf{Z}_{ik}^{\text{start}} = \text{Attention}(\mathbf{Q}_{ik}, \{\mathbf{K}_{jk}\}_j, \{\mathbf{V}_{jk}\}_j), \quad
\mathbf{Z}_{ik}^{\text{end}} = \text{Attention}(\mathbf{Q}_{ik}, \{\mathbf{K}_{ij}\}_j, \{\mathbf{V}_{ij}\}_j)
\end{equation}
These operations allow long-range information flow, helping the model handle extended structural patterns and multi-residue interfaces.

\paragraph{Why Target Attention Layers.}
These attention layers are where AlphaFold2 combines information across sequences, residues, and geometric constraints. For antibodies, the patterns differ from general proteins: CDR variability is structurally meaningful, and antibody-antigen interfaces lack co-evolutionary signals. By applying LoRA to attention layers, we let the model learn antibody-specific patterns while keeping general folding ability in the frozen layers (transition layers, triangle multiplicative updates, outer product mean, and the structure module).

\begin{figure}[htbp]
\centering
\includegraphics[width=\textwidth]{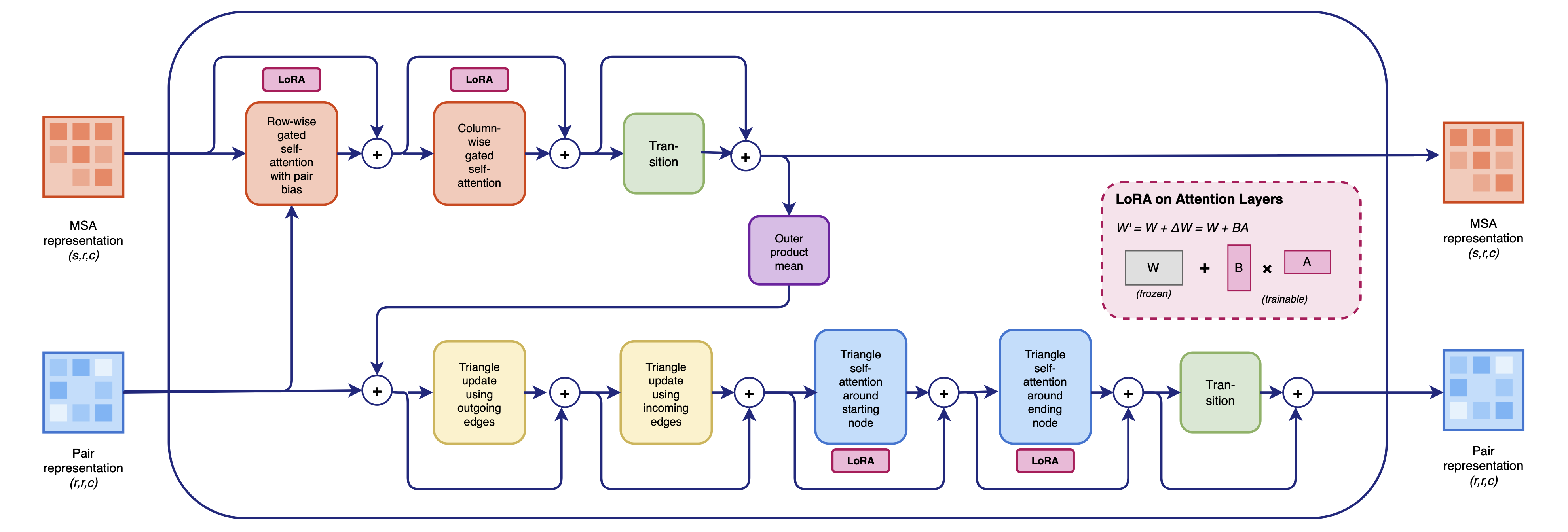}
\caption{LoRA adaptation of AlphaFold2-Multimer Evoformer. Low-rank matrices (pink) are added to attention layers: (1) MSA row-wise attention with pair bias, (2) MSA column-wise attention, (3) Triangle attention around starting node, and (4) Triangle attention around ending node. The inset shows the LoRA formula: $W' = W + BA$, where $W$ stays frozen and only $B$ and $A$ are trainable. Transition layers (green), triangle update layers (yellow), outer product mean (purple), and other components stay frozen. With rank $r=4$, trainable parameters drop from 93M to 2.3M (40× reduction).}
\label{fig:lora_architecture}
\end{figure}

Figure~\ref{fig:lora_architecture} shows our LoRA adaptation. We apply LoRA to the query, key, value, and output projections in all three attention types, while keeping all other components frozen. With rank $r = 4$, this gives about 2.3M trainable parameters compared to 93M for full fine-tuning---a 40× reduction.

\section{Experimental Setup}
\label{sec:ch5_implementation}

\subsection{Dataset}

We collected antibody-antigen structures from SAbDab with resolution $\leq 3.0$ \AA, R-factor $\leq 0.3$ (when available), complete CDR regions without missing residues, and PDB release dates before September 30, 2021 (the AlphaFold2 training cutoff). These filters gave 3,842 antibody-antigen complexes for training.

To increase diversity and avoid redundancy, we used a two-stage clustering strategy. We first clustered structures at 40\% sequence identity using MMseqs2, considering both antigen sequence and IMGT-numbered heavy chain CDR3 sequences. This ensures that antibodies targeting similar epitopes with different paratopes are treated as distinct. During training, we sampled clusters uniformly, then sampled members within clusters. This gives exposure to diverse binding modes while avoiding over-representation of similar complexes.

A validation set of 567 complexes released between September 30, 2021 and January 13, 2023 was used for model selection and hyperparameter tuning. For evaluation, we use the temporally separated test subset described in Section~\ref{sec:ch4_comparison}, with PDB release dates between January 14, 2023 and January 1, 2024 and antigen length $\leq 600$ residues. This temporal split simulates real-world deployment, where models must predict structures of newly discovered complexes that were not available during training. The test targets are selected after MMseqs2 clustering on antigen and CDR-H3 sequences to maintain separation from training clusters, and targets with incomplete CDR or epitope regions are excluded.

\subsection{Training}

We fine-tuned AlphaFold2-Multimer model\_4, selected for its superior validation set performance among the five available model variants. We used the Adam optimizer with $\beta_1 = 0.9$ and $\beta_2 = 0.999$, setting the learning rate to $10^{-4}$ with a cosine annealing schedule. Due to GPU memory constraints, we used a batch size of 1 complex per GPU with gradient accumulation over 8 steps, yielding an effective batch size of 8. Training was performed on 8 NVIDIA A100 GPUs with 80GB memory each, using mixed precision training to reduce memory consumption and improve throughput~\citep{micikevicius2018}.

We use AlphaFold2-Multimer's standard loss function:
\begin{equation}
\mathcal{L} = \mathcal{L}_{\text{FAPE}} + \mathcal{L}_{\text{distogram}} + \mathcal{L}_{\text{masked MSA}}
\end{equation}
where $\mathcal{L}_{\text{FAPE}}$ is the Frame Aligned Point Error loss measuring geometric accuracy of predicted structures, $\mathcal{L}_{\text{distogram}}$ penalizes incorrect distance distributions between residue pairs, and $\mathcal{L}_{\text{masked MSA}}$ trains the model to predict masked MSA positions (maintaining the self-supervised pretraining objective). We retain the default loss weights without modification.

\subsection{Evaluation}

For each test complex, we generate 5 predictions using AlphaFold2-Multimer's default pipeline, including standard MSA generation, inter-chain pairing, and 4 recycling iterations. Predictions are ranked by AlphaFold2-Multimer's ranking score: $0.8 \times \text{ipTM} + 0.2 \times \text{pTM}$, which weights interface confidence higher than global confidence. We report top-1 performance using DockQ as the main evaluation metric, following Chapter~\ref{ch:background}. For antibody-antigen DockQ calculation, antibody chain(s), including heavy and light chains when both are present, are treated as a single receptor group and evaluated against the antigen chain(s). DockQ combines fraction of native contacts, L-RMSD, and I-RMSD into a single score from 0 to 1, with success thresholds at High (DockQ $\geq 0.8$), Medium ($0.49 \leq$ DockQ $< 0.8$), and Acceptable ($0.23 \leq$ DockQ $< 0.49$). We report success rate as the percentage with DockQ $\geq 0.23$ (acceptable or better).

The unadapted AlphaFold2-Multimer model\_4 baseline uses the same evaluation protocol, ranking function, and input sequences as the LoRA-adapted model. The cross-method comparison with other predictors is reported in Section~\ref{sec:ch4_comparison}.

\chapter{Supplementary Material for Chapter 4: Protocols and Algorithms}
\label{app:pipeline_details}

This appendix provides supporting material for Chapter~\ref{ch:pipeline_adaptation}: the evaluation protocols used throughout, validation-set ablations, and full algorithm pseudocode for the method components.

\section{Evaluation Protocols}
\label{sec:appB_protocols}

The chapter uses two evaluation protocols. Each protocol serves a different reporting purpose:

\paragraph{Five-sample comparison protocol.}
Each target is evaluated with a single seed and five prediction candidates. For AlphaFold3-based methods these candidates correspond to five diffusion samples; for other baselines they correspond to the method's native five-candidate sampling or reporting convention. The top-1 prediction is selected within these 5 candidates. This protocol is used in Section~\ref{sec:ch4_comparison} (Table~\ref{tab:main_comparison}) to enable direct comparison with single-seed baselines (AlphaFold2-Multimer, LoRA-AF2, Chai-1, Protenix, Boltz-1) which report 5 predictions per target under the same sample count.

\paragraph{Test-time-scaling protocol.}
Each target is predicted with 20 independent random seeds, with each seed generating 5 diffusion samples (100 predictions per target). The top-1 prediction is selected by the AF3Complex interface confidence score (pIS) across all 100 candidates. This protocol is used in Section~\ref{sec:ch6_results} for the main results.

\paragraph{Test set scope.}
Both protocols are evaluated on the 76-target subset of the temporal test set, restricted to targets with antigen length $\le 600$ residues for computational tractability under the 100-prediction-per-target budget.

\paragraph{DockQ convention.}
DockQ is computed for the antibody-antigen interface by treating antibody chain(s), including heavy and light chains when both are present, as a single receptor group and evaluating them against the antigen chain(s).

\paragraph{Single-seed selection for AlphaFold3-based reporting in Table~\ref{tab:main_comparison}.}
For direct comparison with single-seed baselines, AlphaFold3-based methods report results from seed~0 in Table~\ref{tab:main_comparison}.

\section{Validation Ablations}
\label{sec:appB_ablations}

During method development we validated the contribution of each pipeline component on the validation set, with results summarized in Table~\ref{tab:val_ablation}.

\begin{table}[htbp]
\centering
\caption{Validation-set ablations of the method components. ``Ours'' applies all three components (CDR-focused MSA filtering, depth recovery, convergence-aware recycling); each ``$-$'' row removes one component while keeping the others in place. ``Baseline'' is AlphaFold3 without any of the proposed modifications. Numbers report DockQ-based success rates at the medium and acceptable thresholds as fractions.}
\label{tab:val_ablation}
\begin{tabular}{lcc}
\toprule
\textbf{Configuration} & \textbf{Medium} & \textbf{Acceptable} \\
\midrule
Ours & \textbf{0.52} & \textbf{0.60} \\
$-$ convergence-aware recycling & 0.49 & 0.58 \\
$-$ depth recovery & 0.50 & 0.56 \\
$-$ CDR-focused filtering & 0.50 & 0.57 \\
Baseline & 0.48 & 0.54 \\
\bottomrule
\end{tabular}

\end{table}

These ablations were performed on the validation set and were used for method development and component selection, not for final test-set claims. The relative contribution of the three components is roughly comparable, with the full combination providing the largest gain.

\section{Algorithm Pseudocode}
\label{sec:appB_algorithms}

This section provides the full pseudocode for the two main method components introduced in Chapter~\ref{ch:pipeline_adaptation}. The main text describes the methods in terms of their motivation, criteria, and effects on MSA composition or inference behavior; the procedural details below complement that description.

\subsection{CDR3-Focused MSA Filtering}
\label{sec:appB_cdr3_filter_algo}

Algorithm~\ref{alg:cdr3-filter} formalizes the CDR3-focused MSA filtering procedure described in Section~\ref{sec:ch6_msa_refinement}.

\begin{algorithm}[H]
    \singlespacing
    \caption{CDR3-Focused MSA Filtering for Antibody Chains}\label{alg:cdr3-filter}
    \begin{algorithmic}
    \Require Fold input $\mathcal{I}$ with chains and MSAs; numbering scheme (e.g., IMGT); offsets $[o_\text{start}, o_\text{end}]$
    \Function{FilterAntibodyMSA}{$\mathcal{I}, [o_\text{start}, o_\text{end}]$}
      \State $\mathcal{C}' \leftarrow [\,]$
      \For{each chain $c \in \mathcal{I}.\text{chains}$}
        \If{$c$ is not a ProteinChain}
          \State append $c$ to $\mathcal{C}'$; \textbf{continue}
        \EndIf
        \State $(\text{regions},\_, \text{chain\_type},\_) \leftarrow \textsc{NumberingChain}(c.\text{sequence}, \text{IMGT})$
        \State $\text{is\_ab} \leftarrow \big(\text{chain\_type} \in \{H, L, K\}\big)$
        \If{not $\text{is\_ab}$ or no MSA present}
          \State append $c$ to $\mathcal{C}'$; \textbf{continue}
        \EndIf
        \If{$\text{CDR3} \notin \text{regions}$}
          \State append $c$ to $\mathcal{C}'$; \textbf{continue}
        \EndIf
        \State $s \leftarrow \min(\text{regions}[\text{CDR3}].\text{index})$; \quad $e \leftarrow \max(\text{regions}[\text{CDR3}].\text{index}) + 1$
        \If{$s + o_\text{start} \ge e - o_\text{end}$}
          \State append $c$ to $\mathcal{C}'$; \textbf{continue} \hfill// CDR3 window too small for offset
        \EndIf
        \State $[w_\text{start}, w_\text{end}] \leftarrow [\max(0, s - o_\text{start}),\ e + o_\text{end}]$ \hfill// clamp to sequence length as needed
        \State $M \leftarrow \text{MSA associated with } c$ \hfill// abstract over paired/unpaired
        \If{$M \ne \varnothing$}
          \State $M' \leftarrow [\,]$
          \For{each row (sequence) $r \in M$}
            \State $u \leftarrow r[w_\text{start}:w_\text{end}]$ \hfill// window over CDR3 with offsets
            \If{$\exists\, \text{residue}\ \in u:\ \text{residue} \ne '-'$} \hfill// keep if not all gaps
              \State append $r$ to $M'$
            \EndIf
          \EndFor
          \State replace the chain's MSA with $M'$ in $c$
        \EndIf
        \State append updated $c$ to $\mathcal{C}'$
      \EndFor
      \State \Return input with chains $\mathcal{C}'$
    \EndFunction
    \end{algorithmic}
    \end{algorithm}

\subsection{Convergence-Aware Recycling with Mini Rollout}
\label{sec:appB_adaptive_recycle_algo}

Algorithm~\ref{alg:early-stop} formalizes the convergence-aware recycling procedure with mini rollouts described in Section~\ref{sec:ch6_adaptive_recycling}. Orange-coloured lines highlight modifications relative to standard AlphaFold3 recycling.

\begin{algorithm}[H]
    \singlespacing
    \caption{Early Stop with Mini Rollout}\label{alg:early-stop}
    \begin{algorithmic}
        \State     $\{\mathbf{s}_i^{\text{inputs}}\} = \text{InputFeatureEmbedder}(\{\mathbf{f}^*\})$
        \State $\mathbf{s}_i^{\text{init}} = \text{LinearNoBias}(\mathbf{s}_i^{\text{inputs}})$
        \State $\mathbf{z}_{ij}^{\text{init}} = \text{LinearNoBias}(\mathbf{s}_i^{\text{inputs}}) + \text{LinearNoBias}(\mathbf{s}_j^{\text{inputs}})$
        \State $\mathbf{z}_{ij}^{\text{init}} += \text{RelativePositionEncoding}(\{\mathbf{f}^*\}) $
        \State $\mathbf{z}_{ij}^{\text{init}} += \text{LinearNoBias}(\mathbf{f}_{ij}^{\text{token\_bonds}})$
        \State $\{\hat{\mathbf{z}}_{ij}\}, \{\hat{\mathbf{s}}_i\} = \mathbf{0}, \mathbf{0}$
        \State $\textcolor{orange}{\{\mathbf{z}^{\text{min}}_{ij}\}, \{\mathbf{s}^{\text{min}}_i\} = \mathbf{0}, \mathbf{0}}$
        \State $\textcolor{orange}{d^{\text{prev}}_{ij} = \varnothing,\quad D^{\text{min}} = \infty,\quad C = 1}$
        \While {$C \le N_{\text{recycle}} \land \textcolor{orange}{D^{\text{min}} > D^{\text{threshold}}}$}
            \State $\mathbf{z}_{ij} = \mathbf{z}_{ij}^{\text{init}} + \text{LinearNoBias}(\text{LayerNorm}(\hat{\mathbf{z}}_{ij})) $
            \State $\{\mathbf{z}_{ij}\} += \text{TemplateEmbedder}(\{\mathbf{f}^*\}, \{\mathbf{z}_{ij}\}) $
            \State $\{\mathbf{z}_{ij}\} += \text{MsaModule}(\{\mathbf{f}_{S_i}^{\text{msa}}\}, \{\mathbf{z}_{ij}\}, \{\mathbf{s}_i^{\text{inputs}}\}) $
            \State $\mathbf{s}_i = \mathbf{s}_i^{\text{init}} + \text{LinearNoBias}(\text{LayerNorm}(\hat{s}_i)) $
            \State $\{\mathbf{s}_i\}, \{\mathbf{z}_{ij}\} = \text{PairformerStack}(\{\mathbf{s}_i\}, \{\mathbf{z}_{ij}\}) $
            \State $\{\hat{\mathbf{s}}_i\}, \{\hat{\mathbf{z}}_{ij}\} \leftarrow \{\mathbf{s}_i\}, \{\mathbf{z}_{ij}\}$
            \State $\textcolor{orange}{\{\vec{\mathbf{x}}_l^{\text{rollout}}\} = \text{MiniRollout}(\{\mathbf{f}^*\}, \{\mathbf{s}_i^{\text{inputs}}\}, \{\mathbf{s}_i\}, \{\mathbf{z}_{ij}\})}$
            \State $\textcolor{orange}{d_{ij} = \left\lVert \vec{\mathbf{x}}_{l_{\text{rep}}(i)}^{ \text{rollout}} - \vec{\mathbf{x}}_{l_{\text{rep}}(j)}^{ \text{rollout}} \right\rVert}$
            \If{$\textcolor{orange}{d^{\text{prev}}_{ij} = \varnothing}$}
                \State $\textcolor{orange}{d^{\text{prev}}_{ij} \leftarrow d_{ij}}$ \hfill// first rollout initializes reference
                \State $\textcolor{orange}{\{\mathbf{s}^{\text{min}}_i\}, \{\mathbf{z}^{\text{min}}_{ij}\} \leftarrow \{\mathbf{s}_i\}, \{\mathbf{z}_{ij}\} }$
            \Else
                \State $\textcolor{orange}{D = \text{RMS}(d_{ij} - d^{\text{prev}}_{ij}) }$
                \State $\textcolor{orange}{d^{\text{prev}}_{ij} \leftarrow d_{ij}}$
                \If{$\textcolor{orange}{D < D^{\text{min}}}$}
                    \State $\textcolor{orange}{D^{\text{min}} \leftarrow D}$
                    \State $\textcolor{orange}{\{\mathbf{s}^{\text{min}}_i\}, \{\mathbf{z}^{\text{min}}_{ij}\} \leftarrow \{\mathbf{s}_i\}, \{\mathbf{z}_{ij}\} }$
                \EndIf
            \EndIf
            \State $\textcolor{orange}{C \leftarrow C + 1}$
        \EndWhile
        \State $\{\vec{\mathbf{x}}_l^{\text{pred}}\} = \text{SampleDiffusion}(\{\mathbf{f}^*\}, \{\mathbf{s}_i^{\text{inputs}}\}, \{\textcolor{orange}{\mathbf{s}^{\text{min}}_i}\}, \{\textcolor{orange}{\mathbf{z}^{\text{min}}_{ij}}\}) $
        \State $\{\mathbf{p}_l^{\text{plddt}}\}, \{\mathbf{p}_{ij}^{\text{pae}}\}, \{\mathbf{p}_{ij}^{\text{pde}}\}, \{\mathbf{p}_l^{\text{resolved}}\} = \text{ConfidenceHead}(\{\mathbf{s}_i^{\text{inputs}}\}, \{\textcolor{orange}{\mathbf{s}^{\text{min}}_i}\}, \{\textcolor{orange}{\mathbf{z}^{\text{min}}_{ij}}\}, \{\vec{\mathbf{x}}_l^{\text{pred}}\})$
        \State $\mathbf{p}_{ij}^{\text{distogram}} = \text{DistogramHead}(\textcolor{orange}{\mathbf{z}^{\text{min}}_{ij}})$
    \end{algorithmic}
    \end{algorithm}

\section{Mean DockQ across Inference Budget}
\label{sec:appB_bootstrap_curve}

\begin{figure}[htbp]
\centering
\includegraphics[width=0.7\textwidth]{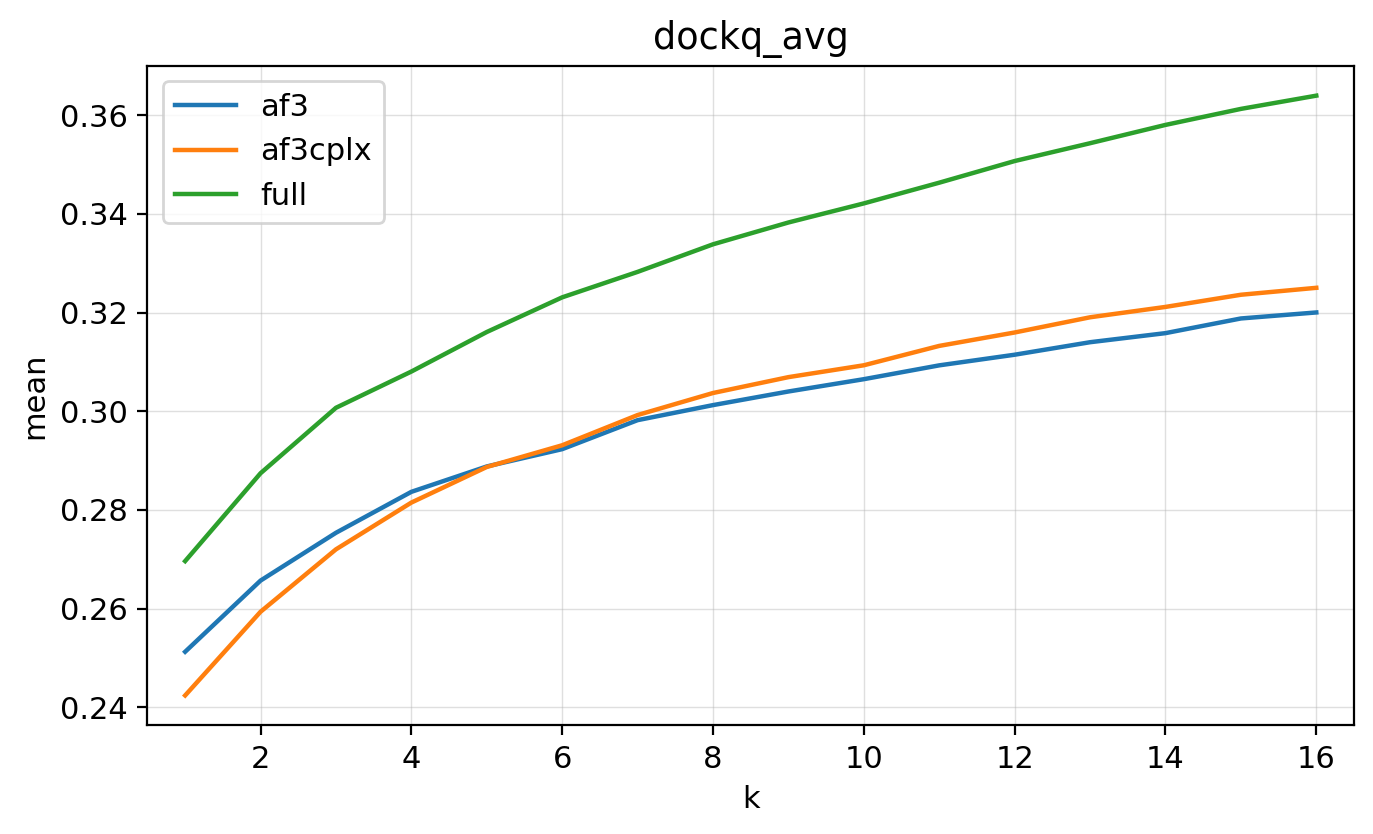}
\caption{Mean DockQ on the 76-target test subset as a function of the seed budget used for top-1 ranking ($k = 1$ to $16$, since at $k = 20$ only one subset---the full set---is possible). Each point is the average over 1{,}000 random samples of $k$ seeds from the 20 available seeds; the top-1 prediction within each sampled seed budget is selected by AF3Complex pIS. \texttt{af3} = AlphaFold3 baseline; \texttt{af3cplx} = AF3Complex; \texttt{full} = our complete method (MSA refinement + convergence-aware recycling).}
\label{fig:appB_avg_dockq}
\end{figure}

Figure~\ref{fig:appB_avg_dockq} complements the main-text seed-ranking comparison (Figure~\ref{fig:seed_ranking_comparison}, which reports success rates) by showing mean DockQ over the 76-target test subset as a function of the seed budget used for top-1 ranking. The same averaging procedure as in the main text is used: each point is the mean over 1{,}000 random samples of $k$ seeds from the 20 available seeds, with the top-1 prediction within each sampled seed budget selected by AF3Complex pIS.

\section{Performance vs CDR Sequence Identity to Training Data}
\label{sec:appB_identity_analysis}

Antibody-antigen prediction quality could in principle depend on how similar a test target is to structures the model has seen during training. We therefore checked whether our method's improvement over the AlphaFold3 baseline is concentrated among targets with high CDR similarity to the training set. For each test target, concatenated CDR identity is the highest MMseqs2~\citep{steinegger2017mmseqs2} hit against the training-set antibody chains after concatenating the IMGT-defined CDR1, CDR2, and CDR3 regions; for two-chain antibodies, the maximum over available antibody chains is used.

Table~\ref{tab:appB_cdr_identity_bins} stratifies the 76-target test subset into three roughly equal-sized groups by concatenated CDR identity to the training set. Mean $\Delta$DockQ remains positive and similar across all three groups, indicating that the gain over AlphaFold3 is not concentrated among targets with the highest CDR similarity to training structures. DockQ is computed for the antibody-antigen interface by treating antibody chain(s) as one receptor group against the antigen chain(s).

\begin{table}[htbp]
\centering
\caption{Mean $\Delta$DockQ (our method minus AlphaFold3 baseline) stratified by concatenated CDR identity to the training set. Identity bins use fixed thresholds on the best MMseqs2 hit over concatenated CDR sequences.}
\label{tab:appB_cdr_identity_bins}
\begin{tabular}{lcc}
\toprule
\textbf{Concatenated CDR identity} & \textbf{Targets} & \textbf{Mean $\Delta$DockQ} \\
\midrule
$<70\%$       & 25 & $+0.033$ \\
$70$--$85\%$ & 24 & $+0.054$ \\
$\ge 85\%$   & 27 & $+0.040$ \\
\bottomrule
\end{tabular}
\end{table}

\bibliographystyle{unsrtnat}
\bibliography{proposal_refs}

\end{document}